\documentclass[12pt,english]{article}
\usepackage{avant}
\usepackage[T1]{fontenc}
\usepackage[utf8]{inputenc}
\usepackage{geometry}
\usepackage{babel}
\usepackage{amsmath}
\usepackage{amssymb}
\usepackage{stmaryrd}
\usepackage[authoryear]{natbib}
\usepackage{algorithm}
\usepackage{graphicx}

\usepackage{float}
\usepackage{booktabs} 
\usepackage{setspace}
\usepackage{subcaption}
\usepackage{multirow}
\usepackage{multicol}
\usepackage{threeparttable}
\usepackage[unicode=true,
bookmarks=false,
breaklinks=false,pdfborder={0 0 1},backref=false,colorlinks=false]
{hyperref}
\usepackage{breakurl}

\makeatletter
\@ifundefined{date}{}{\date{}}

\usepackage{array}
\newcolumntype{L}[1]{>{\raggedright\let\newline\\\arraybackslash\hspace{0pt}}m{#1}}
\newcolumntype{C}[1]{>{\centering\let\newline\\\arraybackslash\hspace{0pt}}m{#1}}
\newcolumntype{R}[1]{>{\raggedleft\let\newline\\\arraybackslash\hspace{0pt}}m{#1}}
\usepackage{float}
\usepackage{multirow}
\usepackage{breakurl}

\providecommand{\tabularnewline}{\\}
\floatstyle{ruled}

\usepackage{xr}
\makeatletter
\newcommand*{\addFileDependency}[1]{
	\typeout{(#1)}
	\@addtofilelist{#1}
	\IfFileExists{#1}{}{\typeout{No file #1.}}
}
\makeatother

\providecommand{\tabularnewline}{\\}
\floatstyle{ruled}

\usepackage{amsthm}\usepackage{bm}\usepackage{rotating}\usepackage{array}\usepackage{booktabs}\usepackage{color}
\usepackage{multirow}\usepackage{enumerate}\usepackage{caption}
\usepackage{float}
\usepackage{multirow}
\providecommand{\tabularnewline}{\\}
\floatstyle{ruled}
\newfloat{algorithm}{tbp}{loa}
\providecommand{\algorithmname}{Algorithm}
\floatname{algorithm}{\protect\algorithmname}

\usepackage{bm}\usepackage{rotating}\usepackage{array}\usepackage{booktabs}\usepackage{color}\usepackage{algorithmic}
\usepackage{algorithm}\usepackage{multirow}\usepackage{enumerate}\usepackage{caption}\usepackage{subcaption}
\floatstyle{ruled}
\newfloat{algorithm}{tbp}{loa}
\providecommand{\algorithmname}{Algorithm}
\floatname{algorithm}{\protect\algorithmname}

\topmargin by -\headheight \advance
\topmargin by -\headsep     

\evensidemargin
\oddsidemargin
\providecommand{\tabularnewline}{\\}

\RequirePackage[displaymath]{lineno}

\usepackage{amsfonts}\usepackage{latexsym}

\usepackage{mathrsfs}\usepackage{amsthm}\usepackage{latexsym}\usepackage{booktabs}

\usepackage{times}

\DeclareMathOperator*{\argmin}{argmin}

\newtheorem{proposition}{{\bf Proposition}}
\newtheorem{lemma}{{\bf Lemma}}

\newtheorem{definition}{{\bf Definition}}
\newtheorem{theorem}{{\bf Theorem}}

\global\long\def\mbA{\mathbf{A}}

\global\long\def\mbB{\mathbf{B}}
\global\long\def\mbc{\mathbf{c}}
\global\long\def\mbC{\mathbf{C}}

\global\long\def\mbE{\mathbf{E}}

\global\long\def\mbG{\mathbf{G}}

\global\long\def\mbH{\mathbf{H}}

\global\long\def\mbI{\mathbf{I}}

\global\long\def\mbJ{\mathbf{J}}

\global\long\def\mbR{\mathbf{R}}

\global\long\def\mbS{\mathbf{S}}

\global\long\def\mbU{\mathbf{U}}

\global\long\def\mbV{\mathbf{V}}

\global\long\def\mbX{\mathbf{X}}

\global\long\def\mbY{\mathbf{Y}}

\global\long\def\mbZ{\mathbf{Z}}

\global\long\def\hatmbB{\widehat{\mathbf{B}}}

\global\long\def\bolalpha{\boldsymbol{\alpha}}

\global\long\def\boldelta{\boldsymbol{\delta}}

\global\long\def\bolmu{\boldsymbol{\mu}}

\global\long\def\bolXi{\boldsymbol{\Xi}}

\global\long\def\bolDelta{\boldsymbol{\Delta}}
\global\long\def\bolTheta{\boldsymbol{\Theta}}

\global\long\def\bolSigma{\boldsymbol{\Sigma}}
\global\long\def\bolPhi{\boldsymbol{\Phi}}

\global\long\def\hatbolmu{\widehat{\boldsymbol{\mu}}}

\global\long\def\hatbolXi{\widehat{\boldsymbol{\Xi}}}

\global\long\def\hatbolSigma{\widehat{\boldsymbol{\Sigma}}}

\global\long\def\hatbolOmega{\widehat{\boldsymbol{\Omega}}}

\global\long\def\tilbolSigma{\widetilde{\boldsymbol{\Sigma}}}

\global\long\def\td{\textsl{t}}

\global\long\def\mbbR{\mathbb{R}}
\global\long\def\mbbP{\mathbb{P}}

\global\long\def\mbbX{\mathbb{X}}
\global\long\def\mbbY{\mathbb{Y}}
\global\long\def\mbbW{\mathbb{W}}

\global\long\def\calA{\mathcal{A}}

\global\long\def\calE{\mathcal{E}}

\global\long\def\calG{\mathcal{G}}

\global\long\def\calM{\mathcal{M}}

\global\long\def\calP{\mathcal{P}}

\global\long\def\calS{\mathcal{S}}

\global\long\def\E{\mathrm{E}}

\global\long\def\APT{\mathrm{APT}}
\global\long\def\OST{\mathrm{OST}}

\global\long\def\AR{\mathrm{AR}}
\global\long\def\APN{\mathrm{APN}}
\global\long\def\APL{\mathrm{APL}}
\global\long\def\REE{\mathrm{REE}}
\global\long\def\TPR{\mathrm{TPR}}
\global\long\def\FPR{\mathrm{FPR}}
\global\long\def\AGPT{\mathrm{AGPT}}

\global\long\def\TN{\mathrm{TN}}
\global\long\def\HOST{\mathrm{HOST}}
\global\long\def\TT{\mathrm{TT}}
\global\long\def\MVT{\mathrm{MVT}}

\global\long\def\L{\mathrm{L}}
\global\long\def\F{\mathrm{F}}

\global\long\def\H{\mathrm{H}}

\global\long\def\Cov{\mathrm{cov}}

\global\long\def\vecc{\mathrm{vec}}

\global\long\def\tr{\mathrm{tr}}

\global\long\def\OLS{\mathrm{OLS}}

\global\long\def\logg{\mathrm{log}}
\global\long\def\exp{\mathrm{exp}}

\global\long\def\tr{\mathrm{tr}}

\global\long\def\N{\mathrm{N}}

\global\long\def\Wis{\mathrm{Wishart}}

\begin{document}
	{\setlength{\baselineskip}{1.25\baselineskip} 

\title{High-dimensional Tensor Response Regression using the $\td$-Distribution}


\author{Ning Wang, Xin Zhang, and Qing Mai\thanks{Ning Wang ({ningwangbnu@bnu.edu.cn}) is Assistant Professor, Center of Statistics and Data Science, Beijing Normal University, Zhuhai, 519807, China; Xin Zhang (xzhang8@fsu.edu) is Associate Professor,  Department of Statistics, Florida State University, Tallahassee, 32312, Florida, USA;Qing Mai (qmai@fsu.edu) is Associate Professor,  Department of Statistics, Florida State University, Tallahassee, 32312, Florida, USA. }}


\maketitle

\begin{abstract}
In recent years, promising statistical modeling approaches to tensor data analysis have been rapidly developed. Traditional multivariate analysis tools, such as multivariate regression and discriminant analysis, are generalized from modeling random vectors and matrices to higher-order random tensors. 
One of the biggest challenges to statistical tensor models is the non-Gaussian nature of many real-world data. Unfortunately, existing approaches are either restricted to normality or implicitly using least squares type objective functions that are computationally efficient but sensitive to data contamination. Motivated by this, we adopt a simple tensor $\td$-distribution that is, unlike the commonly used matrix $\td$-distributions, compatible with tensor operators and reshaping of the data. We study the tensor response regression with tensor $\td$-error, and develop penalized likelihood-based estimation and a novel one-step estimation. We study the asymptotic relative efficiency of various estimators and establish the one-step estimator's oracle properties and near-optimal asymptotic efficiency. We further propose a high-dimensional modification to the one-step estimation procedure and show that it attains the minimax optimal rate in estimation. Numerical studies show the excellent performance of the one-step estimator.
\end{abstract}

  Adaptive lasso, High-dimensional regression, MM algorithm, Robust statistics, Tensor analysis.

\section{Introduction}
A dataset or random variable arranged into the format of multidimensional array is called a tensor. Analysis of tensor data is driven by various modern scientific and engineering problems, including neuroimaging data analysis \citep{zhou2013tensor, karahan2015tensor}, statistical genetics \citep{hore2016tensor}, graphical models \citep{greenewald2019tensor}, recommender systems \citep{bi2018multilayer}, sufficient dimension reduction \citep{li2010dimension}, relational and network data \citep{hoff2015multilinear}, and time series data \citep{chen2019factor,wang2022high}. See \citet{bi2020tensors} for a very recent overview of statistical tensor analysis.

Since the beginning of tensor analysis \citep{hitchcock1927expression, carroll1970analysis, kruskal1977three}, there has been tremendous progress in {tensor decompositions}, from both applied mathematics \citep{kol2009} and machine learning \citep{sidiropoulos2017tensor}. There has also been a rapidly growing literature on building statistical models for the analysis of tensor data, for example, on low-rank decompositions \citep{sun2017provable,zhang2018tensor, zhang2019optimal,han2023guaranteed}, tensor regression \citep{zhou2013tensor,hoff2015multilinear,rte,lock2018tensor,convex2019,chen2019non,hao2021sparse,zhou2023partially}, and tensor classification and clustering \citep{DWD,catch,sun2019dynamic,han2022exact,luo2022tensor,mai2022doubly,cai2021jointly}.
However, an important but less studied research direction is the distributions of random tensors, especially beyond normality. In the past decade, various statistical models and methods have been proposed for characterizing tensor data and for modeling relationships between tensors and other variables (e.g., categorical or multivariate). Many extensions of classical multivariate analysis to high-dimensional tensor analysis, not surprisingly, rely on normality assumptions. A particularly popular choice is the tensor normal distribution, which assumes a separable Kronecker product covariance structure. For example, \citet{hoff2011separable} was one of the earliest works using the tensor normal distribution, with applications to modeling multivariate longitudinal network data; \citet{rte} proposed a parsimonious tensor response regression model with tensor normal errors; \citet{catch} proposed a tensor discriminant analysis model in high dimensions; \citet{he2014graphical} and \citet{sun2015non} studied sparse tensor Gaussian graphical models. The tensor normal distribution is popular in statistics because of its theoretical and computational simplicity, partly due to the parsimonious and interpretable covariance structure. 

In this paper, we introduce a tensor $\td$-distribution that is compatible with the classical multivariate $\td$-distribution and the above-mentioned tensor normal distribution. The separable Kronecker covariance structure is employed to keep the parsimonious and interpretable dependence structure in tensor variables. A single scalar degrees of freedom parameter is used to characterize the heavy-tailed behaviors. This simple tensor $\td$-distribution is fundamentally different from the common matrix-variate $\td$-distributions in the literature \citep{dickey1967, thompson2020, zhang2010multi, mvd}. There are two main challenges with the common matrix $\td$-distribution that are nagging its extensions to higher-order tensor analysis. First, it is incompatible with vectorization operator. If we reshape a matrix-$\td$ into a vector by stacking its columns together, the resulting vector is no longer multivariate-$\td$. Our tensor $\td$-distribution resolves this issue, and is still within the same tensor $\td$-distribution family after various tensor operators such as vectorization, linear transformation, rotation, and sub-tensor extraction. Second, the latent variable representation of matrix $\td$-distribution involves Wishart  distributions that can be computationally expensive since the Expectation-Maximization (EM) algorithm \citep{dempster1977maximum} or other likelihood-based estimates are often used in $\td$-modeling. In contrast, the proposed tensor $\td$-modeling approach, which is scalable to high-dimensional data analysis, is computationally much simpler with a scalar latent variable from the Gamma distribution regardless of the order of the tensor.
\begin{figure}[t]
	\centering
	\makebox{\includegraphics[scale=0.40]{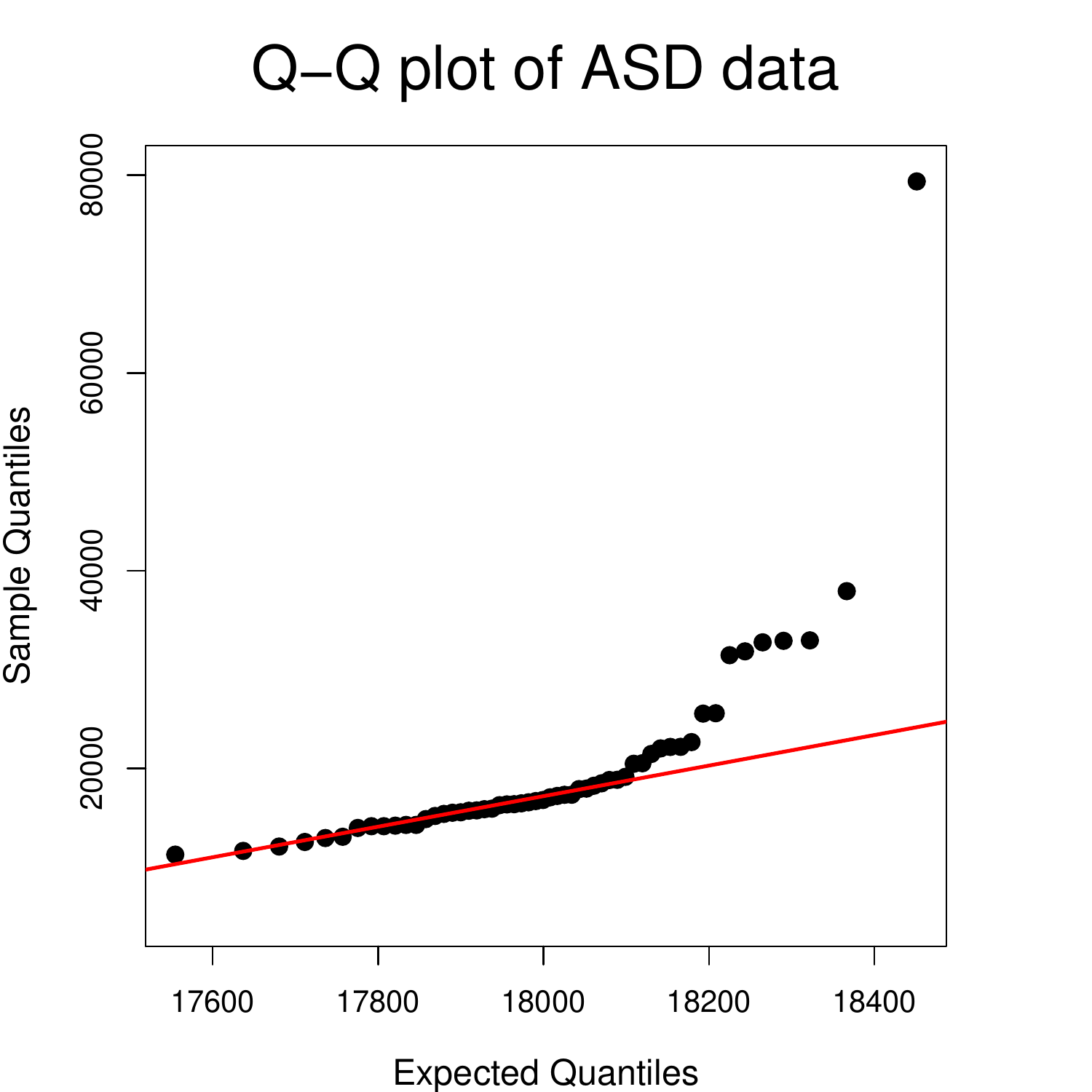}
		\includegraphics[scale=0.40]{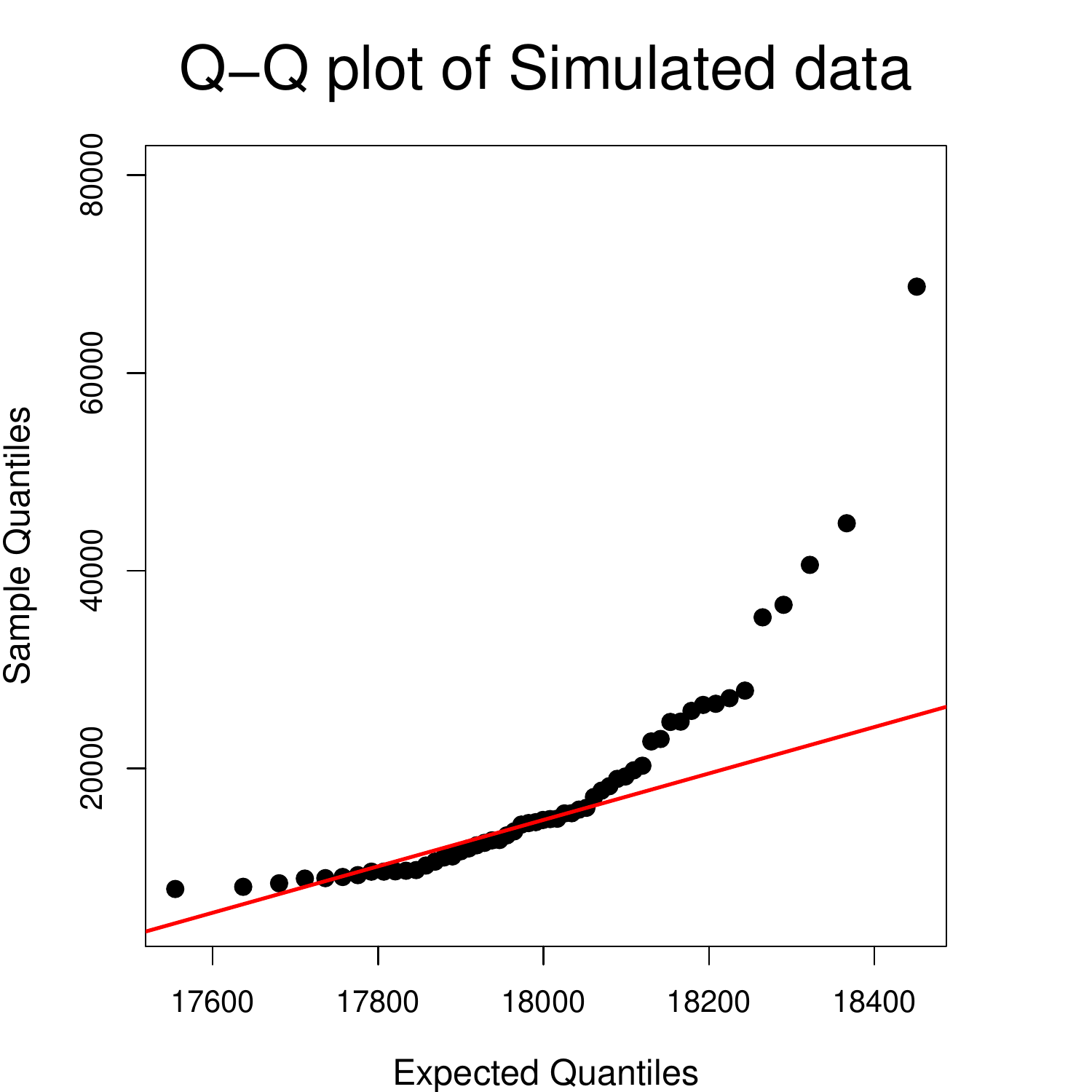}}
	\caption{Quantile-Quantile (Q-Q) plots for a real neuroimaging dataset (left panel) with $n=55$ third-order tensors each of dimension $p_1\times p_2\times p_3=30\times 30\times 20$, and a simulated data of the same dimension and sample size from the tensor $\td$-distribution (right panel) with degrees of freedom $1.5$. }
	\label{QQ}
\end{figure}

As a motivating data example, we consider the functional magnetic resonance
imaging (fMRI) scans from an Autism Spectrum Disorder (ASD) study, where the brain images of 55 subjects are preprocessed and arranged into tensors of dimension $p_1\times p_2\times p_3=30\times 30\times 20$. Details of the study are included in the later real data analysis (Section~\ref{real}). In this study other clinical covariates such as group indicator, age and sex are also provided. It is of great scientific interest to model how the brain structure and functions are affected by the disorder after adjusting for other covariates. A recent approach is the tensor response regression that uses the image as response and all other covariates as predictor \citep{higher, rte,sun2017store}, where normality assumption is either explicitly imposed or implicitly used via the least squares estimation. 
We draw the QQ plot for the ASD data set to check its normality, specifically, we first regress the response tensor in predictors using least squares estimation, and then standardize the residual tensor by its covariance matrices. If we treat the ASD data as normally distributed, the square norm of the standardized residual follows the chi-squared distribution with $p=30\times 30\times 20=18000$ degrees of freedom. In Figure~\ref{QQ}, we plot the empirical quantiles versus the theoretical quantiles (from the $\chi^2$-distribution). The heavy-tailed behavior is clear, and the potential outliers are possibly due to poor imaging quality or problematic image registration.
For comparison, we also simulated data from the proposed tensor $\td$-distribution with the same dimension and sample size. The simulated data mimics the real data behavior well when the degrees of freedom is specified as $1.5$, which corresponds to quite an extremely heavy-tailed distribution.
Although illustrated with this neuroimaging data, the proposed $\td$-modeling strategy is equally applicable to various applications and models.

We apply our tensor $\td$-modeling to the tensor response regression problem with regularization. The new approach is not just robust to heavy tails and outliers, but also equipped with formal likelihood-based inference and consistent variable selection. 
Similar to the classical $\td$-modeling in \citet{little1988robust,robust1989}, the tensor $\td$-modeling has the advantage of an explicit parametric model assumption that is interpretable and therefore an inferential procedure for the estimated parameters. An advantage of such approaches is that, the $\td$-distributional assumption often provides a more accurate standard error estimate than the outlier detection-and-elimination approaches that do not account for the variability in samples. For the same reason, { the $\td$-distribution has its advantages over robust loss function approaches, such as the Huber loss \citep{huber1992robust}, when the $\td$-distributional assumption holds. } After a novel majorization and one-step approximation to the penalized likelihood, our objective function is also as interpretable as the robust loss functions.


{This article includes several significant contributions. 

First of all, we propose a general modeling strategy using the tensor $\td$-distribution and a robust tensor response regression model that is robust to outliers. Although statistical approaches to tensor decomposition \citep{sun2017provable}, clustering and co-clustering \citep{wang2019multiway}, and completion \citep{zhang2019cross} do not necessarily require the normality assumption, they may suffer from heavy-tailed data and outliers. This article thus provides insights for future research in tensor analysis from the $\td$-distributional perspective. 

Secondly, we develop a complete set of estimation methods. We start with the maximum likelihood estimation via the EM algorithm and the penalized likelihood estimation via Majoration-Maximization (MM) algorithm \citep[see][for background]{MM2004}. However, both the EM and MM algorithms are computationally expensive in tensor $\td$-regression and likely converge to local optima due to non-convexity in the optimization. We hence propose a novel one-step estimation (OST) that is motivated by the likelihood-based objective function but is guaranteed to converge globally. This OST algorithm, which essentially solves a weighted least squares problem, is much faster than the EM and MM algorithms. Finally, when the number of parameters in our model is much larger than the sample size, we further develop a high-dimensional one-step estimator (HOST) that is computationally even more efficient and scalable. 

Thirdly, for OST, we establish its oracle properties, asymptotically valid variance estimates, and inference procedures. We study the asymptotic relative efficiency among various feasible solutions such as least squares, tensor normal likelihood-based, and the MLE and OST from the tensor-$\td$ regression; and show that the OST estimator is almost as efficient as the penalized maximum likelihood estimator (which is unattainable due to the non-convexity). For high-dimensional settings, we establish the minimax estimation rates for the tensor response regression and prove that the HOST estimator is optimal in terms of the minimax convergence rate. Theoretical insights on mis-specified $\td$-distribution's degrees of freedom are also provided.

Finally, a special case of our model is the multivariate linear model, i.e.,~when the order of the response tensor is one. For the multivariate linear model, the proposed $\td$-modeling approach is directly applicable and provides a useful addition to the popular dimension reduction approaches such as partial least square regression \citep{chun2010sparse} and reduced-rank regression \citep{izenman1975reduced}, where low-dimensional structures are often assumed. We focus on the settings where the response is non-normal and has a much higher dimension than the predictor so that response selection is critical.
}

The rest of the article is organized as follows. Section \ref{ttd} introduces the definition and some key properties of the tensor $\td$-distribution.
{Section \ref{model} introduces the robust tensor response regression model with tensor $\td$-errors and its (penalized) MLEs.
In Section~\ref{ostep}, we propose the one-step estimator (OST) and high-dimensional one-step estimator (HOST). Theoretical properties are given in Section \ref{theory}. } Extensive simulation studies and real data analysis are presented in Sections \ref{sim} and \ref{real}, respectively, and followed by a short discussion in Section \ref{sec:dis}. The Supplementary Materials contain additional numerical results, proofs, and other technical details.

\section{Tensor notation and tensor $\td$-distribution}
\label{ttd}
\subsection{Notation}

The following notation and (multi-)linear algebra will be used in this article.  Our notion of tensor analysis is different from that in mathematics and physics, although some common operators and techniques  \citep[see,][for example]{kol2009} are used to provide a concise statistical modeling framework and estimation procedures. 
We call a multidimensional array $\mbA \in \mbbR^{p_1\times\cdots\times p_M}$ an $M$-way tensor or $M$-th order tensor, while $M=1$ corresponds to vectors and $M=2$ corresponds to matrices. Some key operators on a general $M$-th order tensor $\mbA$ are defined as follows. 

{
	\begin{itemize}
		\item Vectorization. The vectorization of $\mbA$ is denoted by $\text{vec}(\mbA)\in\mbbR^{\prod_mp_m}$, where the $(i_1,\cdots, i_M)$-th scalar in $\mbA$ is mapped to the $j$-th entry of $\text{vec}(\mbA)$,  $j=1+\sum_{m=1}^{M}\{(i_m-1)\prod_{k=1}^{m-1}p_{k}\}$.
		\item Matricization. The $\textit{mode-$n$ matricization}$, reshapes the tensor $\mbA$ into a matrix denoted by $\mbA_{(n)}\in \mbbR^{p_n\times\prod_{m\neq n}p_m}$, so that the $(i_1,\cdots, i_M)$-th element in $\mbA$ becomes the $(i_n, j)$-th element of the matrix $\mbA_{(n)}$, where $j=1+\sum_{k\neq n}\{(i_k-1)\prod_{l<k, l\neq n} p_{l}\}$. 
		\item Vector product.  The $\textit{mode-$n$ vector product}$ of $\mbA$ and a vector $\mbc\in\mbbR^{p_n}$ is represented by  $\mbA \bar{\times}_{n} \mbc \in \mbbR^{p_1\times\cdots \times p_{n-1}\times p_{n+1}\times\cdots\times p_{M}}$ results in an $(M-1)$-th order tensor. This product is the result of the inner products between every \textit{mode-$n$ fiber} in $\mbA$ with vector $\mbc$.  The mode-$n$ fibers of $\mbA$ are the vectors obtained by fixing all indices except the $n$-th index. 
		\item Matrix product.  The $\textit{mode-$n$ product}$ of tensor $\mbA$ and a matrix $\mbG \in \mbbR^{s\times p_n}$, denoted as $\mbA \times_{n} \mbG$, is an $M$-th order tensor with dimension ${p_1\times\cdots\times p_{n-1}\times s\times p_{n+1}\times\cdots\times p_{M}}$. Similar to the vector product, the product is a result of multiplication between every mode-$n$ fibers of $\mbA$ and the matrix $\mbG$. 
		\item Tucker product. The $\textit{Tucker product}$ of the core tensor $\mbA$ and a series of factor matrices $\mbG_1,\dots,\mbG_M$, is defined as $\mbA\times_{1}\mbG_1\times_{2}\cdots\times_{M}\mbG_M\equiv \llbracket\mbA;\mbG_1,\ldots,\mbG_M\rrbracket$. 
		\item Inner product of two tensors with matching dimensions is $\langle\mbA,\mbB\rangle = \vecc(\mbA)^T\vecc(\mbB)$. 
	\end{itemize} 
	
}

\subsection{Tensor normal distribution and tensor $\td$-distribution}

In this paper, the multivariate $\td$-distribution with location parameter $\bolmu\in\mbbR^p$, symmetric positive definite scale matrix $\bolSigma\in\mbbR^{p\times p}$, and degrees of freedom $\nu>0$, is denoted as $t_p(\bolmu,\bolSigma,\nu)$. We introduce two tensor-variate distributions: The tensor normal distribution denoted as $\TN(\bolmu,\bolXi)$; and the tensor $\td$-distribution denoted as $\TT(\bolmu,\bolXi,\nu)$. In both distributions, the parameter $\bolmu\in\mbbR^{p_1\times\cdots\times p_M}$ characterizes the mean and the set of symmetric positive definite matrices $\bolXi\equiv\{\bolSigma_m \in\mbbR^{p_m\times p_m},\ m=1,\dots,M \}=\{\bolSigma_1,\cdots,\bolSigma_M \}$ characterizes the ``separable'' covariance structure. Moreover,  $\nu>0$ is the degrees of freedom in our tensor $\td$-distribution. The tensor Mahalanobis distance in both distributions is written as $\Vert \mbY-\bolmu \Vert_{\bolXi}^{2}=\langle \llbracket \mbY-\bolmu ; \bolSigma_{1}^{-1},\cdots,\bolSigma_{M}^{-1}\rrbracket,\mbY-\bolmu\rangle=\vecc(\mbY-\bolmu)^T(\bigotimes_{m=M}^1\bolSigma_m^{-1})\vecc(\mbY-\bolmu)$, which generalizes the usual tensor norm $\Vert\mbY-\bolmu\Vert^2 = \langle \mbY-\bolmu,\mbY-\bolmu \rangle$.

For a tensor random variable $\mbY\in\mbbR^{p_1\times\cdots\times p_M}$, the matrix/tensor normal distribution \citep{dawid1981some, mvd, manceur2013maximum} is one of the key statistical modeling approaches of the array and tensor random objects. 
As a generative definition of tensor normal distribution, $\mbY\sim \TN(\bolmu,\bolXi)$ if  $\mbY = \bolmu+\llbracket\mbZ;\bolSigma_{1}^{1/2},\cdots,\bolSigma_{M}^{1/2}\rrbracket $ for some random tensor $\mbZ$ that consists of independent standard normal entries.
The probability density function of $\mbY\sim \TN(\bolmu,\bolXi)$ can be easily obtained from the multivariate normal distribution: $\vecc(\mbY)\sim \N(\vecc(\bolmu), \bolSigma)$, where $\bolSigma = \bolSigma_M\otimes\cdots\otimes\bolSigma_1$ is called the Kronecker separable covariance. Therefore, the distribution of $\mbY$, specifically the probability density function, depends on $\bolXi=\{\bolSigma_1,\dots,\bolSigma_M\}$ only through the Kronecker product of covariances $\bolSigma= \bolSigma_M\otimes\cdots\otimes\bolSigma_1$. Let $\widetilde{\bolXi}=\{a_1\bolSigma_1,\cdots,a_M\bolSigma_{M}\}$, where $a_m>0$, $m=1,\cdots,M$, and $\prod_{m=1}^M a_m = 1$. Then $\TN(\bolmu,\widetilde\bolXi)$ and $\TN(\bolmu,\bolXi)$ are the same distribution. In other words, $\bolXi$ and each $\bolSigma_m$ are not identifiable while $\bolSigma$ is. To ensure parameter identification, we impose the constraint that the first element of $\bolSigma_m$ is one for $m=1,\cdots,M-1$, and then the scaling of the covariance parameters is absorbed into $\bolSigma_M$. 
For the tensor $\td$-distribution, we have the same issue and thus impose the same identification constraint on $\bolXi$ throughout this paper. Other approaches of scaling the covariance parameters $\bolSigma_m$ in the tensor normal distribution are equally applicable to our methodology and theory. 

The formal definition and several key properties of the tensor $\td$-distribution are given as follows. Analogous to the multivariate normal and $\td$-distributions, when the degrees of freedom $\nu\rightarrow\infty$, the tensor $\td$-distribution $\TT(\bolmu,\bolXi,\nu)$  becomes the tensor normal $\TN(\bolmu,\bolXi)$; when $M=1$ the tensor $\td$-distribution reduces to the multivariate $\td$-distribution $t_p(\bolmu,\bolSigma,\nu)$.

\begin{definition}
	A tensor-variate random variable $\mbY\in \mbbR^{p_1\times\cdots\times p_M}$ follows the tensor $\td$-distribution $\TT(\bolmu,\bolXi,\nu)$ if and only if it has probability density function,
	\begin{equation}\label{ttpdf}
		f(\mbY\mid \bolmu,\bolPhi,\nu)=\frac{\Gamma(\frac{\nu+p}{2})\prod_{m=1}^{M}\vert\bolSigma_{m}\vert^{-p_{-m}/2}}{(\pi \nu)^{p/2}\Gamma(\nu/2)}\times(1+\Vert \mbY-\bolmu\Vert_{\bolXi}^{2}/\nu)^{-\frac{\nu+p}{2}},
	\end{equation}
	where $p=\prod_{m=1}^{M}p_m$, $p_{-m}=\prod_{j\neq m}p_j$ and $\Gamma(\cdot)$ is the Gamma function.
	\label{tensort}
\end{definition}
\begin{proposition}
	A tensor $\td$-distributed random variable $\mbY\sim \TT(\bolmu,\bolXi,\nu)$ satisfies the following properties.
	\begin{enumerate}[(a)]
		\item Suppose $\mbX\sim \TN(0,\bolXi)$ and $G\sim \chi_{\nu}^{2}/\nu$ are independent, where $\chi^2_\nu$ is the Chi-square distribution with degree freedom $\nu>0$, then $\mbY\sim\mbX/\sqrt{G}+\bolmu\sim \TT(\bolmu,\bolXi,\nu)$. 
		\item When $\nu>2$, $\E(\mbY_{(m)}\mbC\mbY_{(m)}^{T})=\frac{\nu}{\nu-2}\tr\{\mbC(\bolSigma_M\otimes\cdots\otimes\bolSigma_{m+1}\otimes\bolSigma_{m-1}\otimes\cdots\otimes\bolSigma_1)\}\bolSigma_m+\bolmu_{(m)}\mbC\bolmu_{(m)}^{T}$, where $\mbC\in\mbbR^{p_{-m}\times p_{-m}}$ is a constant matrix. If $\bolmu=0$, then $\E\{\mbY_{(m)}\mbY_{(m)}^{T}\}=\frac{\nu}{\nu-2}\{\prod_{m'\neq m}\tr(\bolSigma_{m'})\}  \bolSigma_{m}$.
		\item Suppose that $\mbA_m\in \mbbR^{p_m\times q_m}$, where $q_m\leq p_m$, and $\mbA_m^{T}\mbA_m>0$. Define $\bolmu_{A}=\llbracket \bolmu;\mbA_{1}^T,\cdots,\mbA_{M}^T\rrbracket$, and $\bolXi_A=\{\mbA_1^{T}\bolSigma_{1}\mbA_1,\cdots,\mbA_M^{T}\bolSigma_{M}\mbA_M\}$. Then
		$\llbracket \mbY;\mbA_{1},\cdots,\mbA_{M}\rrbracket \sim \TT(\bolmu_{A}, \bolXi_{A},\nu)$.
		\item $\mbY_{(m)}\sim \TT(\bolmu_{(m)},\bolXi_m,\nu)$, where $\bolXi_m\equiv\{\bolSigma_m, \bolSigma_M\otimes\cdots\otimes\bolSigma_{m+1}\otimes\bolSigma_{m-1}\otimes\cdots\otimes\bolSigma_1\}$.
		\item $\vecc(\mbY)\sim t_{p} \big( \vecc(\bolmu),\bigotimes_{m=M}^{1}\bolSigma_{m},\nu \big)$.
	\end{enumerate}
	\label{ppt}
\end{proposition}

From the above generative representation of tensor $\td$-distribution, i.e., statement (a) in the proposition, we see that $G\sim \chi^2_\nu/\nu$ induces the heavy tail from $\TN(0,\bolXi)$. The heavy-tailed character is thus spherical and homogeneous across each element, fiber, and mode of the tensor. Future research may extend this definition of tensor $\td$-distribution to incorporate different degrees of heavy-tailedness along each mode of the tensor. This can be achieved by borrowing the idea from the alternative $\td$-model in \citet{finegold2011robust} for vector data. In this article, we focus on the single parameter $\nu$ to explicitly model the heavy tails of the data. As we show later in regression analysis, our tensor $\td$-distribution leads to a computationally efficient and intuitive weighted least squares (WLS) scheme for likelihood-based estimation.

If $\mbY\sim \TT(\bolmu, \bolXi, \nu)$ and $\nu>2$, then $\E(\mbY)=\bolmu$, and $\Cov\{\vecc(\mbY) \}=\frac{\nu}{\nu-2}\bolSigma=\frac{\nu}{\nu-2}\bolSigma_M\otimes\cdots\otimes\bolSigma_1$. 
We also want to make a remark on the choice of the Kronecker product separable covariance structure $\bolSigma=\bigotimes_{m=M}^1\bolSigma_m$ from the shape parameter $\bolXi$. This is a widely used structure in tensor normal analysis \citep[e.g.]{hoff2011separable, catch, rte, he2014graphical}, where the goal is to model the mode-wise dependence structure. Similar to the tensor normal case, this structure also substantially reduces the number of parameters in $\Cov\{\vecc(\mbY)\}$ for tensor $\td$-distribution. 
Specifically, the number of the free parameters in the scale parameter is reduced from $(\prod_{m=1}^Mp_m)(\prod_{m=1}^Mp_m+1)/2$ to $\sum_{m=1}^{M}p_m(p_m+1)/2-M+1$. 
Our tensor $\td$-model framework can be directly generalized to alternative tensor covariance assumptions. For example, \citet{greenewald2019tensor} recently proposed to use Kronecker sum instead of product in tensor graphical models. Then the alternative tensor $\td$-distribution can be generated in the same way as in Proposition~\ref{ppt}. 

Properties (b)--(e) are about quadratic forms, linear transformations, and reshaping (matricization and vectorization) of the tensor $\td$-variable. Specifically,
(b) leads to an easy moment-based sample estimation for $\bolSigma_m$; (c)--(e) implies that tensor $\td$-distribution is preserved after any non-degenerate linear transformation, rotation, sub-tensor extraction, vectorization and matricrization. These nice properties also distinguish our tensor $\td$-distribution from the commonly used matrix $\td$-distribution  \citep[e.g.,][]{mvd}. The matrix $\td$-distribution and, more generally, the matrix elliptical distributions are important topics in multivariate analysis and Bayesian decision theory. See \citet{dawid1977spherical, fang1999bayesian} for some classical results, and \citet{zhang2010multi, thompson2020} for more recent applications. The   matrix $\td$-distribution can be defined as $\mbS^{-1/2}\mbX+\bolmu\sim \MVT(\bolmu,\{\bolSigma_1,\bolSigma_2\},\nu)$, where $\mbX\sim \TN(\bolmu,\{\mbI_{p_1},\bolSigma_2\})$ and $\mbS\sim \Wis(\bolSigma_{1}^{-1},\nu+p_1-1)$ are independent.

Our tensor $\td$-distribution, when $M=2$, is different from this matrix $\td$-distribution. We make a few remarks about the advantages of our $\TT$ distribution over the commonly used $\MVT$ distribution; additional connections and discussion are given in the Supplementary Materials. First of all, while the existing works on matrix $\td$-distribution $\MVT$ focus on the left or right sphericalness of the data matrix or Bayesian inference \citep{ob1988bayesian}, our tensor $\td$-distribution is motivated mainly by the heavy-tailed tensor data in practice. Therefore, when the focus is dealing with heavy-tails and potential outliers, it is more natural to consider the univariate chi-square distribution than the matrix-variate Wishart distribution in generating the matrix-$\td$ variables. Secondly, by comparing the two density functions, our tensor $\td$-distribution is more intuitive. The density in Definitions~\ref{tensort} depends on $\mbY$ only through $\Vert \mbY-\bolmu\Vert_{\bolXi}^{2}$, which intuitively is the tensor Mahalanobis distance. On the other hand, the determinant function in matrix $\td$ (see Definition~\ref{matrixt}), $\vert\mbI_{p_1}+\bolSigma_{1}^{-1}(\mbY-\bolmu)\bolSigma_{2}^{-1}(\mbY-\bolmu)^{T}\vert$, loses such nice interpretation. From a computational perspective, when the dimension is large,  the calculation of matrix determinant (in maximizing the likelihood) and the sampling of Wishart latent variables (in EM algorithm) is far less appealing than calculating the Mahalanobis distance and sampling the univariate chi-squares latent variables. Finally, as shown in Proposition~\ref{ppt}, our tensor-$\td$ random variable remains a tensor-$\td$ random variable if we extract a sub-tensor, or vectorize or matricize it. In tensor analysis, it is crucial to preserve the same distributional characteristics when those tensor operations are performed. This is unfortunately not true for the existing matrix $\td$-distribution: If $\mbY\sim\MVT$ then $\vecc(\mbY)$ does not follow a multivariate $\td$-distribution. To the best of our knowledge, there is no straightforward way of generalizing the matrix $\td$-distribution $\MVT$ to higher-order tensors, while our tensor $\td$-distribution provides an easy and unified modeling approach for an arbitrary order tensor.

\section{Robust tensor response regression}
\label{model}
\subsection{Model}
To model the association between a response tensor $\mbY\in \mbbR^{p_1\times\cdots\times p_M}$ and a covariate vector $\mbX\in\mbbR^q$,  we consider the following tensor response regression model with independent and identically distributed data, 
\begin{equation}\label{ttr}
	\mbY_i=\mbB\bar{\times}_{M+1}\mbX_i+\mbE_i,\ \mbE_i\sim \TT(0,\bolXi,\nu), \quad i=1,\dots,n, 
\end{equation}
where $\mbB\in\mbbR^{p_1\times\cdots\times p_M\times q}$ is the coefficient tensor, and $\mbE_i\in\mbbR^{p_1\times\cdots\times p_M}$ are independent of $\mbX_i$. Without loss of generality, we assume that $\E(\mbY)=0$, $\E(\mbX)=0$, and that the data are centered.
While the tensor response regression is a rapidly developing area of research in recent years, most existing approaches either explicitly assume the error $\mbE$ to be tensor normal  \citep[e.g.,][]{rte} or inexplicitly use the least squares loss that corresponds to isotropic normal distribution \citep[e.g.][]{higher}. Many theoretical results also break down when the sub-Gaussian tail condition is violated \citep[e.g.,][]{sun2017store}. We assume a heavy-tail tensor $\td$-distribution for more robust and flexible modeling. In contrast to the robust loss function approaches, e.g.~using Huber loss function instead of least squares loss \citep[e.g.][]{huber1964,lambert2011}, our Model \eqref{ttr} specifies the $\td$-distribution for $\mbE$ to help derive the maximum likelihood estimation (MLE) and model-based inference procedure. Our focus is the issues of heavy-tail errors and the high-dimensionality in response. Hence our approach is developed for a sparse but not necessarily low-rank $\mbB$, while exiting tensor response regression methods \citep{rte,higher,sun2017store} heavily rely on the low-rankness of $\mbB$. 


To gain some intuition on how the Model \eqref{ttr} is robust to outliers, we consider the MLE of $\mbB$ when $\bolXi$ and $\nu$ are known. 
Let $\mbbX\in\mbbR^{q\times n}$ be the data matrix for the predictor, and $\mbbY\in\mbbR^{p_1\times\cdots\times p_M\times n}$ be the data tensor for the response.
\begin{proposition}
	\label{MLE}
	The MLE to \eqref{ttr} satisfies that $\widehat{\mbB}=\mbbY\times_{M+1}(\mbbX\mbbW\mbbX^{T})^{-1}\mbbX\mbbW$, where $\mbbW\in\mbbR^{n\times n}$ is a diagonal matrix with $w_{i}=(\nu+p)/\big(\nu+\Vert \mbY_i-\hatmbB\bar{\times}_{M+1}\mbX_{i}\Vert_{\bolXi}^{2}\big)$.
\end{proposition} 
From Proposition~\ref{MLE}, if one knows $\bolXi$ and $\nu$, then the MLE $\widehat{\mbB}$ can be viewed as a weighted least squares estimator. The weight $w_i$ for the $i$-th observation is small when the observation is far from the center, i.e.,~when the tensor Mahalanobis distance $\Vert\mbY_i - \hatmbB\bar\times_{M+1}\mbX_i\Vert_{\bolXi}^2$ is large. This weighting scheme thus provides a robust estimator that is also  efficient because it is likelihood-based. When we know $\bolXi$ and $\nu$, the MLE of $\hatmbB$ can then be obtained from an iterative re-weighted procedure, where $\mbbW$ and $\hatmbB$ are iteratively updated based on Proposition~\ref{MLE}. Later, we also consider the penalized likelihood approach where $\mbB$ is sparse and $\bolXi$ is unknown.
{Details of the EM algorithm for the tensor $\td$-distribution, which works well in low-dimensional settings, are given in Section~\ref{distribution} of Supplementary Materials. }


\subsection{Choosing the degrees of freedom}\label{choose}
Before proposing the penalized estimation for tensor $\td$-regression model, we first discuss the choice of the degrees of freedom $\nu$. 
Although the model in \eqref{ttr} can be assumed with any $\nu>0$, in model fitting with unknown $\nu$, we recommend using $\nu=4$. Apparently, such a choice is prone to model mis-specification as the true model may have a different degrees of freedom. However, we have found that setting $\nu=4$ in the estimation works well in numerical studies and applications. The reason is that our estimation is insensitive to the choice of $\nu$. In Proposition \ref{MLE} and the one-step estimator proposed in the next section, the estimation is affected by $\nu$ via the weights $w_{i}=(\nu+p)/({\nu+\Vert \mbY_i-\widehat{\mbB}\bar{\times}_{M+1}\mbX_{i}\Vert_{{\bolXi}}^{2}})$, where the expectation of $\Vert \mbY_i-\widehat{\mbB}\bar{\times}_{M+1}\mbX_{i}\Vert_{{\bolXi}}^{2}$ is in the same order as $p$. In most tensor analysis, the dimension $p=\prod_{m=1}^Mp_m$ is large and then the weights $w_i$ are very insensitive to the choice of $\nu$. Therefore, $\nu$ has a minor influence on the estimation of the model.  In simulation studies, we use an example to give a further demonstration that using $\nu=4$ works as well as using true $\nu$ for a wide range of settings. Note that \cite{robust1989} also illustrated that using $\nu=4$ works well for multivariate $\td$-regression.
If one is particularly interested in estimating the true $\nu$, 
we have developed an algorithm for estimating $\nu$ that works well in low-dimensional settings (details are provided in the Supplementary Materials, Section~\ref{distribution}). However, estimating $\nu$ is not recommended when $p$ is moderately large. Recall that in Figure \ref{QQ}, we generated a simulated data from $\TT(0,\mbI_{p},1.5)$. For this simulated data, the estimated degrees of freedom $\widehat{\nu}=0.402$, which is much smaller than $1.5$. \citet{lucas1997} argued that estimating $\nu$ based on data is not robust to outliers for multivariate $\td$-distribution. For tensor $\td$-distribution, we have the same issue. Another disadvantage for estimating $\nu$ is that we need to using line search method to find the solution iteratively, which is time-consuming. Henceforth, we use $\nu=4$ as a fixed default value unless otherwise specified.

\subsection{The penalized likelihood approach}
\label{pmle}
We consider the following penalized negative log-likelihood  function for the joint estimation of $\mbB$ and $\bolXi=\{\bolSigma_1,\dots,\bolSigma_M\}$,
\begin{equation}\label{ad2}
	\L_n(\mbB,\bolXi)=\frac{n}{2}(\sum_{m=1}^{M}p_{-m} \logg\vert\bolSigma_m\vert)+\frac{\nu+p}{2}\sum_{i=1}^{n}\logg(1+\Vert \mbY_i-\mbB\bar{\times}_{M+1}\mbX_{i}\Vert_{\bolXi}^{2}/\nu)+P_{\lambda}(\mbB),
\end{equation}
where $P_\lambda(\mbB)$ denotes a generic penalty function on $\mbB$ with tuning parameter $\lambda$ such as the lasso penalty \citep{tibshirani1996regression}, elastic net \citep{Zou05regularizationand} and SCAD \citep{Fan01variableselection}. In this article, we consider the adaptive lasso penalty  \citep{ad2006} to induce element-wise sparsity in $\mbB$ and the group adaptive lasso penalty \citep{Wang20085277} to perform response variable selection. 

We impose the adaptive lasso penalty to induce element-wise sparsity in $\mbB$ and define it to be the same as the classical adaptive lasso on $\mbB$: $P_\lambda(\mbB)=\lambda\sum_{j_1\cdots j_{M+1}}r_{j_1\cdots j_{M+1}}\vert b_{j_1\cdots j_{M+1}}\vert$ , where $b_{j_1\cdots j_{M+1}}$ is the $(j_1,\cdots,j_{M+1})$-th element in $\mbB$, $r_{j_1\cdots j_{M+1}}={\widehat{b}_{j_1\cdots j_{M+1}}^{-2}}$, and $\widehat{b}_{j_1\cdots j_{M+1}}$ can be any $\sqrt{n}$-consistent estimator of ${b}_{j_1\cdots j_{M+1}}$. In this paper, we focus on the scenario when the predictor is a low-dimensional vector, i.e. $q\ll n$. Consequently,  the ordinary least squares (OLS) estimator $\widehat\mbB^{\OLS} = \mbbY\times_{M+1}(\mbbX\mbbX^T)^{-1}\mbbX$ is well-defined, $\sqrt{n}$-consistent, and used in the adaptive (group) lasso penalties unless otherwise specified.

In tensor response regression, we are also interested in response variable selection. For example, when $\mbY$ is a brain imaging scan, it is of great interest to identify brain regions that are affected by the predictor. 
For more effective variable selection in $\mbY$, we consider the intrinsic group structure in variable selection. By vectorizing the tensor response Model \eqref{ttr}, we have a multivariate linear regression
\begin{equation}\label{ttr_vec}
	\vecc(\mbY_i) = \mbB_{(M+1)}^T\mbX_i + \vecc(\mbE_i),\quad i=1,\dots,n,
\end{equation}
where each column vector of the regression coefficient matrix $ \mbB_{(M+1)} \in\mbbR^{q\times p}$ can be mapped to a mode-$(M+1)$ fiber of $\mbB$ and correspondingly to a response variable in $\mbY$. Therefore, we also consider the adaptive group lasso penalty with column-wise group structure from $\mbB_{(M+1)}$,  $P_{\lambda}(\mbB)=\lambda\sum_{J=1}^{p}r_{J}\Vert\mbB_{J}\Vert_{2}$,  where $\mbB_{J}\in\mbbR^q$ is the $J$-th column of $\mbB_{(M+1)}$, and $\Vert\mbB_{J}\Vert_{2}^2={\mbB_{J}^{T}\mbB_{J}}$. For the weight $r_J$, we choose $r_J=1/{\Vert\widehat{\mbB}^{\OLS}_{J}\Vert_2^{2}}$ from the OLS estimator.

An important special case of \eqref{ad2} is when $\nu\to \infty$. Then $\L_n(\mbB,\bolXi)$ converges to
\begin{equation}
	\L_{n,\infty}(\mbB,\bolXi)=\frac{n}{2}(\sum_{m=1}^{M}p_{-m} \logg\vert\bolSigma_m\vert)+\frac{1}{2}\sum_{i=1}^{n}\Vert \mbY_i-\mbB\bar{\times}_{M+1}\mbX_{i}\Vert_{\bolXi}+P_{\lambda}(\mbB),
	\label{g2}
\end{equation}
which is the penalized likelihood-based objective function for tensor normal error $\mbE_i$ in \eqref{ttr}. To the best of our knowledge, such an estimator and its properties have not been studied in the literature. As such the global minimum of \eqref{ad2} is denoted as $\hatmbB^{\APT}$ and $\hatbolXi^{\APT}$, and that of \eqref{g2} is denoted as $\hatmbB^{\APN}$ and $\hatbolXi^{\APN}$. The estimator $\hatmbB^{\APN}$ is not robust to outliers.  We will show the advantages of $\hatmbB^{\APT}$ over $\hatmbB^{\APN}$ both numerically and theoretically.


The objective function \eqref{ad2} is clearly not convex and computationally nontrivial to minimize. We develop a Majorize-Minimization (MM) algorithm \citep[e.g.]{MM2004} as a feasible approach. The key in constructing the MM algorithm is to find a convex surrogate function and solve it to update the solution in the majorization and minimization steps. We construct the convex surrogate function in the following. 

\begin{proposition}\label{MMp}
	For any $(\mbB^{(k)},\bolXi^{(k)})$, the following function majorizes $\L_n(\mbB,\bolXi)$ in \eqref{ad2},
	\begin{equation}\label{obj}
		\F_n(\mbB,\bolXi\mid\mbB^{(k)},\bolXi^{(k)})=\frac{n}{2}(\sum_{m=1}^{M}p_{-m} \logg\vert\bolSigma_m\vert)+\frac{1}{2}\sum_{i=1}^{n}\omega_{i}^{(k)}\Vert\mbY_{i}-\mbB\bar{\times}_{M+1}\mbX_{i}\Vert_{\bolXi}^{2}+P_{\lambda}(\mbB),
	\end{equation}
	where the weight is defined as $\omega_{i}^{(k)}=(\nu+p)/\left(\nu+\Vert \mbY_i-\mbB^{(k)}\bar{\times}_{M+1}\mbX_{i}\Vert_{\bolXi^{(k)}}^{2}\right)$.
	Specifically, $\F_n(\mbB,\bolXi\mid\mbB^{(k)},\bolXi^{(k)})\geq \L_n(\mbB,\bolXi)$, and $\F_n(\mbB^{(k)},\bolXi^{(k)}\mid\mbB^{(k)},\bolXi^{(k)})=\L_n(\mbB^{(k)},\bolXi^{(k)})$. 
\end{proposition}

The objective function \eqref{obj} contains two unknown parameters $\mbB$ and $\bolXi$. To solve it, we still need to alternate between $\mbB$ and $\bolXi$. In each iteration, we update $\bolXi$ given $\mbB$, then we update $\mbB$ given the most recently updated $\bolXi$. The detailed MM algorithm for solving our penalized MLE problem is provided in the Supplementary Materials (Section \ref{distribution}). 

Although the MM algorithm always converges, we cannot guarantee its convergence to the global minimum since the objective function \eqref{ad2} is not convex. Also, we need two iterations to solve  \eqref{ad2}. The first iteration is that we find the convex surrogate function iteratively, the second iteration is that to solve the surrogate function, we solve $\mbB$ and $\bolXi$ cyclically. Those two iterations are usually of high computational cost. We will 
show that $\widehat{\mbB}^{\APT}$ has oracle properties. However, the theoretical gap exists since the MM algorithm may converge to a local minimum.  Motivated by these issues, we next develop a novel one-step estimation procedure that is guaranteed to converge to the global solution and much faster than the MM algorithm.


\section{One-step estimation and its high-dimensional modification\label{ostep}}

\subsection{One-step estimation}

The one-step algorithm is motivated by the majorizing function \eqref{obj}, which is convex in $\mbB$. Since our main target in tensor response regression is the sparse tensor coefficient $\mbB$, we propose a one-step estimation procedure that tailors the joint estimation of $\mbB$ and $\bolXi$ for a more targeted estimation of $\mbB$. We define our one-step estimator as the minimizer of the following convex objective function, 
\begin{equation}
	\F_{n}(\mbB)\equiv\F_n(\mbB,\widehat{\bolXi}\mid\widehat{\mbB},\widehat{\bolXi})=\sum_{i=1}^{n}\widehat{\omega}_{i}\Vert\mbY_{i}-\mbB\bar{\times}_{M+1}\mbX_{i}\Vert_{\widehat{\bolXi}}^{2}+P_\lambda(\mbB), \label{onestep1}
\end{equation}
where $\widehat{\mbB}$ and $\hatbolXi$ can be any $\sqrt{n}$-consistent estimators, and $\widehat{\omega}_{i}=(\nu+p)/\big(\nu+\Vert \mbY_i-\widehat{\mbB}\bar{\times}_{M+1}\mbX_{i}\Vert_{\widehat{\bolXi}}^{2}\big)$. Henceforth, we use the previously defined adaptive lasso penalty $P_\lambda(\mbB)=\lambda\sum_{j_1\cdots j_M}r_{j_1\cdots j_{M+1}}\vert b_{j_1\cdots j_{M+1}}\vert$ unless otherwise specified. We choose $\hatmbB$ and $\hatbolXi$ as follows. 

For $\widehat{\mbB}$, we use the following adaptive penalized least squares (APL) estimator, whose theoretical properties are also established in Section~\ref{theory},
\begin{equation}\label{APL}
	\hatmbB^{\APL} = 
	\argmin_{\mbB\in\mbbR^{p_1\times\cdots\times q}} \sum_{i=1}^{n}\Vert \vecc(\mbY_i-\mbB\bar{\times}_{M+1}\mbX_{i})\Vert^{2}+
	P_\lambda(\mbB).
\end{equation}
This APL estimator can be viewed as the na\"{i}ve penalized estimator that is easy to obtain but ignores the tensor and covariance structure in response. Based on this $\hatmbB^{\APL}$, we obtain $\hatbolXi$ by minimizing the plug-in likelihood-based objective function, i.e.~$\L_n(\widehat{\mbB}^{\APL},\bolXi)$ in \eqref{ad2},
\begin{equation}\label{plugin}
	\L_n(\bolXi)\equiv\frac{n}{2}(\sum_{m=1}^{M}p_{-m} \logg\vert\bolSigma_m\vert)+\frac{\nu+p}{2}\sum_{i=1}^{n}\logg(1+\Vert \mbY_i-\widehat{\mbB}^{\APL}\bar{\times}_{M+1}\mbX_{i}\Vert_{\bolXi}^{2}/\nu).
\end{equation}

We denote $\hatmbB^{\OST}$ as the solution of the objective function \eqref{onestep1} given $\hatmbB^{\APL}$ and $\hatbolXi$. 
%
An obvious advantage of the one-step estimator is that it does not need the MM iterations in the penalized MLE. As a result, the one-step estimator is much faster.  In what follows, we further discuss that the one-step estimator should be preferred for theoretical studies as well.

The one-step estimator is not a solution to the original penalized maximum likelihood problem in \eqref{ad2}. However, we will show in Section \ref{aptonestep} that the one-step estimator has an almost identical asymptotic variance as the penalized MLE and, thus, has almost no loss in statistical efficiency. Furthermore, the one-step estimator circumvents the problem of local minima. Since \eqref{ad2} is a nonconvex problem, it is difficult to guarantee that an algorithm achieves the global minimum. There is henceforth an algorithmic gap in the theoretical analysis of the penalized MLE, in the sense that the solution from the MM algorithm might not be the global solution to \eqref{ad2} with the desirable properties. On the other hand, we show that the one-step estimator's algorithm is guaranteed to converge to the global solution of the convex problem \eqref{onestep1} (cf. Theorem \ref{global}).
\subsection{Algorithm for one-step estimator}
\label{alg}

The one-step estimator is obtained in three steps: the initialization for $\mbB$ by solving \eqref{APL}, the estimation of $\bolXi$ by solving  \eqref{plugin}, and the final OST estimator by solving \eqref{onestep1}. The estimation procedure is summarized in Algorithm \ref{onestepalg}.
%

The optimization problem \eqref{APL} can be separated into the following $p$ sub-problems. 
\begin{equation}
	\begin{aligned}
		\argmin_{\mbB_{J}\in\mbbR^{q}} \sum_{i=1}^{n}\Vert \vecc(\mbY_i)_{J}-\mbB_{J}\mbX_{i}\Vert^{2}+\lambda\sum_{k=1}^{q}r_{q(J-1)+k}\vert \beta_{q(J-1)+k}\vert
	\end{aligned}
	\label{APL1}
\end{equation}
with $J=1,\cdots,p$. The separability of \eqref{APL} indicates that to obtain $\widehat{\mbB}^{\APL}$, we regress one element of $\mbY$ a time without considering the correlation between the elements. Each sub-problem can be solved by coordinate descent algorithm used in \cite{JSSv033i01}. 

For the plug-in estimator $\widehat{\bolXi}$, we provide the details of solving objective function \eqref{plugin} in the Supplementary Materials (Section \ref{distribution}). In the proof of Theorem \ref{global}, we showed that when $np_{-m}>p_m$, the estimate $\hatbolSigma_{m}$ obtained by solving \eqref{plugin} is positive definite with a probability of one. This also guarantees that the objective function \eqref{onestep1} is well-defined. Although the estimation for $\bolXi$ is iterative and non-convex, we used the concept and theory of geodesic convexity \citep{geo} to show that the likelihood-base covariance estimation procedure converges to the global solution. This technique is similar to the existing works on this topic \citep[e.g.,][]{global2}.

Objective function \eqref{onestep1} is a penalized weighted least square problem that is strictly convex. When $P_\lambda(\mbB)=\lambda\sum_{j_1\cdots j_{M+1}}r_{j_1\cdots j_{M+1}}\vert b_{j_1\cdots j_{M+1}}\vert$,
we adopt a coordinate descent algorithm to solve it. The main idea is that, in each iteration, we update one element of $\mbB$ while fixing the others. Let $j=(j_1,\cdots,j_{M+1})$, $\mbB_{-j}$ be a tensor that is identical to $\mbB$ except that the $j$-th element is $0$, $(\widehat{\bolSigma}_{m}^{-1})_{i_m i_m}$ be the $(i_m,i_m)$-th element of $\widehat{\bolSigma}_{m}^{-1}$, and $x_{ii_m}$ be the $i_m$-th element of $\mbX_i$. Let $U_1$ be the $j$-th element of $\sum_{i=1}^{n}\big(\widehat{w}_i\llbracket \mbY_i-\mbB_{-j}\bar{\times}_{M+1}\mbX_{i}; \hatbolSigma_{1}^{-1},\cdots,\hatbolSigma_{M}^{-1},\mbX_{i}\rrbracket\big)$, and $U_2=\sum_{i=1}^{n}\big\{\widehat{w}_ix_{ii_m}^2 \prod_{m=1}^{M}(\widehat{\bolSigma}_{m}^{-1})_{i_m i_m} \big\}$. The iteration for the $j$-th element of $\mbB$ is shown in \eqref{coordinate}. 
\begin{algorithm}[ht!]
	\caption{One-step estimation with adaptive lasso penalty }
	\label{onestepalg}
	\begin{algorithmic}
		\STATE \textbf{Input}: Centered data $(\mbX_i, \mbY_i)$, $i=1,\dots,n$, and the tuning parameter $\lambda> 0$.
		\STATE \textbf{Initialization}: 
		\begin{enumerate}
			\item \textbf{Penalty}: Construct the adaptive lasso penalty $P_\lambda(\mbB)$ from $\hatmbB^{\OLS}= \mbbY\times_{M+1}(\mbbX\mbbX^T)^{-1}\mbbX$. 
			\item \textbf{Initial APL estimator $\widehat{\mbB}^{\APL}$:} Solve the $p$ sub-problems in \eqref{APL1} by coordinate descent algorithm.
			\item \textbf{Shape estimator $\widehat{\bolXi}$:} For $m=1,\cdots,M$, cyclically updating the following equation until convergence 
			\begin{equation}
				\hatbolSigma_{m}\longleftarrow \frac{1}{n p_{-m}}
				\sum_{i=1}^{n}
				\widehat w_{i}(\mbY_{i}-\hatmbB^{\APL}\times_{(M+1)} \mbX_i )_{(m)}\hatbolOmega_{m}(\mbY_{i}-\hatmbB^{\APL}\times_{(M+1)} \mbX_i )_{(m)}^{T},
			\end{equation}
			where $\hatbolOmega_{m} = \hatbolSigma_{M}^{-1}\otimes \cdots\otimes\hatbolSigma_{m+1}^{-1}\otimes\hatbolSigma_{m-1}^{-1}\otimes \cdots \otimes\hatbolSigma_{1}^{-1}$. 
			\item \textbf{Weights $\widehat{w}_{i}$:} For $i=1,\dots,n$, calculate $\widehat{w}_{i}=(\nu+p)/\left(\nu+\Vert \mbY_i-\widehat{\mbB}^{\APL}\bar{\times}_{M+1}\mbX_{i}\Vert_{\widehat{\bolXi}}^{2}\right)$
		\end{enumerate}
		\STATE \textbf{Coordinate descent updates for one-step estimator:}
		\STATE {\textbf{repeat} For $j_m=1,\cdots,p_m$, $m=1,\cdots, M$, and  $j_{M+1}=1,\cdots,q$, } 
		\begin{itemize}
			\item Calculate $U_1$ as the $j$-th element of $\sum_{i=1}^{n}\big(\widehat{w}_i\llbracket \mbY_i-\hatmbB_{-j}^{\OST}\bar{\times}_{M+1}\mbX_{i}; \hatbolSigma_{1}^{-1},\cdots,\hatbolSigma_{M}^{-1},\mbX_{i}\rrbracket\big)$, and $U_2=\sum_{i=1}^{n}\big\{\widehat{w}_ix_{ii_m}^2 \prod_{m=1}^{M}(\widehat{\bolSigma}_{m}^{-1})_{i_m i_m} \big\}$.
			\item For each elements in $\hatmbB^{\OST}$, we update $b_{j_1\cdots j_{M+1}}$ as,
			\begin{equation}
				b_{j_1\cdots j_{M+1}}^{(new)}\leftarrow
				\begin{cases}
					\dfrac{U_1}{U_2+{\lambda U_2}/({\vert U_1\vert-\lambda}) },& \text{if } \vert U_1\vert-\lambda>0\\
					0,              & \text{otherwise}
				\end{cases}\label{coordinate}
			\end{equation}
		\end{itemize}
		{\textbf{until} the largest absolute value of the elements between two updates is smaller than $10^{-4}$.}
		
		\STATE \textbf{Output}: $\hatmbB^{\OST}$ and $\hatbolXi$.
	\end{algorithmic}
\end{algorithm}

Algorithm \ref{onestepalg} is much faster than the penalized MLE approach (i.e.~the MM algorithm). In a Windows 10 laptop computer with Intel(R) Core(TM) i7-6700 CPU@3.4GHz, the running time is 3.8s  for one-step algorithm and 101.8s for the MM algorithm, both including cross validation under Model M1 in our simulation studies. 
The one-step estimator is the solution of a regularized weighted least square problem which is strictly convex. For this weighted least square problem, we can find the largest tuning parameter $\lambda$ such that all coefficients of $\mbB$ are zero, then use the warm start method in \cite{hastie2015statistical} to speed up the computation.  For the selection of tuning parameter, we use five-fold cross validation. Tuning parameter $\lambda$ with the smallest cross validation prediction error is selected.

In parallel to Algorithm~\ref{onestepalg}, we also develop the algorithm for one-step estimation with adaptive group lasso penalty $P_{\lambda}(\mbB)=\lambda\sum_{J=1}^{p}r_{J}\Vert\mbB_{J}\Vert_{2}$. For this scenario, the main difference is in the coordinate descent steps, where we adopt the groupwise-majorization-descent algorithm proposed by \citep{groupalg}. Details are provided in Section \ref{groupp} of the Supplementary Materials. We demonstrate such response variable selection using the group adaptive lasso penalty in our numerical studies.

\subsection{High-dimensional modification for one-step estimator}\label{Sec:high}
In Section \ref{theory}, we consider the asymptotic properties of $\hatmbB^{\APT}$ and $\hatmbB^{\OST}$  when the dimensions $p_m$ are fixed and $n \rightarrow \infty$. However, in the recent two decades, statisticians have been keenly interested in high-dimensional problems. In our context, one may be curious about the estimation of Model \eqref{ttr} when the number of parameters in $\mbB$, i.e, $(\prod_{m=1}^Mp_m)q$, is much larger than $n$. To solve this problem, we propose a high-dimensional one-step estimator (HOST) in this section. Similar to OST, HOST re-weights the observations to mitigate the impact of heavy tails, and uses the adaptive LASSO penalty to achieve sparsity. However, to tackle the high dimensionality, HOST adopts a slightly different approach from OST to estimate the nuisance parameter $\bolXi$ and evaluate the weights $\omega_i$. As will be seen later, HOST is computationally more efficient in high-dimensional problems, and is minimax rate optimal in estimation.

To estimate Model \eqref{ttr} in high dimensions, we consider the following HOST estimator. For ease of theoretical studies, we split the data into two equal-size batches. Without loss of generality, we assume that we have $2n$ i.i.d.~observations in total, and split them into
$\calE_1=\{(\widetilde{\mbY}_i,\widetilde{\mbX}_i), i=1,\ldots,n \}$, and $\calE_2=\{({\mbY}_i,{\mbX}_i), i=1,\ldots,n \}$. \begin{enumerate}
	\item With the first batch of data $\calE_1$, we stack the data $\widetilde{\mbY}_i,\widetilde{\mbX}_i$ to $\widetilde{\mbbY}\in\mathbb{R}^{p_1\times\cdots\times p_M\times n},\widetilde{\mbbX}\in\mathbb{R}^{q\times n}$ and calculate
	$r_{jk}=1/(\widehat{b}_{jk}^{\OLS})^2$, where  
	$
	\widehat{\mbB}^{\OLS}=\widetilde{\mbbY}\times_{M+1}(\widetilde{\mbbX}\widetilde{\mbbX}^T)^{-1}\widetilde{\mbbX}
	$. We further calculate 
	\begin{eqnarray}
		\hatmbB^{\APL}&=&\argmin_{\mbB\in\mbbR^{p_1\times\cdots\times q}} \sum_{i=1}^{n}\Vert \vecc(\widetilde{\mbY}_i-\mbB\bar{\times}_{M+1}\widetilde{\mbX}_{i})\Vert^{2}+
		\lambda\sum_{j=1}^{pq}r_{j}\vert \beta_{j}\vert,\label{APLsplit1}\\
		\hatbolSigma_{m}&=&\frac{1}{np_{-m}}\sum_{i=1}^n\widehat{\omega}_i(\widetilde{\mbY}_i-\widehat{\mbB}^{\APL}\bar{\times}_{(M+1)}\widetilde{\mbX}_i)_{(m)}(\widetilde{\mbY}_i-\widehat{\mbB}^{\APL}\bar{\times}_{(M+1)}\widetilde{\mbX}_i)_{(m)}^T,
		\label{sigmam}
	\end{eqnarray}
	where $\widehat{\omega}_{i}=p/\Vert\widetilde{\mbY}_i-\widehat{\mbB}^{\APL}\bar{\times}_{(M+1)}\widetilde{\mbX}_i\Vert_{\mbI_{p}}^2$.
	\item With the second batch of data $\calE_2$, we calculate
	\begin{equation}
		\hatmbB^{\HOST}=\argmin_{\mbB\in\mbbR^{p_1\times\cdots\times q}}\sum_{i=1}^{n}\widehat{\omega}_{i}\Vert\mbY_{i}-\mbB\bar{\times}_{M+1}\mbX_{i}\Vert_{\widehat{\bolXi}}^{2}+\lambda\sum_{j=1}^{pq}r_{j}\vert \beta_{j}\vert, 
		\label{stage2}
	\end{equation}
	where $\widehat{\omega}_{i}=p/\Vert\mbY_i-\widehat{\mbB}^{\APL}\bar{\times}_{(M+1)}\mbX_i\Vert_{\mbI_{p}}^2$ and $\widehat{\bolXi}=\{\widehat{\bolSigma}_1,\ldots,\widehat{\bolSigma}_M\}$.
\end{enumerate}
We make some remarks for $\hatmbB^{\HOST}$ from the above estimation procedure. 

First of all, we split the dataset such that the first batch $\calE_1$ is used for the initialization including $r_{jk}$, $\widehat{\mbB}^{\APL}$, and $\hatbolSigma_m$, $m=1,\cdots,M$, while the second batch $\calE_2$ is used to obtain $\hatmbB^{\HOST}$. The splitting procedure makes $\hatbolXi$ and $\calE_2$ independent, which facilitates the theoretical studies for the convergence rate of $\hatmbB^{\HOST}$. 

Second, we are considering high-dimensional problems with diverging $p_m$ while $\hatmbB^{\OLS}$ is still well-defined. This is because we assume that the number of predictors, $q$, is fixed, and hence $\widetilde{\mbbX}\widetilde{\mbbX}^T$ is invertible in $\hatmbB^{\OLS}$. 

Third, in Step 1 of HOST, we estimate the nuisance parameter $\bolXi$ differently from that in OST. 
HOST estimates $\bolXi$ with an explicit formula in \eqref{sigmam}, instead of the iterative estimate in OST. By avoiding the iterations, HOST is faster than OST, especially when the dimension is high. Moreover, the explicit formula makes it easier to show the desirable theoretical properties for HOST. If we continue to employ the iterative estimator in OST, the theoretical studies are expected to be very challenging. \citet{lyu2019tensor} considered a much simpler problem of tensor graphical model, where the data are drawn from a tensor normal model without covariates, and all the inverses of the covariance matrices are sparse. In our model, the precision matrices are nuisance parameters, and we hope to avoid sparsity assumptions on them; our sparsity assumption is solely imposed on the parameter of interest, $\mbB$. Consequently, it is difficult to establish consistency for the iterative estimator. On the contrary, we show that our modified estimator $\hatbolSigma_m$ in HOST is close to $c_m\bolSigma_m$ under our model assumptions, where $c_m>0$ is a constant, and eventually leads to a high convergence rate of our estimate for B.

Fourth, compared to OST, we evaluate the weight $\omega_i$ differently in HOST. Instead of using the Mahalanobis distance $\Vert\mbY_i-\widehat{\mbB}^{\APL}\bar{\times}_{(M+1)}\mbX_i\Vert_{\hatbolXi}^2$, we use the Euclidean distance $\Vert\mbY_i-\widehat{\mbB}^{\APL}\bar{\times}_{(M+1)}\mbX_i\Vert_{\mbI_{p}}^2$.  The two distances are generally not equal, but both of them represent how the data $\mbY_{i}$ is far away from the center and controls the robustness of the estimation. Moreover, they are closely related to each other in expectation under our model. Recall that $\mbE_{i}=\mbY_{i}-{\mbB}\bar{\times}_{(M+1)}\mbX_i$ can be written as $\mbZ_i/\sqrt{G_i}$, where $\mbZ_i$ follows tensor normal distribution, independent of $G_i\sim \chi_{\nu}^2/\nu$. We have $\E(\Vert\mbY_i-{\mbB}\bar{\times}_{(M+1)}\mbX_i\Vert_{\mbI_{p}}^2/p\mid G_i)\propto\E(\Vert\mbY_i-{\mbB}\bar{\times}_{(M+1)}\mbX_i\Vert_{\bolXi}^2/p\mid G_i)=1/G_i$. Therefore, just as the Mahalanobis distance, the weight $\widehat{\omega}_i$ in HOST can be viewed as an imputation for the latent variable $G_i$. By re-weighting the observations with $\widehat{\omega}_{i}=p/\Vert\mbY_i-\widehat{\mbB}^{\APL}\bar{\times}_{(M+1)}\mbX_i\Vert_{\mbI_{p}}^2$, we are still able to reduce the impact of the outliers.

Finally, an especially interesting phenomenon is that, in evaluating $\omega_i$, the high dimensionality is beneficial. Since all the elements in $\mbY_{i}$ share the same latent variable $G_i$, having more of such elements gives us more information about $G_i$, and improves the accuracy in our imputation of $G_i$. Also, as we show in Section \ref{choose}, when $p$ is large, $\nu$ has a minor numerical effect on $\widehat{\omega}_i$. So $\nu$ is not included in the modified one-step estimation.

\section{Theoretical properties}
\label{theory}

In this section, we establish oracle properties and make asymptotic efficiency comparisons among various estimators when the tensor response dimension is fixed. It is not difficult to see that the estimators of $\mbB$ and $\bolXi$ are asymptotically independent in the regression model \eqref{ttr}. We thus focus on the asymptotic analysis for various feasible estimators of $\mbB$. To avoid redundancy, all penalized estimators use the adaptive lasso penalty to encourage elementwise sparsity in $\mbB$; similar results (oracle properties, asymptotic efficiency, etc.) for adaptive group lasso penalty are relegated to the Supplementary Materials (Section \ref{groupp}). { To shed light on the asymptotic effects of mis-specifying the $\td$-distribution degrees of freedom, we let $\nu^*$ denote the true degrees of freedom, $\hatmbB^{\APT}$ be the penalized MLE using $\nu^*$, and $\widetilde{\mbB}^{\APT}$ be the penalized MLE using degrees of freedom $\nu$ which may be different from $\nu^*$. Note that the estimators $\hatmbB^{\APN}$ and $\hatmbB^{\APL}$ and their asymptotics are not affected by the mis-specification of $\nu$. }

\subsection{Asymptotic properties for the penalized MLEs}
\label{apt}
For element-wise sparsity, $\calA=\{(i_1,\cdots,i_{M+1})\mid b_{i_1\cdots i_{M+1}}\neq 0, i_m=1,\cdots,p_m,\ m=1,\cdots,M+1\}$ is the set of nonzero elements in the true parameter $\mbB$. Let $\widehat{\mbB}$ be the minimizer of a certain adaptive penalized objective function, and $\calA_{n}=\{(i_1,\cdots,i_{M+1})\mid \widehat{b}_{i_1\cdots i_{M+1}}\neq 0, i_m=1,\cdots,p_m,\ m=1,\cdots,M+1\}$ be the estimated sparsity set. Denote $\mbB_{\calA}$ as the collection of elements of $\mbB$ corresponding to set $\calA$.  Without loss of generality, we let $\vecc(\mbB)=\big(\vecc(\mbB_{\calA}), \vecc(\mbB_{\calA^c}) \big)$. For a covariance matrix $\mbJ$, $\mbJ_{\calA}$ represents the sub covariance matrix corresponding to set $\calA$.


First of all, we establish the oracle properties (i.e., variable selection consistency and asymptotically normal distribution) for various penalized likelihood-based estimators. Without the penalty term, the APL estimator is the MLE if we assume independent isotropic normal errors (i.e., elements in $\mbE_i$ are all independently drawn from $N(0,\sigma^2)$); the APN estimator is the MLE if we assume tensor normal error; the APT estimator is the MLE when the error follows the tensor $\td$-distribution. 
\begin{theorem}
	Under model \eqref{ttr}, if $\lambda n^{-1/2}\to 0$, $\lambda n^{1/2}\to \infty$, $\nu^*>2$, and $\lim_{n\to\infty}\sum_{i=1}^{n}\mbX_i \mbX_i^{T}/n=\bolSigma_{\mbX}$, then,\\
	1. $\lim_{n\to\infty} P(\calA_{n}^{\APL}=\calA)=1=\lim_{n\to\infty} P(\calA_{n}^{\APN}=\calA)=\lim_{n\to\infty} P(\calA_{n}^{\APT}=\calA)$\\
	2. $\sqrt{n}\{\vecc(\widehat{\mbB}_{\calA}^{\APL}-\mbB_{\calA})\}\to N(0,\mbV_L)$, where $\mbV_L=\frac{\nu^*}{\nu^*-2}\{(\bolSigma_{\mbX}\bigotimes\mbI)_{\calA}\}^{-1}(\bolSigma_{\mbX}\bigotimes\bolSigma)_{\calA}\{(\bolSigma_{\mbX}\bigotimes\mbI)_{\calA}\}^{-1}$.\\
	3.
	$\sqrt{n}\{\vecc(\widehat{\mbB}_{\calA}^{\APN}-\mbB_{\calA})\}\to N(0,\mbV_N)$, where $\mbV_N=(\widetilde{\mbJ}_{\calA})^{-1}$, and $\widetilde{\mbJ}=\frac{\nu^*-2}{\nu^*}\bolSigma_{\mbX}\bigotimes\bolSigma^{-1}$.\\
	4. $\sqrt{n}\{\vecc(\widehat{\mbB}_{\calA}^{\APT}-\mbB_{\calA})\}\to N(0,\mbV_T)$
	,where $\mbV_T=(\mbJ_{\calA})^{-1}$, and $\mbJ=\frac{\nu^*+p}{\nu^*+p+2}\bolSigma_{\mbX}\bigotimes\bolSigma^{-1}$.\\
	5. For the asymptotic covariance $\mbV_T,\mbV_N$, and $\mbV_L$, we have $\mbV_L\geq\mbV_N>\mbV_T$.
	\label{thero1}
\end{theorem}

The conditions $\lambda n^{-1/2}\to 0$ and $\lambda n^{1/2}\to \infty$ are also used in \cite{ad2006}, which indicates that the tuning parameter should be moderately large and the order is between $n^{-1/2}$ and $n^{1/2}$. The condition $\lim_{n\to\infty}\sum_{i=1}^{n}\mbX_i \mbX_i^{T}/n=\bolSigma_{\mbX}$ is mild, which indicates the existence of the second moment for $\mbX$.  We can show that the unpenalized MLE has asymptotic covariance $\mbJ^{-1}$, and under element-wise sparsity model assumption, the non-sparse part of the unpenalized MLE has asymptotic covariance $(\mbJ^{-1})_{\calA}$. Using the inverse property of the block matrix, we have  $(\mbJ^{-1})_{\calA}\geq (\mbJ_{\calA})^{-1}$. The equivalence can only be obtained when $\mbJ$ is a block-wise diagonal matrix after rearranging the sub-matrix corresponding to set $\calA$, which is generally not true. Under Model $\eqref{ttr}$, APT is the most efficient estimation for $\mbB$. The asymptotic covariance $\mbV_T$ reaches the Cramer-Rao lower bound with known set $\calA$. 

All the estimators $\widehat{\mbB}^{\APT}$, $\widehat{\mbB}^{\APN}$, and $\widehat{\mbB}^{\APL}$ have variable selection consistency property. We compared the asymptotic efficiency of the three estimators $\widehat{\mbB}^{\APT}$, $\widehat{\mbB}^{\APN}$, and $\widehat{\mbB}^{\APL}$ in property 5 of Theorem \ref{thero1}.
It shows the advantages of our proposed estimator $\widehat{\mbB}^{\APT}$ over $\widehat{\mbB}^{\APN}$ and $\widehat{\mbB}^{\APL}$. Under Model \eqref{ttr}, $\widehat{\mbB}^{\APT}$ achieves the highest asymptotic efficiency among all the three estimators. Comparing $\widehat{\mbB}^{\APN}$ and $\widehat{\mbB}^{\APL}$, we know that the covariance matrix $\bolSigma$ helps us to obtain more asymptotic efficient results. The equality between $\mbV_N$ and $\mbV_L$ can be obtained when $\bolSigma$ is a diagonal matrix which means that when the error in the model has no correlation, APT and APL have the same performance. Our estimator $\widehat{\mbB}^{\APT}$ further improves $\widehat{\mbB}^{\APN}$  by considering the heavy tail behavior of the error in Model \eqref{ttr}. 


{
\subsection{Effects of mis-specifying $\nu$}
While the true degrees of freedom $\nu^*$ is unknown, we recommend setting $\nu=4$ in APT for practical considerations (see Section \ref{choose}). This means that APT estimator in practice may suffer from model mis-specification if $\nu^*\neq 4$. In the following theorem, we show that, in general, when the degrees of freedom is mis-specified, the asymptotic efficiency loss in APT estimator is related to $\Delta_\nu=\vert\nu^*-\nu\vert$ and is ignorable if $p$ is sufficiently large. 
\begin{theorem}\label{the1}
	Under the same assumptions as in Theorem~\ref{thero1}, if $p\geq \max\{16, 32\Delta_\nu\}$, then $\sqrt{n}\{\vecc(\widetilde{\mbB}_{\calA}^{\APT}-\mbB_{\calA})\}\to N(0,\widetilde{\mbV}_T)$, where ${\mbV}_T<\widetilde{\mbV}_T\leq (1+24\Delta_\nu/{p})\mbV_T$.
\end{theorem}
The explicit form of the asymptotic covariance for $\vecc(\widetilde{\mbB}_{\calA}^{\APT})$ is difficult to express. In the proof of the above theorem, we show that $\widetilde{\mbB}_{\calA}^{\APT}$ is a penalized M-estimator that equals to $\widehat{\mbB}_{\calA}^{\APT}$ if and only if $\Delta_\nu=0$. Nevertheless, we are able to show the asymptotic normality results of  $ \widetilde{\mbB}_{\calA}^{\APT}$ with an efficient loss at most  ${24\Delta_\nu}/{p}\mbV_T$, where $\mbV_T$ is the asymptotic covariance of APT estimator $\hatmbB^{\APT}$ using the true degrees of freedom. When $p$ is much larger than $\Delta_\nu$, the efficient loss is ignorable. For example, in the real data analysis, $p=18000$ and $\nu^*$ seems to be smaller than $4$, then we have ${24\Delta_\nu}/{p}<0.005$. This explains why in numerical studies, setting $\nu=4$ works as well as using true $\nu^*$: $\Delta_\nu$ is a small number when the true degrees of freedom is small. 

If $\nu^*$ is large, then setting $\nu=4$ leads to a large $\Delta_\nu$. One may naturally wonder if this means large efficiency loss for $\widetilde{\mbB}^{\APT}$.   From our numerical studies, we noticed that $\widetilde{\mbB}^{\APT}$ can perform as well as APN even when the data is normally distributed, i.e., $\Delta_\nu=\infty$. We next provide a theoretical explanation for this empirical findings.

\begin{theorem}\label{the2}
 Under the same assumptions as in Theorem~\ref{thero1} and suppose $p> 4$, then, $\sqrt{n}\{\vecc(\widetilde{\mbB}_{\calA}^{\APT}-\mbB_{\calA})\}\to N(0,\widetilde{\mbV}_N)$, where $\mbV_N<\widetilde{\mbV}_N\leq \{1+{14p^3}/{(p-4)^4}\}\mbV_N$.
\end{theorem}
In tensor response model the total dimension $p$ is typically large, the efficient loss is then roughly $14\mbV_N/p$ and thus ignorable. This agrees with our simulation results that the APT estimator with mis-specified $\nu=4$ is as efficient as the APN estimator when the error distribution is normal. Based on the results in Theorems~\ref{the1} and \ref{the2}, we now have theoretical guarantee that using $\nu=4$ is practically almost as good as using $\nu^*$, with a bounded and often small efficiency loss especially when $p$ is large. }

\subsection{Properties of the one-step estimator}
\label{aptonestep}
Although we demonstrated that $\hatmbB^{\APN}$ and $\hatmbB^{\APT}$ have oracle properties, we have a prerequisite that they are the global solutions of \eqref{ad2} and \eqref{g2}, correspondingly. However, since the objective functions are not convex, the prerequisite may not be guaranteed. The following theorem shows the advantage of   $\hatmbB^{\OST}$ in terms of the convergence as the global solution.
\begin{theorem}
	Suppose $np_{-m}>p_m$, $m=1,\dots,M$. With probability one, the estimators $\hatmbB^{\OST}$, $\hatmbB^{\APL}$ and $\hatbolXi$ from Algorithm \ref{onestepalg} are the global solutions to \eqref{onestep1}, \eqref{APL} and \eqref{plugin}, respectively. 
	\label{global}
\end{theorem}
Theorem \ref{global} indicates that the estimator $\hatmbB^{\OST}$ obtained by Algorithm \ref{onestepalg} is always the global solution, which eliminates the gap between the oracle properties of the global and algorithmic solutions in $\hatmbB^{\APN}$ and $\hatmbB^{\APT}$. Then condition $np_{-m}>p_m$ guarantees that the solutions $(\hatbolSigma_1,\cdots,\hatbolSigma_{M})$ of \eqref{plugin}, are all positive definite with probability 1. The proof of the theorem is based on the fact that \eqref{onestep1} and \eqref{APL} are strictly convex and \eqref{plugin} is geodesic convex \citep[e.g.]{rapcsak1991geodesic,liberti2004class,wiesel2012geodesic}. Relatedly, for matrix normal case (e.g. $M=2$ and $\nu=\infty$), see \cite{Drton2020ExistenceAU} for a more precise formula of the necessary of the sample size to guarantee a unique MLE and global convergence of Kronecker separable covariance.   

With algorithmic guarantee, we next establish the oracle properties for the OST.

\begin{theorem}
	Under the tensor t-regression Model \eqref{ttr}, if $\lambda_{n}/{\sqrt{n}}\to 0$, $\lambda_{n}n^{1/2}\to \infty$, $\nu^*>4$, and $\lim_{n\to\infty}\sum_{i=1}^{n}\mbX_i \mbX_i^{T}/n=\bolSigma_{\mbX}$, then we have the following properties.\\
	1. $\lim_{n\to\infty}P(\calA_{n}^{\OST}=\calA)=1$. \\
	2. $\sqrt{n}(\vecc(\widehat{\mbB}_{\calA}^{\OST}-\mbB_{\calA}))\to N(0,\mbV)$,
	where $\mbV=\mbV_T+4(\mbV_L-\mbV_T)/{(\nu^*+p+2)^{2}}$.
	\label{theor2}
\end{theorem}
The asymptotic covariance $\mbV$ is a weighted average of $\mbV_T$ and $\mbV_L$.  Though this one-step estimator is not the most asymptotic efficient one, it is almost the most asymptotic efficient. When $p$ is moderately large, $4/{(\nu^*+p+2)^{2}}$ will be very small. For example, suppose that $\mbX$ and $\mbY$ have been standardized, the difference between $\mbV$ and $\mbV_T$ will be less than $0.1\%$ when $p$ is just $100$. For most tensor response regression applications and our numerical studies, the dimension $p$ often exceeds $1000$. Then we can safely claim that the one-step estimator is almost as efficient as the penalized MLE.

\subsection{High-dimensional theory for HOST}
We study the statistical properties for $\widehat{\mbB}^{\HOST}$ when $p=\prod_{m=1}^M p_m\gg n$. The dimension of the predictor $q<n$ is assumed to be fixed. Let $s$ be the number of non-zero elements in $\mbB$. We first introduce some technical conditions. Throughout the rest of this section, we use $c$ and $C$ to denote generic positive constants that can vary from line to line.
\begin{enumerate}[(A)]
	\item The true degrees of freedom $\nu^*>4$.
	\item The design matrix $\mbbX$ is fixed, the eigenvalues of $\mbbX^T\mbbX/n$ are bounded below by some constant $c$, and $\mbX_{i}$ is bounded for all $i$. \label{assumptionB}
	\item The eigenvalues of $\bolSigma_m$, $m=1,\cdots,M$, are between $1/c$ and $c$.
	\item $\dfrac{\sum_{m=1}^M\log{p_m}}{n}=o(1)$, $\dfrac{\logg(n)}{\prod_{m=1}^M p_m}=o(1)$, and $\dfrac{(\sum_{m=1}^M\log{p_m})p_m}{np_{-m}}=o(1)$ for all $m$.
	\item $n\prod_{m=1}^M p_m^{1/2}\gg s\sqrt{\logg(n)}(\sum_{m=1}^M\log{p_m})$.
\end{enumerate}


With these assumptions, we present our theoretical results for the estimation error of $\mbB$. Define the tensor $k$-norm as $\Vert \mbA\Vert_{k}=(\sum_{j_1,\cdots,j_M}\vert a_{j_1\cdots j_M}\vert^k)^{1/k}$.
The following theorem shows the rate of convergence for $\hatmbB^{\HOST}$.
\begin{theorem}
	Under Model \eqref{ttr} and Assumptions A-E,
	there exists a generic constant $C$ such that, if $\lambda/n=C\sqrt{n^{-1}\sum_{m=1}^M\log{p_m}}$, we have
	\begin{equation*}
		\Vert\widehat{\mbB}^{\HOST}-\mbB\Vert_{1}=O\left(\sqrt{s^2n^{-1}\sum_{m=1}^M\log{p_m}}\right),\quad \Vert\widehat{\mbB}^{\HOST}-\mbB\Vert_{2}=O\left(\sqrt{sn^{-1}\sum_{m=1}^M\log{p_m}}\right),
	\end{equation*} 
	with probability at least $1-C_1\prod_{m=1}^M p_m^{-C_2}-C_3/n-2n\exp(-C_4\prod_{m=1}^M p_m)$.
	\label{hdt}
\end{theorem} 

Theorem \ref{hdt} implies that the HOST estimator is consistent even when the dimension of each mode grows at an exponential rate of the sample size. When $s^2(\sum_{m=1}^M\log{p_m})=o(n)$ or $s(\sum_{m=1}^M\log{p_m})=o(n)$, the estimation error in $\ell_1$ norm or $\ell_2$ norm converges to zero in probability, respectively. We will later show in Theorem~\ref{minimax} that these rates are sharp. Also, Theorem \ref{hdt} is derived without sub-Gaussian or sub-exponential assumptions, indicating that HOST is suitable for high-dimensional heavy-tailed tensor data. Moreover, note that we consider a problem where the number of predictors is fixed, but the response is a high-dimensional tensor with $\prod_{m=1}^M p_m$ elements in total. Thus, our results are fundamentally different from those in linear regression with a diverging number of predictors \citep[e.g]{bickel2009simultaneous}.

Theorem \ref{hdt} requires Assumptions A--E. All of these assumptions are very mild. We discuss them one by one.
Assumption A guarantees the existence of the fourth moment of $\mbE_{i}$. This assumption replaces the popular sub-Gaussian or sub-exponential assumptions in the high-dimensional statistics literature. Note that the $t$-distribution is neither sub-Gaussian nor sub-exponential. Thus, we need innovative techniques to derive concentration inequalities. Recall that $\mbE_{i}$ can be written as $\mbZ_i/\sqrt{G_i}$, where $\mbZ_i$ follows tensor normal distribution, independent of $\mbG_i\sim \chi_{\nu}^2/\nu$. To obtain the tail bound for $\mbE_{i}$, we need to bound quantities related to $\mbZ_i$ and $1/\sqrt{G_i}$. The condition $\nu>4$ guarantees the existence of the second moment of $1/\sqrt{G_i}$, which is sufficient for bounding the one-dimensional random variable $1/\sqrt{G_i}$.

Assumption B adds some restrictions for the design matrix $\mbbX$. The design matrix being fixed is a common assumption in regression problems, because the main interest in regression is to estimate the conditional distribution of $\mbbY$ given $\mbbX$. For random design cases, we can still get the same rate of convergence if $\mbX_{i}$ is sub-Gaussian following our line of proof. Also, recall that we consider the case where $q$ is smaller than $n$. Hence, the assumption that the eigenvalues of $\mbbX^T\mbbX/n$ are bounded below by $c$ is easily satisfied. Assumption C implies that the largest and smallest eigenvalues of $\bolSigma_m$ do not change with dimension $p$, which is a common assumption in the literature \citep[eg.]{catch}. It guarantees that the inverse of the covariance matrix is well-conditioned. 

Assumption D includes a set of requirements on the dimension and the sample size. The assumption $\frac{\sum_{m=1}^M\log{p_m}}{n}=o(1)$ is commonly used in high-dimensional theoretical studies. It controls the growth rate of $p$ with respect to $n$. The assumption $\frac{\logg(n)}{\prod_{m=1}^M p_m}=o(1)$ guarantees the convergence of the weights $\widehat{\omega}_i$, $i=1,\cdots,n$; recall that we discussed in the last paragraph of Section~\ref{Sec:high} that high-dimensionality is beneficial here. This assumption is trivially true for high-dimensional data with $\prod_{m=1}^M p_m\gg n$. The assumption $\frac{(\sum_{m=1}^M\log{p_m})p_m}{np_{-m}}=o(1)$ for all $m$ is imposed to
guarantee that the eigenvalues of $\widehat{\bolSigma}_m$ are bounded by $c$ and $1/c$ with high probability. 
Note that this assumption is very mild for tensor data set. For example, if the dimensions $p_m$, $m=1,\cdots,M$ grow at the same rate, this assumption is implied by the assumption $(\sum_{m=1}^M\log{p_m})/n=o(1)$ even when $M=2$.
This assumption also reveals a fundamental difference between tensor and vector data. For $p$-dimensional vector data, this assumption becomes $\sqrt{\logg(p)p/n}$, which is a very strong assumption. The difference results from the estimation of the covariance matrix. For vector data, it is challenging to estimate the conditional covariance of $\mbY$ in high dimensions, but in tensor data, we can aggregate the information from different modes to achieve consistent estimation. 

Assumption E is concerned with $s$ relative to $n,p_m,m=1,\ldots,M$, and can be viewed as a sparsity assumption. Note that in high dimensions it is often assumed that $s=o(n)$. In this case, Assumption E is implied by $\prod_{m=1}^M p_m>\log{n}$, which overlaps with Assumption D. Since we have a plug-in estimator $\widehat{\mbB}^{\APL}$ in $\widehat{\omega}_i$, controlling the growth rate of $s$ helps to bound the difference between $\widehat{\mbB}^{\APL}$ and $\mbB$, and further guarantees the concentration properties of $\widehat{\omega}_i$.

Next, we show that the rate of convergence we obtained in Theorem \ref{hdt} is optimal in a minimax sense by finding lower bounds for the estimation errors. To state our results, we introduce our parameter space as follows. Let $\calM(\mbA)$ represent the number of non-zero elements of any tensor $\mbA$. We define the set of sparse tensors $\calG=\{\mbA\in\mbbR^{p_1\times\cdots\times p_M \times q}\mid \calM(\mbA)\leq s \}$, and the set of well-conditioned covariance matrices $\calS_m=\{\bolDelta_m\in \mbbR^{p_m\times p_m}\mid \bolDelta_m$ is symmetric and positive definite with the smallest eigenvalue bounded below by a constant $c_m^*\}$.
In Model \eqref{ttr}, the parameters we consider include the coefficient tensor $\mbB$, the covariance $\bolXi$, and the degrees of freedom $\nu$. Let $\bolTheta=(\mbB,\bolSigma_1,\cdots,\bolSigma_M,\nu)$. Define the parameter space
\begin{equation*}
	\calP(s,p_m,c_m^*,m=1,\cdots,M)=\{\bolTheta: \mbB\in\calG,\bolSigma_1\in\calS_1,\cdots,\bolSigma_m\in\calS_M, \nu\in (4,\infty] \}.
\end{equation*}
This parameter space is very general. We only assume that the coefficient tensor is sparse, the covariance matrices have lower-bounded eigenvalues, and the degrees of freedom is greater than 4. We allow the degrees of freedom to be infinity such that the tensor normal distributed noise is a special case of our parameter space. We have the following theorem for this parameter space.

\begin{theorem}\label{minimax}
	Under the tensor t-regression Model \eqref{ttr}, if  Assumption B holds, then there exists a positive constant $\overline{c}$ such that 
	\begin{equation*}
		\inf_{\widehat{\mbB}}\sup_{\bolTheta\in \calP}\mbbP\Big(\Vert\widehat{\mbB}-\mbB\Vert_{1}\geq \phi_2(n,p,s)\Big) \geq\overline{c},\quad
		\inf_{\widehat{\mbB}}\sup_{\bolTheta\in \calP}\mbbP\Big(\Vert\widehat{\mbB}-\mbB\Vert_{2}\geq \phi_1(n,p,s)\Big) \geq\overline{c},
	\end{equation*}
	where $\phi_1(n,p,s)=\sqrt{s\logg(ep/s)/n}$, $\phi_2(n,p,s)=s\sqrt{{\logg(ep/s)}/n}$, and $\calP=\calP(s,p_m,c_m^*,m=1,\cdots,M)$. 
\end{theorem}
In Theorem \ref{minimax}, we show the minimax lower bound for the $l_1$-norm and $l_2$-norm of $\hatmbB-\mbB$ under  Model \eqref{ttr} with the sparsity assumption. This is the first time the minimax lower bound is derived for tensor response regression for arbitrary $M$ and potentially heavy-tailed errors. Moreover, we see that the upper bounds on the rates of convergence obtained in Theorem \ref{hdt} match the lower bounds in Theorem \ref{minimax}. Thus, HOST is optimal in a minimax sense (up to a logarithmic factor). 

We further obtain two byproducts of our study that could be of independent interest. In Supplement Materials, we show the rates of convergence for $\widehat{\omega}_{i}$ and $\hatbolSigma_{m}$. More specifically, we prove that 
$\max_i \vert\widehat{\omega}_i/G_i-\prod_{m=1}^M p_m/\tr(\bolSigma)\vert=O(\sqrt{\logg(n)/\prod_{m=1}^M p_m})$ with high probability. Since $\prod_{m=1}^M p_m\gg n$ , the weight $\widehat{\omega}_i$ is very close to the latent variable $G_i$ times a constant $\prod_{m=1}^M p_m/\tr(\bolSigma)$ for all $i$. 
Also, we showed that the spectral norm of  $\hatbolSigma_m-p_m/\tr(\bolSigma_m)\bolSigma_m$ is of the order of $O(\sqrt{s\logg(p)p_m/(np_{-m})})$, $m=1,\cdots,M$, with high probability. As we discussed for Assumption D, this rate of convergence is sharp for tensor data. This result provides a theoretical guarantee for the estimation of the Kronecker covariance structure in the tensor normal distribution and tensor $\td$-distribution. The two conclusions make the objective function \eqref{onestep1} close to the objective function for the penalized MLE with known $G_i$ and $\bolXi$. Hence, we can expect that $\hatmbB^{\HOST}$ is close to $\hatmbB^{\APT}$ numerically.

\section{Simulation studies}
\label{sim}

\subsection{Simulation set-up}

We carefully investigate the empirical performances of the four estimators that are theoretically studied in the previous sections: APL from solving the adaptively penalized least squares problem \eqref{APL}, APN from maximizing the adaptively penalized tensor normal likelihood \eqref{g2}, APT from maximizing the adaptively penalized tensor-$\td$ likelihood \eqref{ad2}, and finally the one-step estimator OST that solves the weighted least squares problem \eqref{onestep1} and its high-dimensional modification HOST. Some existing methods from the literature are included: the OLS,  the robust reduce rank regression \citep[R4;][]{she2017robust} applied to vectorized tensor response, the sparse tensor response regression \citep[STORE;][]{sun2017store}, and the higher-order low-rank regression \citep[HOLRR;][]{higher}. { We also include a truncated OLS estimator, tOLS, whose elements are set to zero based on element-wise $\chi^2$ test with Bonferroni correction.} All these methods are implemented in {\sc R}. More details about the implementation can be found in the Supplementary
Materials.
For APL, APN, APT, OST, and HOST we use five-fold cross-validation to select the tuning parameter $\lambda$.  {The data splitting procedure of HOST is not employed for our numerical studies (i.e., the two batches of data $\calE_1=\calE_2$ in the estimation procedure).} The tuning parameter with the smallest cross validation prediction error is selected. 
For R4 and HOLRR, the default tuning methods in their {\sc R} packages were used. 
For STORE, the rank is either the true rank (e.g. Model M2) or the one that corresponds to the lowest estimation error.
For each of the following simulation models and settings, we generate 100 independent data sets and report the averages and standard errors of the relative estimation error (REE), true positive rate (TPR), false positive rate (FPR):
\begin{equation*}
	\REE=\dfrac{\Vert \widehat{\mbB}-\mbB\Vert_{F}^{2}}{\Vert \mbB\Vert_{F}^{2}}\times 100,\quad
	\TPR=\frac{\vert\widehat{\calA}\bigcap\calA\vert}{\vert\calA\vert }\times 100,\quad
	\FPR=\frac{\vert\widehat{\calA}\bigcap\calA^{c}\vert}{\vert\calA^c\vert}\times 100,
\end{equation*}
where $\calA$ and $\widehat\calA$ are the true and estimated index sets of the nonzero entries in $\mbB$. 

Let $\AR(\rho):=[\rho^{\vert i-j\vert}]_{ij}$ represent auto-regressive correlation matrix with correlation $\rho$. In all simulation models, we generate data from the tensor response regression Model \eqref{ttr}, where we set $\mbY_i\in\mbbR^{32\times 32}$, $\mbE_i\sim \TT(0,\bolXi,\nu)$, and $\bolXi=\{\bolSigma_1,\bolSigma_2\}=\{\AR(\rho),\AR(\rho)\}$. In the following models, $\rho$ and $\nu$ are set to be 0.5 and 4, correspondingly, unless there are further descriptions.

\begin{itemize}
	\item[M1] Consider that $X_{i}$ is scalar generated from the standard normal distribution. The sample size $n$ is 100.  we generate the true signal $\mbB$ as $b\mbS$, where $b$ controls the signal strength, and $\mbS\in\mbR^{32\times 32}$ is a randomly generated indicator matrix. The proportion of non-zero elements in $\mbS$ is $s$. 
	Parameter $s$ controls the sparsity of $\mbB$. We will show the performance of different methods for different settings of covariance $\rho$, degrees of freedom $\nu$, signal strength $b$, and sparsity $s$. The default values of the parameters are $\rho=0.5$, $\nu=4$, $b=1$, and $s=0.03$. We change each of these parameters while fixing the others at the default value.
	
	\item[M2] Suppose that $\bolalpha_1\in \mbbR^{32}$, $\bolalpha_2\in \mbbR^{32}$, and $\bolalpha_3\in \mbbR^5$. In this model, we assume that $\mbB=\bolalpha_1\circ\bolalpha_2\circ\bolalpha_3$, where the operator $\circ$ is the outer product. Each entry of $\bolalpha_1$ and $\bolalpha_2$ is generated from normal distribution with mean $0$ and variance $0.5$. Then the entries of $\bolalpha_1$ and $\bolalpha_2$ whose absolute values are not among the largest $20\%$ entires are cut to zero. The sample size is set to be 100.
	
	\item[M3] Consider that $\mbX_{i} \in \mbbR^{4}$, where $\mbX_i$ is generated from standard multivariate normal distribution. The sample size $n=50$. We assume that the second slice of $\mbB$ has a cross shape, the third slice has a diagonal shape, the fourth slice has a bat shape, and the other elements of $\mbB$ are zero. The signal strength is 0.8. 
	
	\item[M4] Consider that $\mbX_{i} \in \mbbR^{10}$, and $\mbX_i$ are generated from standard multivariate normal distribution. The sample size is 100. We assume that the variables in every slice of $\mbY$ which have association with $\mbX$ are chosen as a rhombus shape, a bat shape, or a cross shape. The nonzero elements in $\mbB$ have a rhombus shape, a bat shape or a cross shape, respectively, in each $mode$-3 slice. The three sub-models are considered for this model.
	
\end{itemize}

Model M1 is designed to study the performance of proposed estimates under different model settings including the degrees of freedom, the sparsity and signal strength of $\mbB$, and the correlation of the error. The model assumptions for STORE and HOLRR are violated in Model M1. We design Model M2 whose settings satisfy the model assumptions of STORE. More specifically, the coefficient tensor $\mbB$ has sparse tensor low-rank structure \citep{sun2017store}. Model M3 is designed to study the sparsity pattern recovery performances of different methods. For Model M4, we redefine $\calA$ and $\widehat\calA$ as the index sets of the nonzero mode-$(M+1)$ fibers. We aim to show the performance of proposed methods in variable selection for $\mbY$. 


\subsection{Simulation Results}
In all simulations, the results of APT, OST, and HOST are almost identical. This is expected as we have discussed previously. In the following analysis, we will not distinguish these three methods and only show the results of OST in the figures.

Table \ref{tab: table2} indicates that OST performs better than other methods for M1 and M3 in terms of REE and TPR. FPR of OST is not the smallest but is very small. APN and OST perform better than APL for considering covariance information. And OST performs better than APN by further taking account of heavy tail behavior. For M1 and M3, the low-rank structure for STORE and HOLRR is violated, which explains the poor performance of these methods.

\begin{table}[ht!]
		\renewcommand\arraystretch{0.8}
	\resizebox{\textwidth}{!}{
		\centering
		{\scriptsize\begin{tabular}{ccccccc}
				\toprule
				&\multirow{1}{*}{M1}&\multirow{1}{*}{M2}&\multirow{1}{*}{M3}&\multirow{1}{*}{M4 rhombus}&\multirow{1}{*}{M4 bat}&\multirow{1}{*}{M4 cross}\\
				\toprule
				{OLS}&70.15 (2.40)&44.03 (1.11)&171.94 (4.57)&35.97 (0.77)&76.74 (1.64)&63.39 (1.46)\\	
				\midrule
				{tOLS}&3.87 (0.73)&8.21 (0.28)&41.03 (1.09)&3.14 (0.24)&4.11 (0.45)&3.87 (0.40)\\	
				\midrule
				{R4}&48.81(1.09)&10.35(0.70)&91.12(2.45)&5.96(0.49)&12.72(1.04)&10.68(0.83)\\
				\midrule
				{HOLRR}&52.36(1.87)&6.35(0.50)&72.84(0.80)&3.32(0.17)&7.66(0.69)&6.20(0.16)\\		
				\midrule
				{STORE}&67.19(0.42)&1.17(0.13)&68.38(2.58)&12.64(0.48)&26.98(0.89)&34.42(3.41)\\
				\midrule
				{APL}&3.65(0.24)&6.11(0.12)&21.87(1.22)&4.00(0.18)&4.72(0.26)&4.63(0.28)\\
				\midrule
				{APN}&1.36(0.10)&3.75(0.11)&19.81(0.94)&0.89(0.02)&1.61(0.05)&1.83(0.07)\\	
				\midrule
				{OST}&0.61(0.04)&2.64(0.05)&10.05(0.39)&0.49(0.01)&0.92(0.02)&0.94(0.02)\\
				\midrule
				{HOST}&0.60(0.04)&2.60(0.05)&10.27(0.41)&0.49(0.01)&0.95(0.02)&0.96(0.02)\\
				\midrule
				{APT}&0.60(0.04)&2.32(0.05)&9.70(0.36)&0.48(0.01)&0.89(0.02)&0.92(0.02)\\
				\toprule
		\end{tabular}}
	}
	\resizebox{\textwidth}{!}{
		\renewcommand\arraystretch{0.8}
		\centering
		\begin{tabular}{ccccccccccccc}
			\toprule
			&$\quad$TPR&FPR &$\quad$TPR&FPR&$\quad$TPR&FPR&$\quad$TPR&FPR &$\quad$TPR&FPR&$\quad$TPR&FPR$\ $\\
			\midrule
			\multirow{2}{*}{tOLS}&$\quad$99.9&0.16&$\quad$33.3&0.08&$\quad$56.4&0.13&$\quad$99.8&1.33&$\quad$99.8&1.31&$\quad$99.7&1.29$\ $\\
			&$\quad$(0.04)&(0.06)&$\quad$(0.15)&(0.03)&$\quad$(0.36)&(0.23)&$\quad$(0.03)&(0.27)&$\quad$(0.04)&(0.26)&$\quad$(0.01)&(0.25)$\ $\\		
			\midrule
			\multirow{2}{*}{STORE}&$\quad$92.40&50.52&$\quad$100&0&$\quad$51.15&14.94&$\quad$29.0&14.0&$\quad$99.0&10.6&$\quad$100&8.0$\ $\\
			&$\quad$(1.39)&(1.54)&$\quad$(0)&(0)&$\quad$(0.36)&(0.23)&$\quad$(0.85)&(0.26)&$\quad$(0.20)&(0.20)&$\quad$(0)&(0.28)$\ $\\		
			\midrule
			\multirow{2}{*}{APL}&$\quad$100&0.64&$\quad$51.84&0.90&$\quad$90.28&1.89&$\quad$99.9&1.71&$\quad$99.8&1.01&$\quad$99.7&1.12$\ $\\
			&$\quad$(0)&(0.04)&$\quad$(0.55)&(0.04)&$\quad$(1.46)&(0.06)&$\quad$(0)&(0.06)&$\quad$(0)&(0.03)&$\quad$(0)&(0.04)$\ $\\
			\midrule
			\multirow{2}{*}{APN}&$\quad$100&0.72&$\quad$70.56&4.15&$\quad$95.35&4.03&$\quad$100&1.07&$\quad$100&0.72&$\quad$100&1.44$\ $\\
			&$\quad$(0)&(0.16)&$\quad$(0.47)&(0.14)&$\quad$(0.43)&(0.15)&$\quad$(0)&(0.03)&$\quad$(0)&(0.10)&$\quad$(0)&(0.12)$\ $\\
			\midrule
			\multirow{2}{*}{OST}&$\quad$100&0.88&$\quad$75.39&5.59&$\quad$98.36&4.56&$\quad$100&0.54&$\quad$100&0.56&$\quad$100&3.94$\ $\\
			&$\quad$(0)&(0.17)&$\quad$(0.40)&(0.19)&$\quad$(0.20)&(0.13)	&$\quad$(0)&(0.02)&$\quad$(0)&(0.14)&$\quad$(0)&(0.26)$\ $\\
			\midrule
			\multirow{2}{*}{HOST}&$\quad$100&0.84&$\quad$74.60&4.44&$\quad$98.35&4.58&$\quad$100&0.32&$\quad$100&0.56&$\quad$100&3.78$\ $\\
			&$\quad$(0)&(0.16)&$\quad$(0.34)&(0.18)&$\quad$(0.21)&(0.15)	&$\quad$(0)&(0.02)&$\quad$(0)&(0.19)&$\quad$(0)&(0.29)$\ $\\
			\midrule
			\multirow{2}{*}{ APT}&$\quad$100&0.95&$\quad$81.03&6.60&$\quad$98.76&5.09&$\quad$100&1.25&$\quad$100&0.67&$\quad$100&4.65$\ $\\
			&$\quad$(0)&(0.18)&$\quad$(0.38)&(0.26)&$\quad$(0.17)&(0.16)&$\quad$(0)&(0.04)&$\quad$(0)&(0.18)&$\quad$(0)&(0.32)$\ $\\
			\toprule
		\end{tabular}
	}
	\caption{The top half shows the REE, and the bottom half shows the TPR and FPR. The results are based on 100 replicates. Reported numbers are average and standard error (in parentheses). For settings of all these models, correlation $\rho=0.5$, and degrees of freedom $\nu=4$. For M1, the sparsity $s=0.03$, and the signal strength $b=1$. For M4, the signal strength of $\mbB$ is 1. APT, OST, APN, and APL are obtained through objective functions with adaptive group lasso penalty.}
	\label{tab: table2}
\end{table} 

In Figure \ref{dof}, we check the influence of the degrees of freedom $\nu$ on our algorithms. Using the recommended $\nu=4$ returns almost identical results to using the true degrees of freedom in the algorithm. We also show results of using some other degrees of freedom in Algorithm \ref{onestepalg} including 400, $4000$ and $\infty$. When we use relatively large $\nu$, we will lose some accuracy in estimation especially when the data is generated from a tensor $\td$ distribution with small degrees of freedom. If the data is normally distributed, using all the degrees of freedom returns almost identical results. We also try some small $\nu$ in our algorithms such as 10 and 20. The results of using the small $\nu$ are almost the same as using $\nu=4$. The reason is that in the weight of one-step algorithm, the expectation of the Mahalanobis distance is greater than $p$, although we choose $\nu=20$, the weight is still dominated by the Mahalanobis distance. So we can expect that using $\nu=10$ or $\nu=20$ has the same result as using $\nu=4$ or the true $\nu$.

\begin{figure}[ht!]
	\centering
	\includegraphics[scale=0.455]{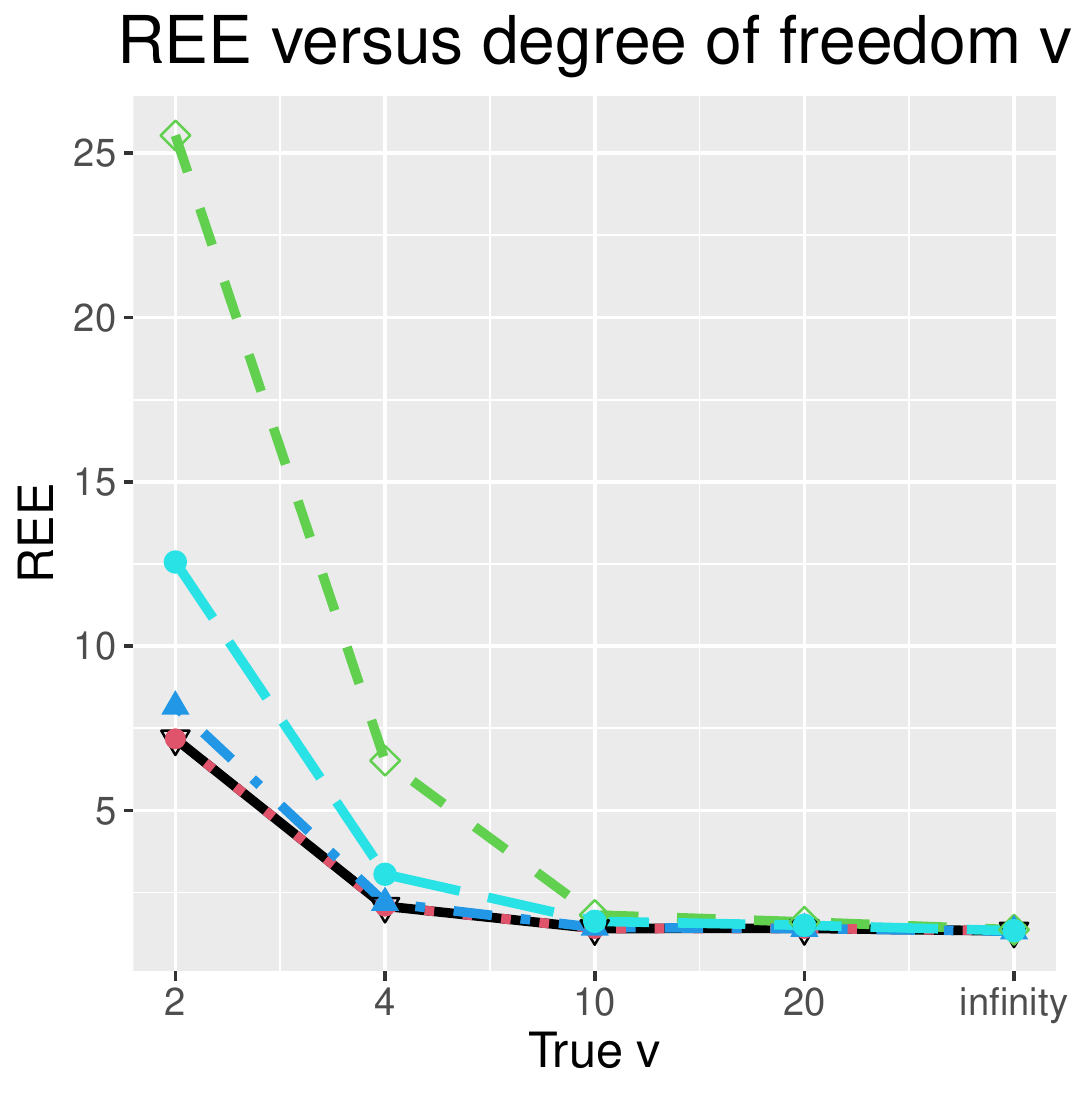}
	\includegraphics[scale=0.47]{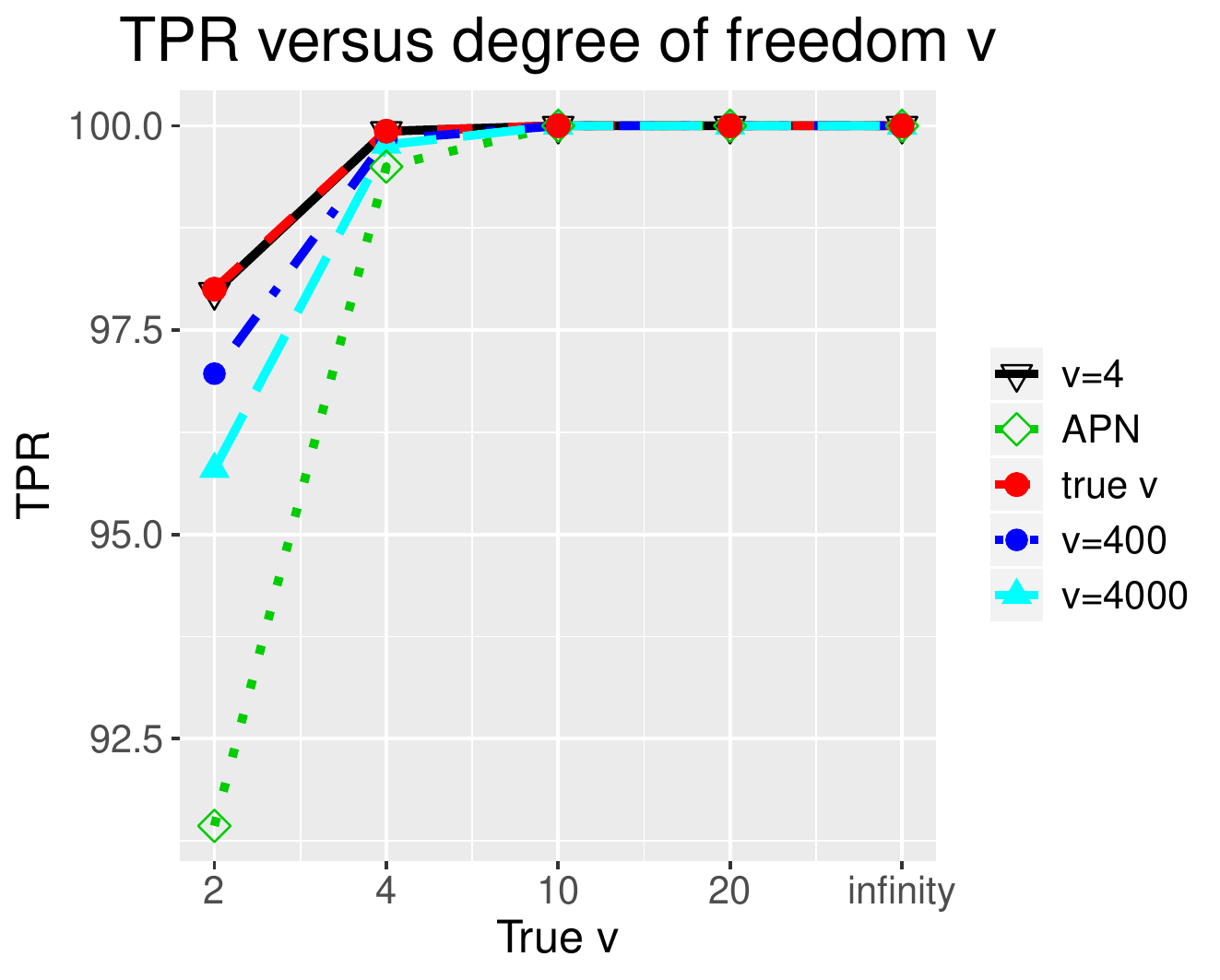}
	\caption{Using different $\nu$ in one step algorithm for M1 given $n=50$, correlation $\rho=0.5$, $b=1$, and $s=0.03$. The setting for degrees of freedom $\nu$ is 2, 4, 10, 20, $\infty$ (normal). The result of using true degrees of freedom is almost identical to using $\nu=4$.}
	\label{dof}
\end{figure}


\begin{figure}[ht!]
	\centering
	\includegraphics[scale=0.45]{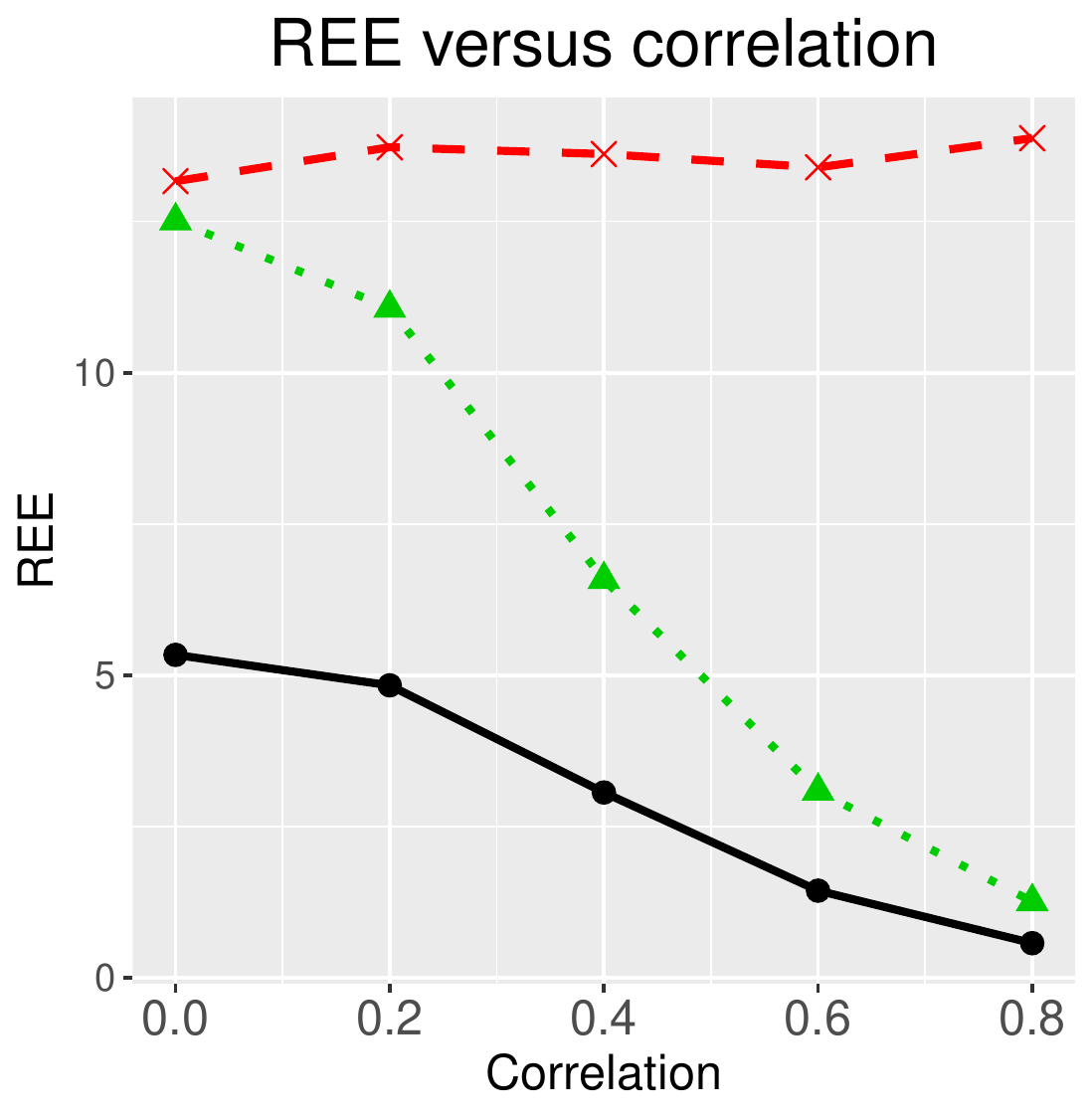}
	\includegraphics[scale=0.45]{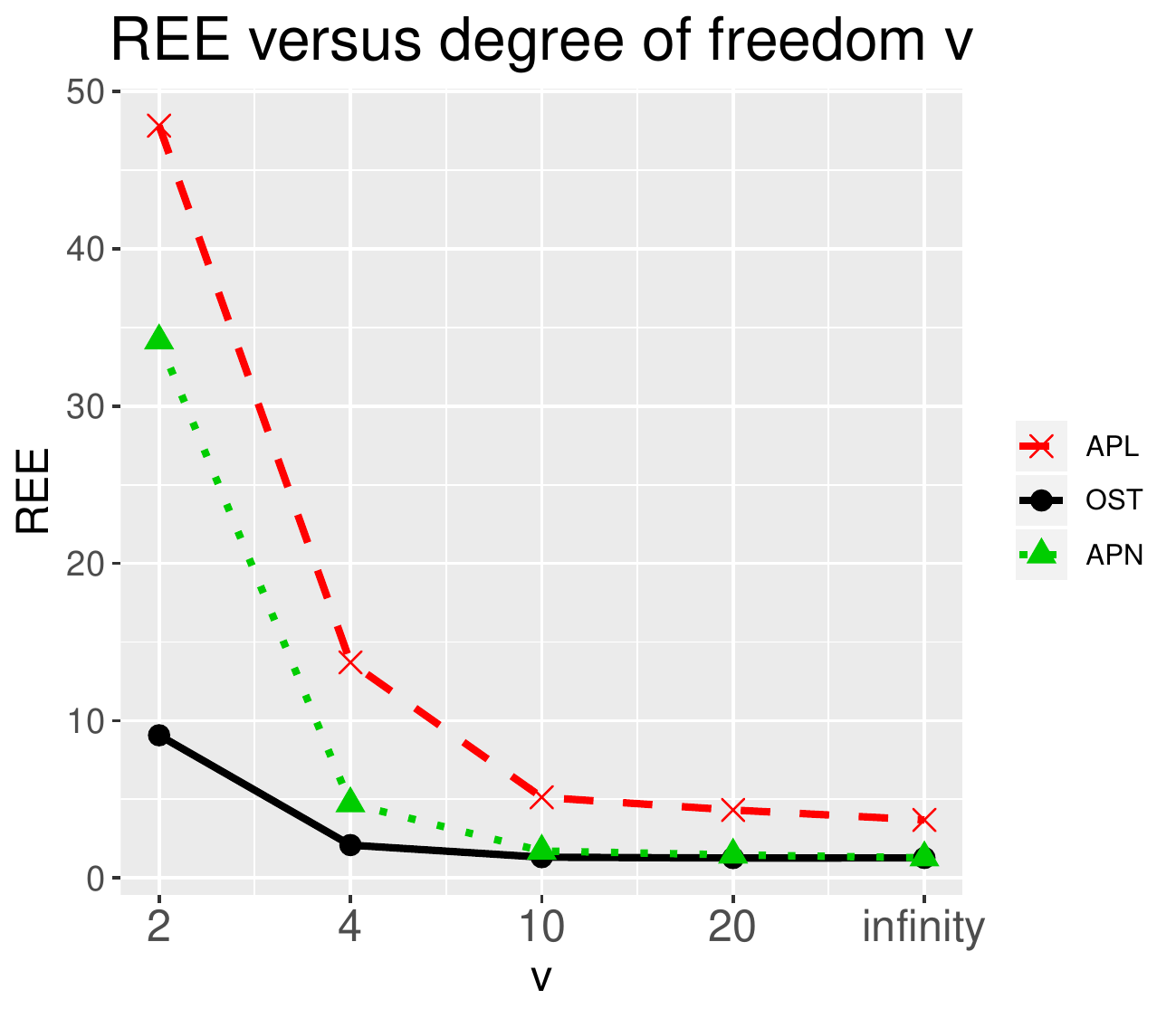}\\
	\includegraphics[scale=0.45]{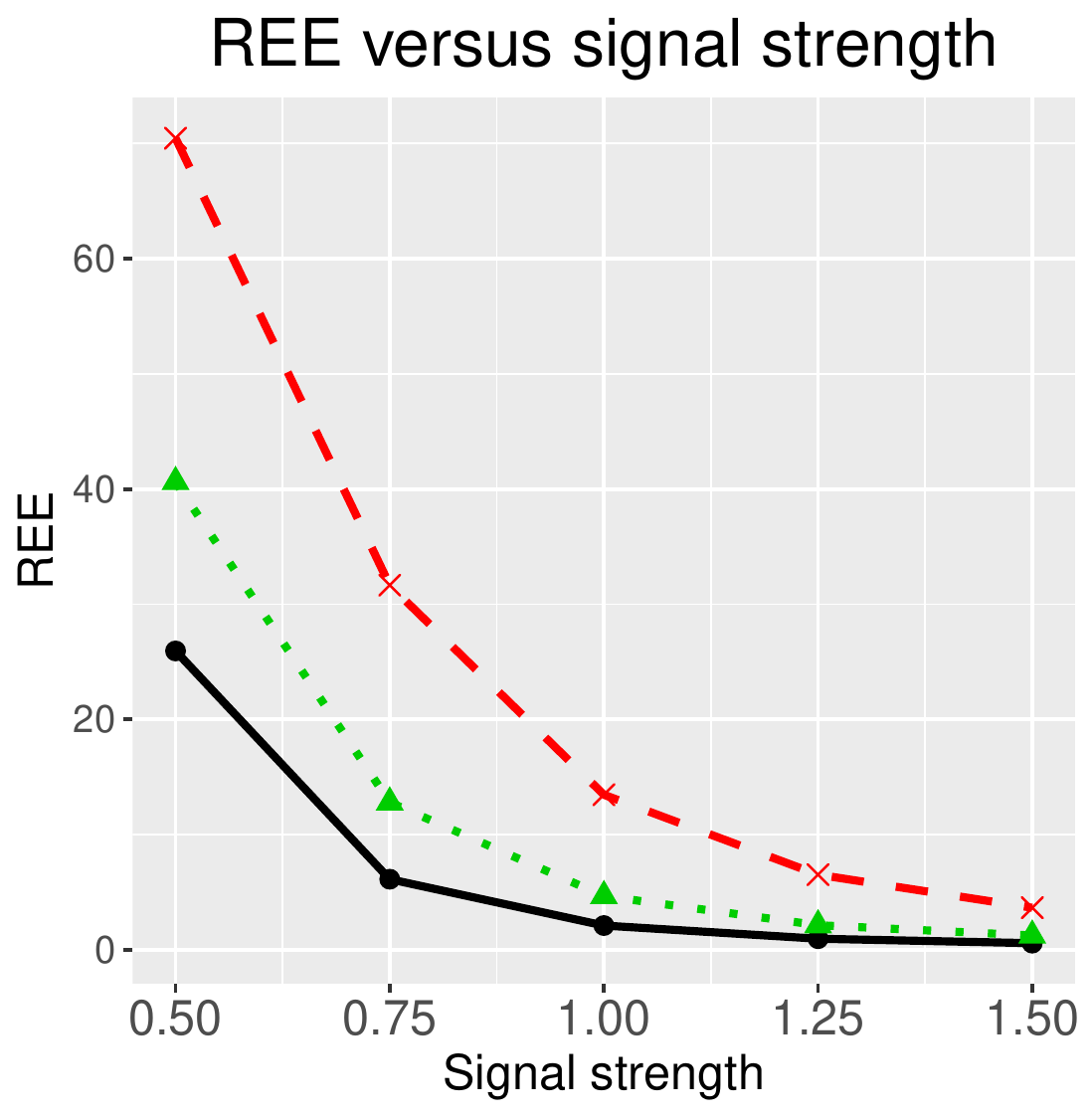}
	\includegraphics[scale=0.45]{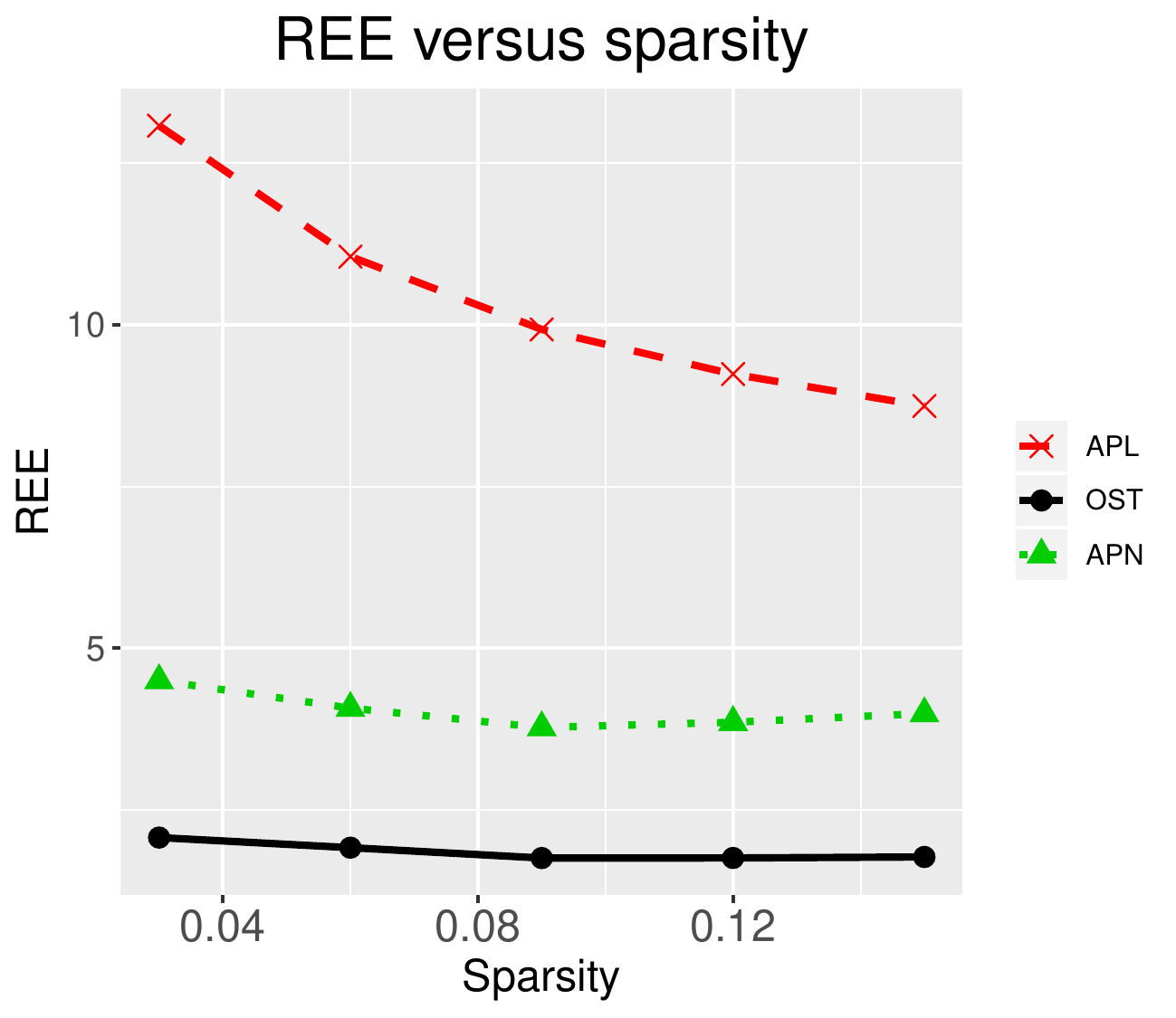}
	\caption{REE for Model M1. The setting for correlation $\rho$ is 0, 0.2, 0.4, 0.6, 0.8, for degrees of freedom $\nu$ is 2, 4, 10, 20, normal distribution, for signal strength $b$ is 0.5, 0.75, 1, 1.25, 1.5, and for sparsity $s$ is 0.03, 0.06, 0.09, 0.12, 0.15.\label{ree}}
\end{figure}


In Figures \ref{ree}, we show the influence of different parameters on REE and TPR.  
We only show results of APL, APN, and OST for their out-performance compared with other methods. Most FPR is smaller than $4\%$ in our simulation studies, we omit the figures for it. With the increase of the correlation $\rho$, APN and OST gain some improvement because of the usage of covariance information. As a comparison, the results of APL do not change much. When the degrees of freedom $\nu$ is small, OST performs much better than other methods as a consequence of taking account of heavy tail issues. When $\nu\to\infty$ which corresponds to the normal distribution case, OST has the same performance as APN and performs better than APL by using the covariance information. With the increase of signal strength, all the methods gain some improvement, but OST still performs best.


In Figure \ref{recover1}, we show the sparsity pattern recovery results of several methods. OST gives the best sparsity recovery result. OLS fails to select any coefficients and gives a vague recovery of the true signal. APL fails to select some true signals, especially for the second and third slices. OST improves the results of APL by considering the heavy tail issue and the covariance information. HOLRR and STORE fail to recover the sparsity pattern because of the violation for the low-rank assumption in $\mbB$. The results of R4 and APN are similar to OLS and APL, respectively. We show them in the Supplementary Materials (Section \ref{add:numerical}).

In Table \ref{tab: table2}, we also display the performance of different methods for variable selection of $\mbY$. Similar to element-wise sparsity case, OST still performs better than other methods in terms of REE and TPR. Although the FPR for OST is not the smallest, it is very small ($<5$). APN performs better than APL by using covariance information. OST further improves the results of APN by taking account of the heavy-tail behavior of the data.

\section{Real data Analysis}
\label{real}
Autism Spectrum Disorder (ASD) is a developmental disability that affects an individual's ability to communicate (e.g., the ability to use language to express one's needs) and the ability to engage in social interactions (e.g., the ability to engage in joint attention). Additionally, the individual may have a restricted range of interests or repetitive behavior. We aim to study the brain area that may be related to ASD through neuroimaging studies.
For this dataset, we have a tensor response from structural functional magnetic resonance imaging (fMRI), a one-dimensional predictor which indicates if the observation has ASD or not, and
additional covariates (age, sex, handed score, IQ). The original data is from the Autism Brain Imaging Data Exchange (ABIDE). We use the dataset from Kennedy Krieger Institute that contains 55 samples, with
22 observations of ASD subjects and 33 normal controls. For each subject, we have a $96 \times 96 \times 47 \times 156$ tensor with the last dimension representing scan time. We first make an average of all the scan time for each individual and then downsize the data set to $30\times 30\times 20$. This downsizing step is to facilitate estimation,
and results in a reduced resolution in images. This is a compromise given the limited sample size and a large number of unknown parameters. 

We first draw the Q-Q plot for the ASD data set to check its normality. Specifically, we first regress the response tensor in predictors using ordinary least square estimation to get the residuals. Then we standardize the residuals by its covariance matrices on each mode. If the data is normally distributed, the standardized residuals should follow $\chi_{p}^{2}$ distribution. For this dataset $p=18000$. The sample quantiles versus the $\chi_{p}^{2}$ quantiles is shown in Figure $\ref{QQ}$. We calculate the weights defined in Proposition 5 when plugging in the OLS estimator. The smallest 5 weights corresponding to the five points most far away from the Q-Q line are 0.297, 0.485, 0.586, 0.599, and 0.639. Most of the other weights are greater than 1. By assigning small weights for the outliers, we can obtain a more robust estimation. 
\begin{figure}[h]
	\begin{center}
		\includegraphics[scale=0.35]{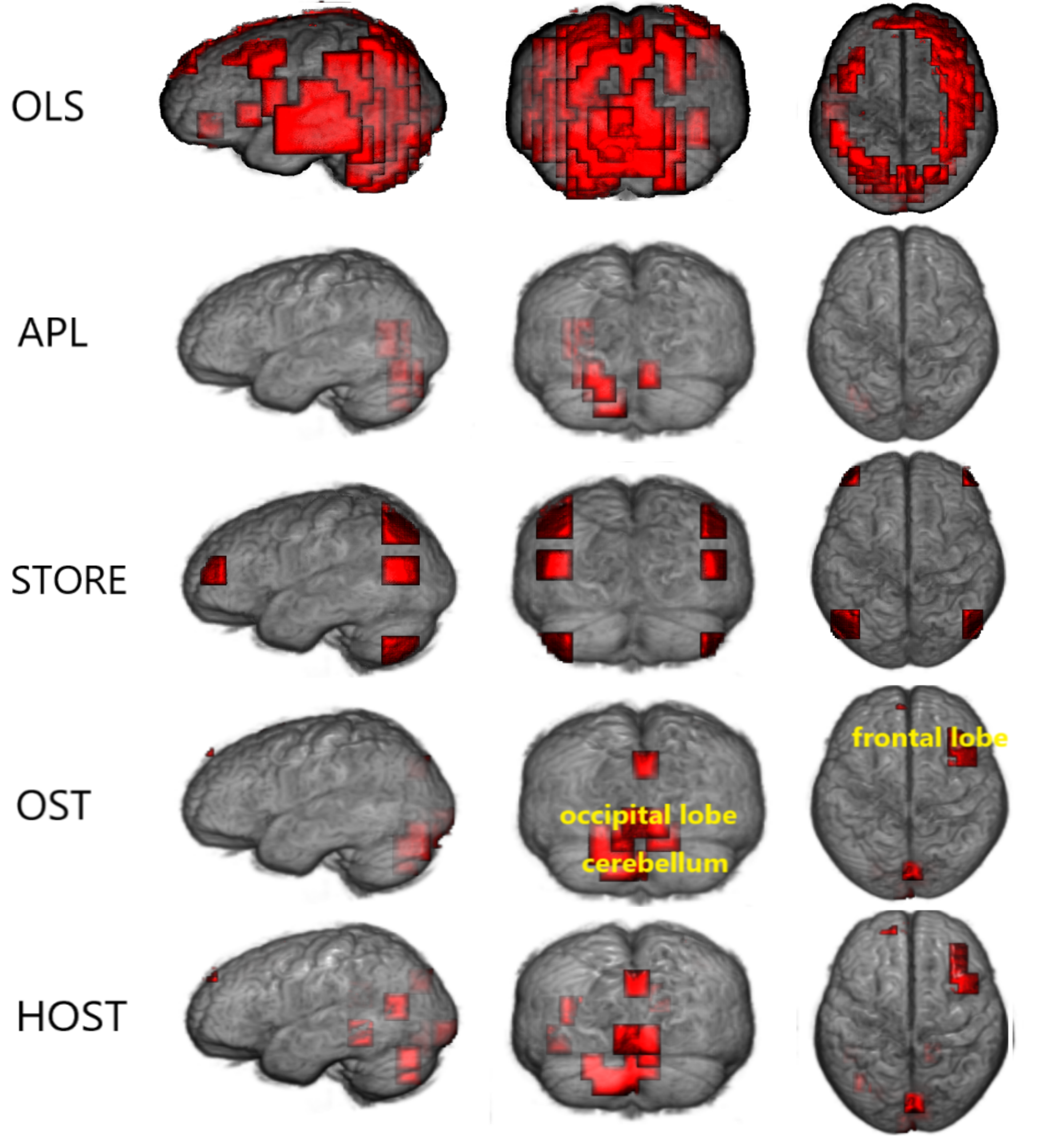}
	\end{center}
	\caption{For OLS, the figure shows the significant areas of the brain for distinguishing if a patient has ASD or not at level 0.05 after Bonferroni Correction. For APL and OST,  the figure shows the significant areas of the brain at level 0.05. For HOST and STORE, the figure shows the nonzero areas.}
	\label{ASD2}
\end{figure}

We showed the significant coefficients selected by OLS, APL, and OST, and the absolute value of non-zero coefficients selected by HOST and STORE in Figure \ref{ASD2}. Without Bonferroni Correction, almost all the areas of the brain are selected. Because we are actually dealing with a multiple testing problem, the significant level 0.05 can be for individual voxels can be conservative in terms of region selection. However, most areas of the brain are selected by OLS even after Bonferroni Correction. It fails to provide meaningful results.
The regions selected by STORE seem symmetric. We use $(s_1=3,s_2=3,s_3=6)$ and $K=5$ for the results shown in the figure. We found that the tunning parameters for STORE influenced the selected area a lot. Slightly Changing the tunning parameters for STORE can result in different selected areas. We see that HOST has a similar performance with OST. More brain areas are selected by OST and HOST. The significant coefficients selected by OST are more concentrated. OST and HOST clearly select the occipital lobe. Recall that in Figure \ref{QQ}, we showed the heavy tail issue of this dataset, we have reason to believe that OST and HOST give us more robust and reliable results for this dataset. Additional figures are provided in Supplementary Materials (Section \ref{add:numerical}).

Several brain regions are identified by OST and HOST including cerebellum, occipital lobe, and frontal lope of the right hemisphere. The regions selected by OST and HOST are consistent with those identified in the literature. 
Cerebellum is primarily responsible for coordinating motor activities such as posture, balance, coordination and eye movement, and is also believed to play a role in language, mental imagery, attention and learned sequences of movements. Cerebellum is shown to be responsible for the poor motor control of ASD patients \citep{stoodley2016cerebellum}. Occipital lobe controls vision, \cite{ha2015characteristics}
showed that people with ASD exhibited greater activity in the bilateral occipital cortex. Frontal lobe controls emotional expression, problem-solving, memory, language, judgment, and sexual behaviors, the area we found in frontal lobe is consistent with the location identified by \cite{margari2018frontal}. 

\section{Discussion}
\label{sec:dis}
	In this paper, we study the tensor response regression with a new tensor $\td$-distribution. The tensor $\td$-modeling approach provides a natural and general strategy for extending popular tensor normal-based statistical models and methods. The proposed robust tensor response regression method simultaneously performs variable selection and estimation in regression mean and covariance functions. We develop a complete set of penalized estimation and algorithms. In particular, we devise a novel one-step estimation approach that is computationally efficient, guaranteed to global optimality, asymptotically nearly as efficient as the oracle-MLE,  and is further modified to ultrahigh dimensional settings, where we establish the minimax estimation rates for tensor response regression prove the optimality of the modified one-step estimator. 
	
	A wide range of tensor problems are solved by (alternating) least squares, where our weighted least squares formulation from tensor $\td$-distribution can be immediately adopted. For instance, the current tensor response regression model can be extended to tensor-on-tensor regression \citep{lock2018tensor, convex2019, llosa2022reduced, luo2022tensor}, where we can assume the error term is tensor $\td$-distributed. Because the response and predictor are both tensors, the regression coefficient in tensor-on-tensor regression is often an even higher order tensor than in the tensor response regression. It is thus more desirable to incorporate tensor low-rank structures and use existing low-rank estimation methods as the initialization \citep[e.g.,][]{luo2022tensor, si2022efficient} to our HOST procedure. Then, we can use the initial estimator to construct weights in our weighted least square formulation. Because the tensor-on-tensor model is more complex and often involves non-convex optimization from the additional low-rank assumption, we expect the theoretical studies of this extension to be an interesting and challenging future work.
\baselineskip=15pt
\bibliography{ref_tensor1}
\bibliographystyle{agsm}

\appendix

\section{Additional properties of matrix t- and tensor t-distributions}\label{add:distribution}

In this section, we show some additional propositions of matrix t- and tensor t- distributions.
The matrix $\td$-distribution and, more generally, the matrix elliptical distributions are important topics in multivariate analysis and Bayesian decision theory. See \citet{dawid1977spherical, anderson1993nonnormal, fang1999bayesian} for some classical results, and \citet{zhang2010multi, thompson2020} for more recent applications in statistical machine learning. The most popular definition of the matrix $\td$-distribution \citep[e.g.,][]{mvd} is given as follows. 
\begin{definition}
	Let $\bolmu \in  \mbbR^{p_1\times p_2}$, and $ \bolSigma_m\in\mbbR^{p_m\times p_m}$, $m=1, 2$, and $\nu>0$. 
	A random variable $\mbY\in \mbbR^{p_1\times p_2}$ follows the matrix $\td$-distribution $\MVT(\bolmu,\{\bolSigma_1,\bolSigma_2\},\nu)$ if it has pdf $f(\mbY)$ as,
	\begin{equation*}
		\frac{\Gamma_{p_1}[\frac{1}{2}(\nu+p_1+p_2-1)]}{\pi^{\frac{1}{2} p_1 p_2}\Gamma_{p_1}[\frac{1}{2}(\nu+p_1-1)]}\vert\bolSigma_{1}\vert^{-p_2/2}\vert\bolSigma_{2}\vert^{-p_1/2}\vert\mbI_{p_1}+\bolSigma_{1}^{-1}(\mbY-\bolmu)\bolSigma_{2}^{-1}(\mbY-\bolmu)^{T}\vert^{-\frac{1}{2}(\nu+p_1+p_2-1)},
	\end{equation*}
	where $\Gamma_{p_1}$ is the $p_1$-dimensional multivariate Gamma function.
	\label{matrixt}
\end{definition}
Clearly, our tensor $\td$-distribution, when $M=2$, is different from this matrix $\td$-distribution. To reveal the connections between the two, consider the following two equivalent representations of the multivariate $\td$-distribution.
\begin{proposition}[\citealt{lin1972}]
	Let $\mbX\sim \N(\mathbf{0},\bolSigma)$ and $G\sim \chi_{\nu}^{2}/\nu$ be independent, $\mbX\in\mbbR^{p}$, then $\mbX/\sqrt{G}+\bolmu\sim t_{p}(\bolmu,\bolSigma,\nu)$. Let  $\mbS\sim \Wis(\bolSigma^{-1},\nu+p-1)$, and $\mbZ\sim \N(\mathbf{0},\mbI_p)$ be independent , then $\sqrt{\nu}\mbS^{-1/2}\mbZ+\bolmu\sim t_{p}(\bolmu,\bolSigma,\nu)$.
	\label{pp1}
\end{proposition}
Our matrix/tensor $\td$-distribution is motivated from the first representation, while the matrix $\td$-distribution in Definition~\ref{matrixt} is obtained from the second representation in Proposition \ref{pp1}.
\begin{proposition}[\citealt{mvd}]
	Let $\mbX\sim \TN(\bolmu,\{\mbI_{p_1},\bolSigma_2\})$, and $\mbS\sim \Wis(\bolSigma_{1}^{-1},\nu+p_1-1)$ be independent, then $\mbS^{-1/2}\mbX+\bolmu\sim \MVT(\bolmu,\{\bolSigma_1,\bolSigma_2\},\nu)$.
	\label{pn:mvt}
\end{proposition}
Proposition \ref{pn:mvt} and the property (a) in Proposition \ref{ppt} result in different probability density functions. One is for matrix $\td$-distribution, while the other one is for our tensor $\td$-distribution. As is shown in the main paper, our $\TT$ distribution is more intuitive and computation appealing. Also, our tensor-$\td$ random variable stays a tensor-$\td$ random variable if we extract a sub-tensor, vectorize or matricize it. Unfortunately, a matrix-$\td$ random variable does not have these properties.

\section{Fixed $p$ statistical inference}
\label{inf}


In this section, we propose the testing procedures for $\mbB$ based on the oracle properties in Section 5. The statistical inference procedures for APT and OST are analogous. We use OST as an example to show the procedure. Based on Theorem \ref{theor2}, we can easily construct Wald-type confidence intervals for $\hatmbB^{\OST}$ that are asymptotically valid. We focus on the testing procedure for $\hatmbB^{\OST}$. 
Let $\widehat{\bolSigma}=\bigotimes_{m=M}^{1}\widehat{\bolSigma}_{m}$, where $\widehat{\bolSigma}_{m}$, $m=1,\cdots,M$ are estimated by Algorithm \ref{onestepalg}, $\widehat{\mbV}_T=\frac{\nu^*+p+2}{\nu^*+p}\{(\hatbolSigma_{\mbX}\bigotimes\hatbolSigma^{-1})_{\calA_{n}} \}^{-1}$, $\widehat{\mbV}=\widehat{\mbV}_T+\frac{4}{(\nu^*+p+2)^{2}}(\widehat{\mbV}_L-\widehat{\mbV}_T) $ and $\widehat{\mbV}_L=\frac{\nu^*}{\nu^*-2}\{(\hatbolSigma_{\mbX}\bigotimes\mbI)_{\calA_n}\}^{-1}(\hatbolSigma_{\mbX}\bigotimes\hatbolSigma)_{\calA_n}\{(\hatbolSigma_{\mbX}\bigotimes\mbI)_{\calA_n}\}^{-1}$. 

We consider the following type of tests,
\begin{equation}
	H_0:h(\vecc(\mbB_{\calA}))=h(\vecc(\mbB_{\calA}^{*})),\quad H_{1,n}(\boldelta): h(\vecc(\mbB_{\calA}))=h(\vecc(\mbB_{\calA}^{*})+\frac{\boldelta}{\sqrt{n}}),
\end{equation}\label{eq:inf}
where $h:\mbbR^{s}\to \mbbR^{k}$, $s=\vert\calA\vert$, $k\leq s$, is a continuously differentiable function, and $\mbB^{*}$ is a constant hypothesized tensor value. Note that $\calA$ can be any index set of interests.
The local alternatives, where $\boldelta$ is a fixed vector in $\mbbR^{s}$, is used for power assessment \citep[see, e.g.][]{Xu2016, kim2020}. 


Let $\mbH(\mbB)=\partial h^{T}(\mbB_{\calA})/\partial \vecc(\mbB_{\calA})$,  $\mbH=\mbH(\mbB_{\calA}^{*})$, and $\widehat{\mbH}=\mbH(\widehat{\mbB}^{\OST}_{\calA})$. We assume that $\mbH$ is a non-singular matrix. Define test statistic
\begin{equation*}
	T=\sqrt{n}\{h(\widehat{\mbB}^{\OST}_{\calA})-h(\mbB_{\calA}^{*})\}^{T}(\widehat{\mbH}^{T}(\widehat{\mbV}_{\calA})^{-1}\widehat{\mbH})^{-1}\sqrt{n}\{h(\widehat{\mbB}^{\OST}_{\calA})-h(\mbB_{\calA}^{*})\}.
\end{equation*}
The following theorem shows the asymptotic properties of $T$.
\begin{theorem}	\label{thero3}
	As $n\to \infty$,  $T\xrightarrow[d]{H_0}\chi_{k}^{2}$, and $T\xrightarrow[d]{H_{1,n}}\chi_{k}^{2}(\boldelta^{T}\mbH(\mbH^{T}(\mbV_{\calA})^{-1}\mbH)^{-1}\mbH^{T}\boldelta)$.
\end{theorem}
Based on Theorem \ref{thero3}, we can test for each element of $\mbB$, linear combinations of $\mbB$, and more generally, a differentiable function of $\mbB$ by test statistic $T$. When $T>\chi_{k}^{2}(1-\alpha)$, we reject the null hypothesis at significant level $\alpha$. The p-value can be calculated by $\mbbP(Z\geq T)$, where $Z\sim \chi_{k}^{2}$.

The proposed testing procedure works asymptotically. For a small finite sample (relative to $p$), we use $\widehat{\mbB}^{\APL}$ in the weight of adaptive lasso penalty instead of using the OLS estimator. The estimator $\widehat{\mbB}^{\APL}$ is $\sqrt{n}$-consistent which satisfies the requirement of adaptive lasso penalty. When we use $\widehat{\mbB}^{\APL}$, the elements in the sparse set will be given a large weight that goes to infinity. In this way, we can weaken the bias of the estimator, especially when $n$ is not large enough.

In practice, since the true degree of freedom $\nu^*$ is unknown, a practitioner can use the recommended value $\nu=4$ to construct the p-value for high dimensions, and use the proposed method (Section S.2.4) to estimate $\nu$ for low-dimensions. In the additional simulation studies (Section S.4.2), we show that this strategy works well. From our Theorems 6 and 7, we know that the APT estimator is still $\sqrt{n}$-consistent and asymptotically normal even under mis-specified $\nu$. Therefore, one may also apply bootstrap methods to obtain the asymptotically valid p-value.

\section{Additional algorithms and implementation details}\label{distribution}
This section contains additional algorithms and implementation details for the one-step estimation, tensor $\td$-distribution, the penalized MLE, and a method for estimating the degree of freedom $\nu$. 

\subsection{Implementation details for the one-step algorithm}
\label{algo}
The one-step estimator is obtained in three steps: the initialization for $\mbB$ by solving \eqref{APL}, the estimation of $\bolXi$ by solving  \eqref{plugin}, and the final OST estimator by solving \eqref{onestep1}. The estimation procedure is summarized in Algorithm \ref{onestepalg}.
%

The optimization problem \eqref{APL} can be separated into the following $p$ sub-problems. 
\begin{equation}
	\begin{aligned}
		\argmin_{\mbB_{J}\in\mbbR^{q}} \sum_{i=1}^{n}\Vert \vecc(\mbY_i)_{J}-\mbB_{J}\mbX_{i}\Vert^{2}+\lambda\sum_{k=1}^{q}r_{q(J-1)+k}\vert \beta_{q(J-1)+k}\vert
	\end{aligned}
	\label{APL1}
\end{equation}
with $J=1,\cdots,p$. The separability of \eqref{APL} indicates that to obtain $\widehat{\mbB}^{\APL}$, we regress one element of $\mbY$ a time without considering the correlation between the elements. Each sub-problem can be solved by coordinate descent algorithm used in \cite{JSSv033i01}. 

For the plug-in estimator $\widehat{\bolXi}$, we provide the details of solving objective function \eqref{plugin} in the Supplementary Materials (Section \ref{distribution}). In the proof of Theorem \ref{global}, we showed that when $np_{-m}>p_m$, the estimate $\hatbolSigma_{m}$ obtained by solving \eqref{plugin} is positive definite with a probability of one. This also guarantees that the objective function \eqref{onestep1} is well-defined. Although the estimation for $\bolXi$ is iterative and non-convex, we used the concept and theory of geodesic convexity \citep{geo} to show that the likelihood-base covariance estimation procedure converges to the global solution. This technique is similar to the existing works on this topic \citep[e.g.,][]{global2}.

Objective function \eqref{onestep1} is a penalized weighted least square problem that is strictly convex. When $P_\lambda(\mbB)=\lambda\sum_{j_1\cdots j_{M+1}}r_{j_1\cdots j_{M+1}}\vert b_{j_1\cdots j_{M+1}}\vert$,
we adopt a coordinate descent algorithm to solve it. The main idea is that, in each iteration, we update one element of $\mbB$ while fixing the others. Let $j=(j_1,\cdots,j_{M+1})$, $\mbB_{-j}$ be a tensor that is identical to $\mbB$ except that the $j$-th element is $0$, $(\widehat{\bolSigma}_{m}^{-1})_{i_m i_m}$ be the $(i_m,i_m)$-th element of $\widehat{\bolSigma}_{m}^{-1}$, and $x_{ii_m}$ be the $i_m$-th element of $\mbX_i$. Let $U_1$ be the $j$-th element of $\sum_{i=1}^{n}\big(\widehat{w}_i\llbracket \mbY_i-\mbB_{-j}\bar{\times}_{M+1}\mbX_{i}; \hatbolSigma_{1}^{-1},\cdots,\hatbolSigma_{M}^{-1},\mbX_{i}\rrbracket\big)$, and $U_2=\sum_{i=1}^{n}\big\{\widehat{w}_ix_{ii_m}^2 \prod_{m=1}^{M}(\widehat{\bolSigma}_{m}^{-1})_{i_m i_m} \big\}$. The iteration for the $j$-th element of $\mbB$ is shown in \eqref{coordinate}.

Algorithm \ref{onestepalg} is much faster than the penalized MLE approach (i.e.~the MM algorithm). In a Windows 10 laptop computer with Intel(R) Core(TM) i7-6700 CPU@3.4GHz, the running time is 3.8s  for one-step algorithm and 101.8s for the MM algorithm, both including cross validation under Model M1 in our simulation studies. The one-step estimator is the solution of a regularized weighted least square problem which is strictly convex. For this weighted least square problem, we can find the largest tuning parameter $\lambda$ such that all coefficients of $\mbB$ are zero, then use the warm start method in \cite{hastie2015statistical} to speed up the computation.  For the selection of tuning parameter, we use five-fold cross validation. Tuning parameter $\lambda$ with the smallest cross validation prediction error is selected.

In parallel to Algorithm~\ref{onestepalg}, we also develop the algorithm for one-step estimation with adaptive group lasso penalty $P_{\lambda}(\mbB)=\lambda\sum_{J=1}^{p}r_{J}\Vert\mbB_{J}\Vert_{2}$. For this scenario, the main difference is in the coordinate descent steps, where we adopt the groupwise-majorization-descent algorithm proposed by \citep{groupalg}. Details are provided in Section \ref{groupp} of the Supplementary Materials. We demonstrate such response variable selection using the group adaptive lasso penalty in our numerical studies.

\subsection{EM algorithm for tensor $\td$-distribution}
For multivariate $\td$-distribution, EM algorithm is widely used to estimate the unknown parameters, for example \cite{robust1989} and \cite{liu1995ml}. In this section, we propose the EM procedure of estimating parameters in tensor $\td$-distribution with known $\nu$ . Suppose that we have independent and identical distributed data $\mbY_i\sim TT(\bolmu,\bolXi,\nu)$, $i=1,\cdots,n$.

In E step of the EM algorithm, we update the weight as $\omega_i^{(k)}=(\nu+p)/(\nu+\Vert\mbY_i-\bolmu^{(k)}\Vert_{\bolXi^{(k)}}^{2})$. 
In M step, similar to the multivariate case, the solution of $\bolmu^{(k+1)}$ is given by $(\sum_{i=1}^{n} \omega_i^{(k)}\mbY_i)/(\sum_{i=1}^{n}\omega_i^{(k)})$.
The scale parameter estimation is different from the multivariate case. We use a method that is a generalization to the flip-flop method \citep[e.g.,]{matrixff,manceur2013maximum} to solve it. The optimization in M step for $\bolXi$ can be summarized as solving the following objective function.
\begin{equation}
	\H_n(\bolXi\mid\mbY_1,\cdots,\mbY_n,\bolmu^{(k)},\omega^{(k)})=\sum_{m=1}^{M}\frac{np_{-m}\logg\vert \bolSigma_m\vert}{2}+\frac{1}{2}\sum_{i=1}^{n}\omega_{i}^{(k)}\Vert\mbY_{i}-\bolmu^{(k)}\Vert_{\bolXi}^{2}.
	\label{fff}
\end{equation}
Let $\widetilde{\bolXi}=\argmin_{\bolXi}\H_n(\bolXi\mid\mbY_1,\cdots,\mbY_n,\bolmu^{(k)},\omega^{(k)})$. We have the following lemma.
\begin{lemma}
	The minimizer  $\widetilde{\bolXi}=\{\widetilde{\bolSigma}_1,\cdots,\widetilde{\bolSigma}_M\}$ satisfies the following equalities.
	\begin{equation*}
		\tilbolSigma_{m}=\frac{1}{n p_{-m}}\sum_{i=1}^{n}\omega_{i}^{(k)}(\mbY_{i}-\bolmu^{(k)})_{(m)}(\bigotimes_{j\neq m}\tilbolSigma_{j}^{-1})(\mbY_{i}-\bolmu^{(k)})_{(m)}^{T},\ m=1,\cdots,M.
	\end{equation*}
	where $\bigotimes_{j\neq m}\tilbolSigma_{j}^{-1}=\tilbolSigma_{M}\bigotimes\cdots\bigotimes\tilbolSigma_{m+1}\bigotimes\tilbolSigma_{m-1}\bigotimes\cdots\bigotimes\tilbolSigma_{1}.$
	\label{ffff}
\end{lemma}
Numerically, we can cyclically update $\tilbolSigma_m$, $m=1,\cdots,M$ until convergence to obtain that $\bolXi^{(k+1)}=\{\widetilde{\bolSigma}_1,\cdots,\widetilde{\bolSigma}_M\}$. If the weights $\omega_{i}^{(k)}$ are set to be 1, then our proposed flip-flop method reduces to the flip-flop method \citep[e.g.,]{matrixff,manceur2013maximum}. 

The algorithm for the estimation of $(\bolmu,\bolXi)$ is summarized in Algorithm \ref{fullmle}.
\begin{algorithm}[t]
	\begin{flushleft}Choose $\bolmu=\overline{\mbY}$, $\bolSigma_m=\mbI_{p_m}$ as initialization.
		For $k=1,2,\cdots$, do
	\end{flushleft}
	\begin{enumerate}
		\item[(E1)] $\omega_i^{(k)}=(\nu+p)(\nu+\Vert \mbY_i-\bolmu^{(k)}\Vert_{\bolXi^{(k)}}^{2})$.
		\item[(M1)] $\bolmu^{(k+1)}=(\sum_{i=1}^{n} \omega_i^{(k)}\mbY_i)/(\sum_{i=1}^{n}\omega_i^{(k)})$.
		\item[(M2)] For $m=1,\cdots,M$, starting with $\bolXi^{(k)}$, cyclically updating the following equation until convergence. 
		\begin{equation}
			\tilbolSigma_{m}=\frac{1}{n p_{-m}}\sum_{i=1}^{n}\omega_{i}^{(k)}(\mbY_{i}-\bolmu^{(k)})_{(m)}(\bigotimes_{j\neq m}\tilbolSigma_{j}^{-1})(\mbY_{i}-\bolmu^{(k)})_{(m)}^{T}.\label{flip}
		\end{equation}
		Then let $\bolXi^{(k+1)}=\{\tilbolSigma_{1},\cdots,\tilbolSigma_M \}$.
	\end{enumerate}
	\begin{flushleft}
		until convergence. At the convergence, output $\hatbolXi=\bolXi^{(k+1)}$ and $\hatbolmu=\bolmu^{(k+1)}$.
	\end{flushleft}
	\caption{Algorithm for obtaining MLE of $(\bolmu,\bolXi)$ given $\nu$}
	\label{fullmle}
\end{algorithm}
The algorithm is a reweighed version compared with the MLE for tensor normal distribution. If all the weights $\omega_i^{(k)}$ in the algorithm are set to be 1, Algorithm \ref{fullmle} reduces to tensor normal MLE algorithm given in \cite{manceur2013maximum}. From Algorithm \ref{fullmle}, we know that the samples far away from the center have relatively small weights in estimation, which is more robust compared with setting all the weights to be 1.

For the algorithm that solves unpenalized MLE of Model \eqref{ttr},
we only need to change the (M1) step of Algorithm \ref{fullmle} into $\mbB^{(k+1)}=\mbbY\times_{M+1}(\mbbX\mbbW^{(k)}\mbbX^{T})\mbbX\mbbW$, where $\mbbW^{(k)}$ is a diagonal matrix with the $i$-th diagonal element being $\omega_i^{(k)}$, then replace all the $\bolmu^{(k+1)}$ by $\mbB^{(k+1)}\bar{\times}_{M+1}\mbX_{i}$.
For the objective function \eqref{plugin}:
\begin{equation*}
	\L_n(\bolXi)\equiv\frac{n}{2}(\sum_{m=1}^{M}p_{-m} \logg\vert\bolSigma_m\vert)+\frac{\nu+p}{2}\sum_{i=1}^{n}\logg(1+\Vert \mbY_i-\widehat{\mbB}^{\APL}\bar{\times}_{M+1}\mbX_{i}\Vert_{\bolXi}^{2}/\nu),
\end{equation*}
we are solving a simplified case where $\bolmu$ is known. To solve it, we only need to omit step (M1) in Algorithm \ref{fullmle}. The detailed algorithm is summarized in Algorithm \ref{xi}.
\begin{algorithm}[ht!]
	\begin{flushleft}
		Choose $\bolSigma_m=\mbI_{p_m}$ as initialization.
		For $k=1,2,\cdots$, do
	\end{flushleft}
	\begin{enumerate}
		\item[(E1)] $\omega_i^{(k)}=(\nu+p)/(\nu+\Vert \mbY_i-\hatmbB^{\APL}\bar{\times}_{M+1}\mbX_{i}\Vert_{\bolXi^{(k)}}^{2})$.
		\item[(M1)] For $m=1,\cdots,M$, starting with $\bolSigma^{(k)}$, cyclically updating the following equation until convergence.
		\begin{equation}
			\hatbolSigma_{m}=\frac{1}{n p_{-m}}\sum_{i=1}^{n}\omega_{i}^{(k)}(\mbY_{i}-\hatmbB^{\APL}\bar{\times}_{M+1}\mbX_{i})_{(m)}(\bigotimes_{j\neq m}\hatbolSigma_{j}^{-1})(\mbY_{i}-\hatmbB^{\APL}\bar{\times}_{M+1}\mbX_{i})_{(m)}^{T}.
			\label{flip1}
		\end{equation}
		Then let $\bolXi^{(k+1)}=\{\hatbolSigma_{1},\cdots,\hatbolSigma_M \}$.
	\end{enumerate}
	\begin{flushleft}
		until convergence. At the convergence, output $\hatbolXi=\bolXi^{(k+1)}$.
	\end{flushleft}
	\caption{Algorithm for solving objective function \eqref{plugin}}
	\label{xi}
\end{algorithm}


%
%

\subsection{Algorithm for penalized MLE}
As is shown in the paper, we use the MM algorithm to solve the penalized MLE. In Section \ref{pmle} of the paper, we have proposed the majorization function $\F_n(\mbB,\bolXi\mid\mbB^{(k)},\bolXi^{(k)})$ for the penalized negative log-likelihood function \eqref{ad2}. In majorization step, we find the majorization function $\F_n(\mbB,\bolXi\mid\mbB^{(k)},\bolXi^{(k)})$ for $k=1,\cdots$, iteratively. In minimization step, we solve $\F_n(\mbB,\bolXi\mid\mbB^{(k)},\bolXi^{(k)})$ to obtain $\mbB^{(k+1)}$ and $\bolXi^{(k+1)}$. In $\F_n(\mbB,\bolXi\mid\mbB^{(k)},\bolXi^{(k)})$. We have two unknown parameters $\mbB$ and $\bolXi$. To achieve the minimization of $\F_n$, we need to solve them back and forth until convergence, which is of high computational cost. In our algorithm, instead of solving them iteratively in each majorization step, we update them only once through the following optimizations. This strategy is similar to the first order EM algorithm shown in \cite{balakrishnan2017}. 
\begin{flalign}
	\bolXi^{(k+1)}=\argmin_{\bolXi}\frac{n}{2}(\sum_{m=1}^{M}p_{-m} \logg\vert\bolSigma_m\vert)+\frac{1}{2}\sum_{i=1}^{n}\omega_{i}^{(k)}\Vert\mbY_{i}-\mbB^{(k)}\bar{\times}_{M+1}\mbX_{i}\Vert_{\bolXi}^{2}, \label{sigma}\\
	\mbB^{(k+1)}=\argmin_{\mbB\in\mbbR^{p_1\times\cdots\times p_{M+1}}}\frac{1}{2}\sum_{i=1}^{n}\omega_{i}^{(k)}\Vert\mbY_{i}-\mbB\bar{\times}_{M+1}\mbX_{i}\Vert_{\bolXi^{(k+1)}}^{2}+P_{\lambda}(\mbB). \label{b}
\end{flalign}

To solve optimization problem \eqref{sigma}, we cyclically do the following iterations,
\begin{equation}
	\hatbolSigma_{m}=\frac{1}{n p_{-m}}\sum_{i=1}^{n}\omega_{i}^{(k)}(\mbY_{i}-\mbB^{(k)}\bar{\times}_{M+1}\mbX_{i})_{(m)}(\bigotimes_{j\neq m}\hatbolSigma_{j}^{-1})(\mbY_{i}-\mbB^{(k)}\bar{\times}_{M+1}\mbX_{i})_{(m)}^{T},\label{flipf}
\end{equation}
until convergence. 

The optimization problem \eqref{b} is parallel to that for one-step estimation in Section \ref{alg}. We can solve it similarly. The detailed algorithm for penalized MLE is summarized in Algorithm \ref{fullalg}. The convergence of the  algorithm is guaranteed since, in each iteration, the objective function value is decreasing.
\begin{algorithm}[ht!]
	\begin{enumerate}
		\item  Let $\mbB^{(0)}$ be the OLS estimator, and $\bolSigma_{m}^{(0)}=\mbI_{p_m}$.
		\item For $k=0,1,\cdots,$ repeat the following updates until convergence.\\
		(a) Update $\omega_i^{(k)}=(\nu+p)/(\nu+\Vert \mbY_i-\mbB^{(k)}\bar{\times}_{M+1}\mbX_{i}\Vert_{\bolXi^{(k)}}^{2})$.\\
		(b) Update $\bolXi^{(k+1)}$ by cyclically updating
		\begin{equation*}
			\hatbolSigma_{m}=\frac{1}{n p_{-m}}\sum_{i=1}^{n}\omega_{i}^{(k)}(\mbY_{i}-\mbB^{(k)}\bar{\times}_{M+1}\mbX_{i})_{(m)}(\bigotimes_{j\neq m}\hatbolSigma_{j}^{-1})(\mbY_{i}-\mbB^{(k)}\bar{\times}_{M+1}\mbX_{i})_{(m)}^{T}.
		\end{equation*}
		\quad(c) Update $\mbB^{(k+1)}$ using the most recently updated  $\bolXi^{(k+1)}$. The solution can be found in the coordinate descent step of Algorithm \ref{onestepalg}.
		\item At convergence, output $\widehat{\mbB}^{\APT}$ and $\hatbolXi^{\APT}$.
	\end{enumerate}
	\caption{Algorithm for penalized MLE}
	\label{fullalg}
\end{algorithm}

\subsection{A method for estimating $\nu$}
\label{df}

We consider two ways for choosing $\nu$. The first one is viewing $\nu$ as a tunning parameter that controls the robustness of the model. As is shown in the main paper, $\nu=4$ is recommended for our methods. In this subsection, we propose a method that can be used to estimate $\nu$. 
In Theorem \ref{thero1}, we show that the adaptive lasso estimator $\widehat{\mbB}^{\APL}$ is a $\sqrt{n}$-consistent estimator for $\mbB$. Meanwhile, $\widehat{\mbB}^{\APL}$ does not depend on $\nu$. Plugging $\widehat{\mbB}^{\APL}$ into the log-likelihood function of model \eqref{ttr}, we obtain 
\begin{equation}
	\begin{aligned}
		\L_n(\nu,\bolXi\mid \widehat{\mbB}^{\APL})&=n\{\logg(\Gamma(\frac{\nu+p}{2}))-\frac{p}{2}\logg(\nu)-\frac{n}{2}\logg\vert\bolSigma\vert-\logg(\Gamma(\frac{\nu}{2}))\}\\
		&-\frac{\nu+p}{2}\sum_{i=1}^{n}\logg(1+\Vert \mbY_i-\widehat{\mbB}^{\APL}\bar{\times}_{M+1}\mbX_{i}\Vert_{\bolXi}^{2}/\nu), \label{plikelihood1}
	\end{aligned}
\end{equation}
where $\bolSigma=\bigotimes_{m=M}^{1}\bolSigma_{m}$. 
We use the ECME algorithm proposed by \citep{liu1995ml} to get the solution. 

The main idea of the algorithm is that given $\nu^{(k)}$ and $\bolXi^{(k)}$, we use the generalized flip-flop method in Lemma \ref{ffff} to solve $\bolXi^{(k+1)}$. And given $\bolXi^{(k+1)}$, we solve the following objective function to get $\nu^{(k+1)}$,
\begin{equation*}
	\begin{aligned}
		\L_n(\nu\mid \bolXi^{(k+1)},\widehat{\mbB}^{\APL})&=n\{\logg(\Gamma(\frac{\nu+p}{2}))-\frac{p}{2}\logg(\nu)-\frac{n}{2}\logg\vert\bolSigma^{(k+1)}\vert-\logg(\Gamma(\frac{\nu}{2}))\}\\
		&-\frac{\nu+p}{2}\sum_{i=1}^{n}\logg(1+\Vert \mbY_i-\widehat{\mbB}^{\APL}\bar{\times}_{M+1}\mbX_{i}\Vert_{\bolXi^{(k+1)}}^{2}/\nu),
	\end{aligned}
\end{equation*}
which is equivalent to finding the solution of following equation
\begin{equation}
	0=-\phi(\nu/2)+\logg(\nu/2)+\frac{1}{n}\sum_{i=1}^{n}\{\logg(\omega_i)-\omega_i \}+1+\frac{1}{n}\sum_{i=1}^{n}\{\phi(\frac{\nu+p}{2})-\logg(\frac{\nu+p}{2})  \},\label{sol}
\end{equation}
where $\omega_{i}=(\nu^{(k)}+p)/(\nu^{(k)}+\Vert\mbY_{i}-\widehat{\mbB}^{\APL}\bar{\times}_{M+1}\mbX_{i}\Vert_{\bolXi^{(k+1)}}^{2})$. The solution to \eqref{sol} can be obtained by the line search method. 
\begin{algorithm}[ht!]
	\begin{enumerate}
		\item  Given the initial value of $\nu^{(0)}$ and $\bolSigma^{(0)}=\mbI_p$
		\item for $k=0,1,\cdots,$ do the following steps until convergence.\\
		(a) Update $\bolXi^{(k+1)}$ using $\nu^{(k)}$ through generalized flip-flop method. More specifically, for $m=1,\cdots,M$, do
		\begin{equation*}
			\tilbolSigma_{m}=\frac{1}{n p_{-m}}\sum_{i=1}^{n}\omega_{i}^{(k)}(\mbY_{i}-\widehat{\mbB}^{\APL}\bar{\times}_{M+1}\mbX_{i})_{(m)}(\bigotimes_{j\neq m}\tilbolSigma_{j}^{-1})(\mbY_{i}-\widehat{\mbB}^{\APL}\bar{\times}_{M+1}\mbX_{i})_{(m)}^{T}.
		\end{equation*}
		until convergence.\\
		(b) $\omega_{i}^{(k+1)}=(\nu^{(k)}+p)/(\nu^{(k)}+\Vert\mbY_{i}-\widehat{\mbB}\bar{\times}_{M+1}\mbX_{i}\Vert_{\bolXi^{(k+1)}}^{2})$.\\
		(c) Update $\nu^{(k+1)}$ using the most recently updated  $\bolXi^{(k+1)}$ by getting the solution for \eqref{sol}.
		\item At convergence, output $\widehat{\nu}$ and $\widetilde{\bolXi}$.
	\end{enumerate}
	\caption{ECME algorithm to get an estimation for $\nu$}
	\label{ECME}
\end{algorithm}
Algorithm \ref{ECME} gives an estimation of degree freedom $\nu$ and a bi-product $\widetilde{\bolXi}$. We can use the estimated $\widehat{\nu}$ in Algorithm \ref{onestepalg}, then get the estimation for $\mbB$ based on this $\widehat{\nu}$. However, estimating $\nu$ through data is not recommended for our methods. Recall that in Figure \ref{QQ}, we generate a simulated data from $TT(0,\mbI_{p},1.5)$. For this simulated data, the estimated degree of freedom $\widehat{\nu}=0.402$, which is much smaller than $1.5$. The estimation for $\nu$ is not robust. Another disadvantage for estimating $\nu$ is that we need to use the line search method to find the solution of \eqref{sol} iteratively, which is time-consuming.

\section{Algorithm and theorem for one-step estimator with adaptive group lasso penalty}
\label{groupp}
\subsection{Algorithm for one-step estimator with adaptive group lasso penalty}
When $P_{\lambda}(\mbB)=\lambda\sum_{J=1}^{p}r_{J}\Vert\mbB_{J}\Vert_{2}$, we use groupwise-majorization-descent algorithm proposed by \citep{groupalg} to solve the objective function \eqref{onestep1}. The main idea of the algorithm is that in each iteration, we fix $p-1$ mode-$(M+1)$ fibers of $\mbB$ and update the $J$-th fiber $\mbB_J$ by minimizing a majorize function. Let $\widetilde{\mbB}$ be the current solution of $\mbB$, $\mbU_{J}$ be the $J$-th row of $\mbU=2\sum_{i=1}^{n}\big(\widehat{\omega}_i\llbracket \mbY_i-\mbB\bar{\times}_{M+1}\mbX_{i}; \bolSigma_{1}^{-1},\cdots,\bolSigma_{M}^{-1},\mbX_{i}\rrbracket\big)$, and
$h_{J}$ be the largest eigenvalue of $\mbH=2(\sum_{i=1}^{n}\widehat{\omega}_{i}\mbX_{i}\mbX_{i}^{T})(\hatbolSigma^{-1})_{JJ}$, where $\hatbolSigma^{-1}=\bigotimes_{m=M}^1\hatbolSigma_m^{-1}$.
More specifically, in each groupwise-majorization iteration, we find the solution for the following objective function,
\begin{equation}
	\F_n(\widetilde{\mbB})-(\mbB_J-\widetilde{\mbB}_{J})^T\mbU_{J}+\frac{1}{2}h_J(\mbB_J-\widetilde{\mbB}_{J})^T(\mbB_J-\widetilde{\mbB}_{J})+\lambda\sum_{J=1}^{p}r_{J}\Vert\mbB_{J}\Vert_{2},
	\label{gmm}
\end{equation}
which majorizes $\F_n(\mbB)$ when we fix the other $p-1$ mode-$(M+1)$ fibers. The solution for \eqref{gmm} is give by
\begin{equation}
	\mbB_{J}^{(new)}=\frac{1}{h_{J}}(\mbU_{J}+h_{J}\widetilde{\mbB}_J)\bigg(1-\frac{\lambda}{\Vert\mbU_{J}+h_{J}\widetilde{\mbB}_J\Vert_2}\bigg)_{+}.
	\label{gos}
\end{equation}
We cyclically update $\mbB_J$ until convergence and output the final result as $\widehat{\mbB}^{\mathrm{OSGT}}$. In each iteration of the algorithm, the function value decreases, so the convergence of the algorithm is guaranteed. The detailed algorithm is parallel to Algorithm \ref{onestepalg}, except that the coordinate descent step for one-step estimation is substituted by \eqref{gos}.

\subsection{Theorem for adaptive group lasso penalized estimation }
\label{grouptheo}
In the paper, we have shown the theoretical results for adaptive lasso penalized estimation. In this subsection, we will show the oracle properties for the group lasso penalized estimation.

Let $\mbG=\mbB_{(M+1)}^{T}$.
For adaptive group lasso penalized case, let $\mbG_J$ be the $J$-th row of $\mbG$. Let $\calG=\{J\mid \mbG_J=0, J=1,\cdots,p\}$, $\calG_n=\{J\mid \widehat{\mbG}_{J}=0, J=1,\cdots,p\}$, and $\mbG_{\calG}$ represent the rows of $\mbG$ corresponding to set $\calG$. 
Suppose that $\widehat{\mbG}^{\AGPT}$ is the global minimizer of the following objective function
\begin{equation*}
	\L(\mbG,\bolSigma)=\frac{n}{2}\logg\vert\bolSigma\vert+\frac{\nu+p}{2}\sum_{i=1}^{n}\logg(1+\Vert \vecc(\mbY_{i})-\mbG\mbX_{i}\Vert_{\bolXi}^{2}/\nu)+\lambda\sum_{J=1}^{p}r_{J}\Vert\mbG_{J}\Vert_{2},\label{adglikeli}
\end{equation*}
where $r_J=\frac{1}{\Vert\widetilde{\mbG}_{J}\Vert_2^{l}}$, and $\widetilde{\mbG}_{J}$ is a $\sqrt{n}$-consistent estimator of $\mbG_{J}$. 
\begin{theorem}
	Suppose $\lambda_{n}/{\sqrt{n}}\to 0$ and $\lambda_{n}n^{(l-1)/2}\to \infty$, $\lim_{n\to\infty}\frac{1}{n}\sum_{i=1}^{n}\mbX_i \mbX_i^{T}=\bolSigma_{\mbX}$. Then we have\\ 
	1. Variable selection consistency: $\lim_{n\to\infty} P(\calG_{n}=\calG)=1$.\\
	2. Asymptotic normality : $\sqrt{n}(\vecc(\widehat{\mbG}_{\calG}^{	\AGPT}-\mbG_{\calG}))\to N(0,\mbV_G)$
	,where $\mbV_G=\mbJ^{-1}$, and $\mbJ=\frac{\nu+p}{\nu+p+2}\bolSigma_{\mbX}\bigotimes(\bolSigma^{-1})_{\calG}$.
	\label{cor1}
\end{theorem}
We make a remark for Theorem \ref{cor1}. Theorem \ref{cor1}  can be viewed as is a special case of Theorem \ref{thero1}. When we have group sparsity, the set $\calA$ will be the collection of all the elements in the rows contained in $\calG$. We can show that $\mbV_{T}=\mbV_G$.

\section{Additional numerical results}\label{add:numerical}
\label{numerical}
\subsection{Additional numerical results for M1}
In Model M1 of the paper, we generate the true signal $\mbB$ as $b\mbS$, where $b$ controls the signal strength, and $\mbS\in\mbR^{32\times 32}$ is a randomly generated indicator matrix. The proportion of non-zero elements in $\mbS$ is $s$.
The indicator matrix $\mbS$ for $s=0.03$ is displayed in Figure \ref{trues}. 
\begin{figure}[ht!]
	\begin{center}
		\includegraphics[scale=0.25]{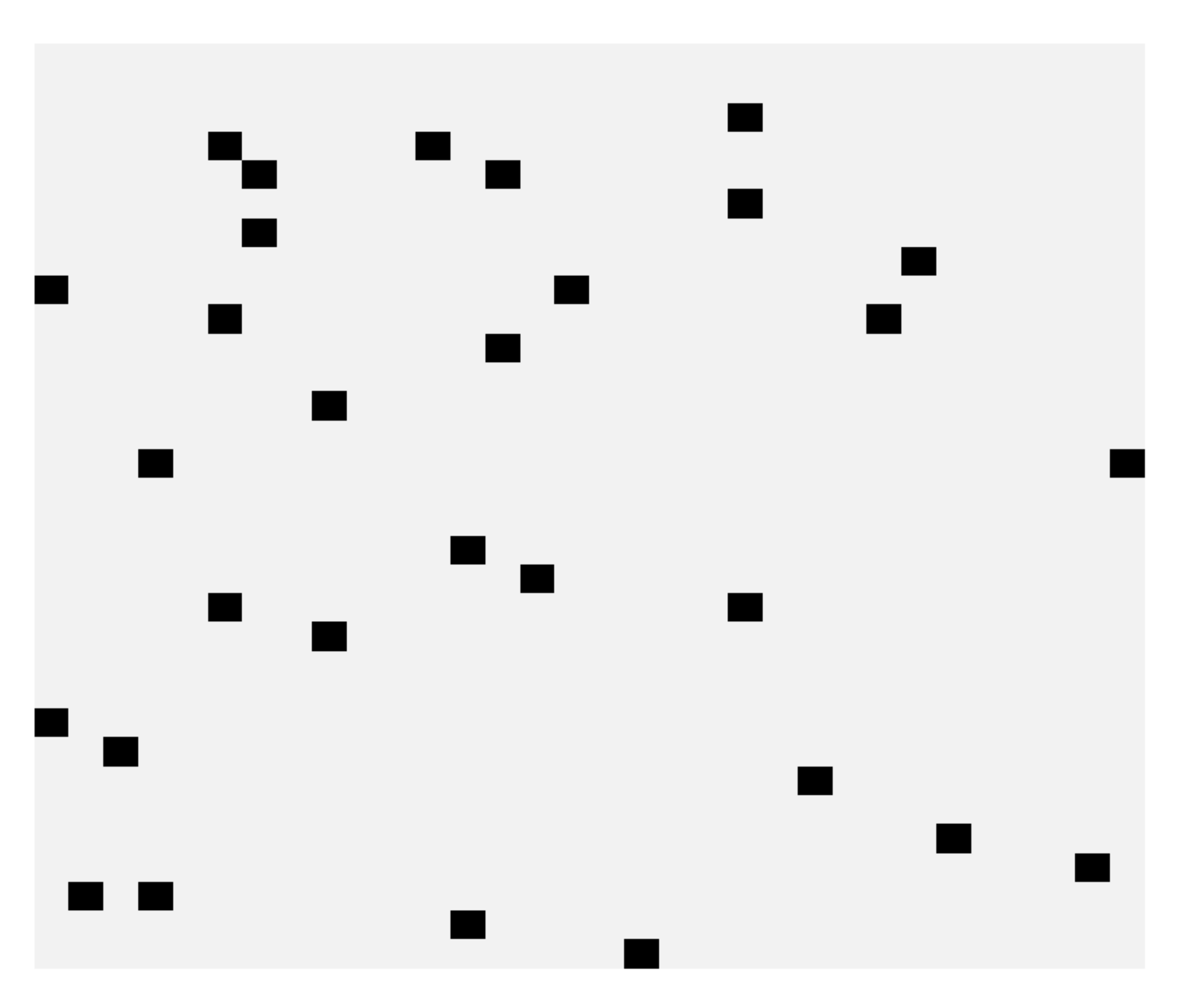}
	\end{center}
	\caption{Indicator matrix $\mbS$ for M1. For hypotheses testing, we test if the first non-zero, namely the 9-th element in the first column, equals to 1. The light gray pixel represents 0 and the dark pixel represents 1. \label{trues}}
\end{figure}
\begin{figure}[ht!]
	\centering
	\includegraphics[scale=0.46]{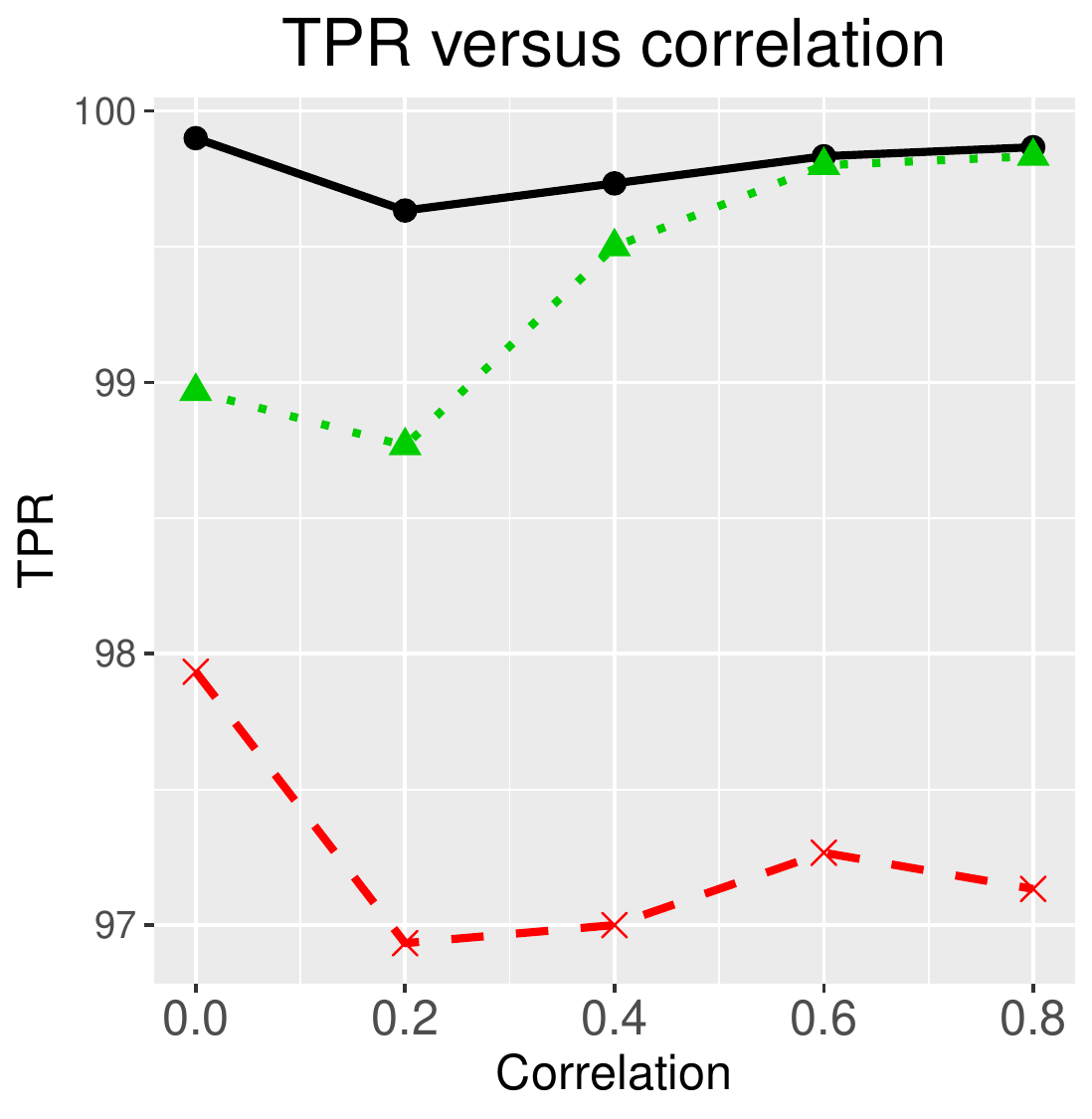}
	\includegraphics[scale=0.46]{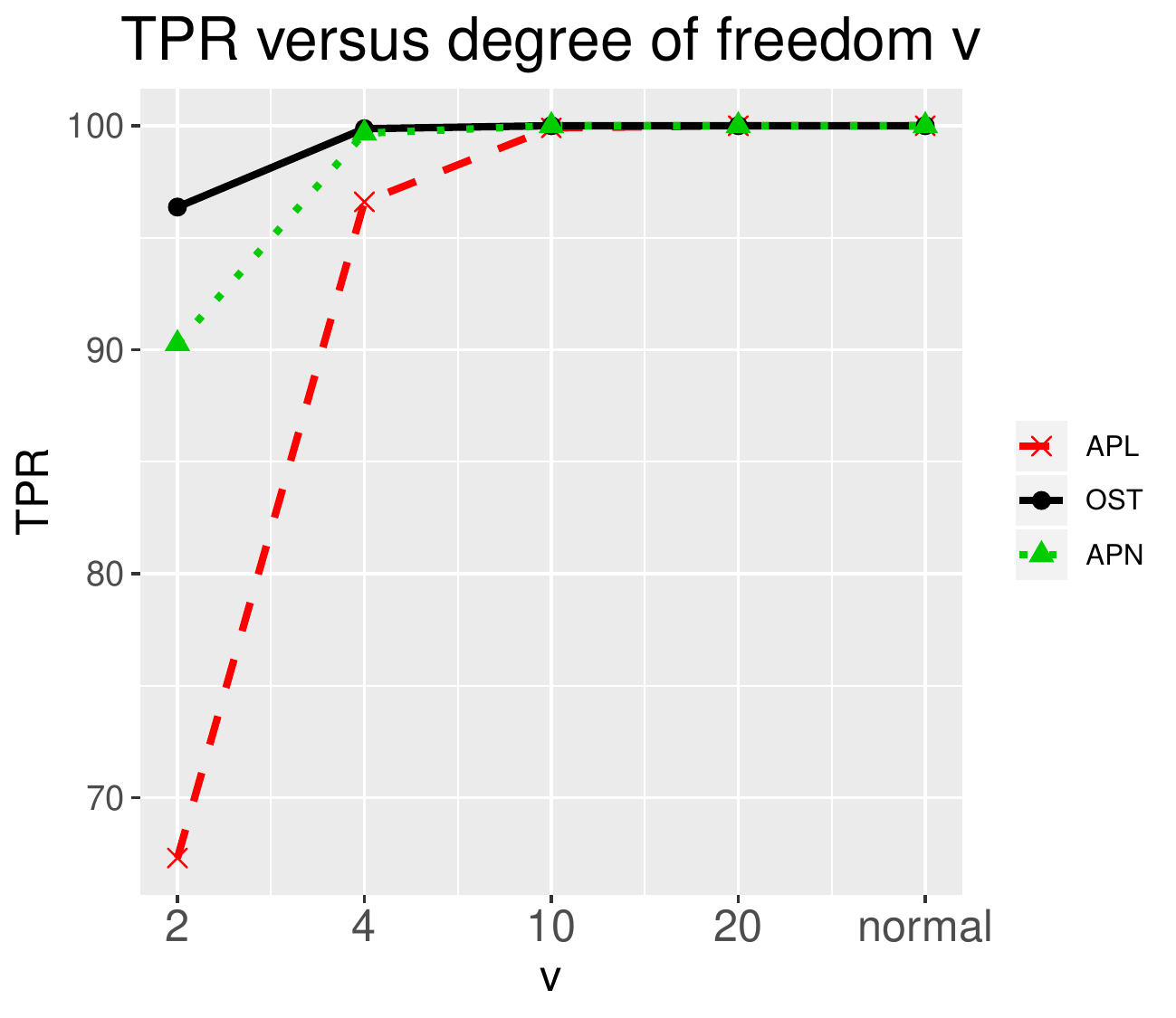}
	\includegraphics[scale=0.46]{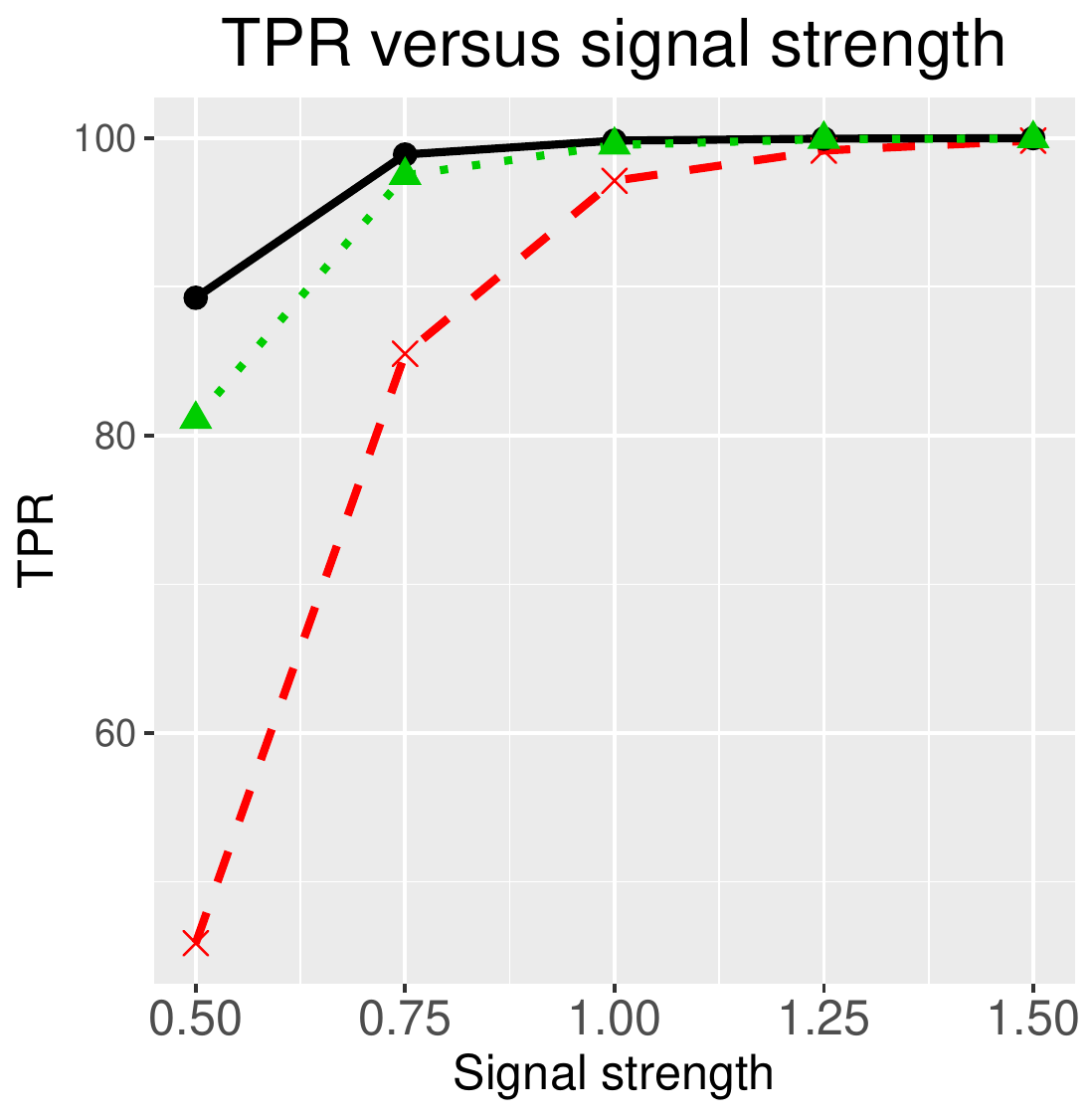}
	\includegraphics[scale=0.46]{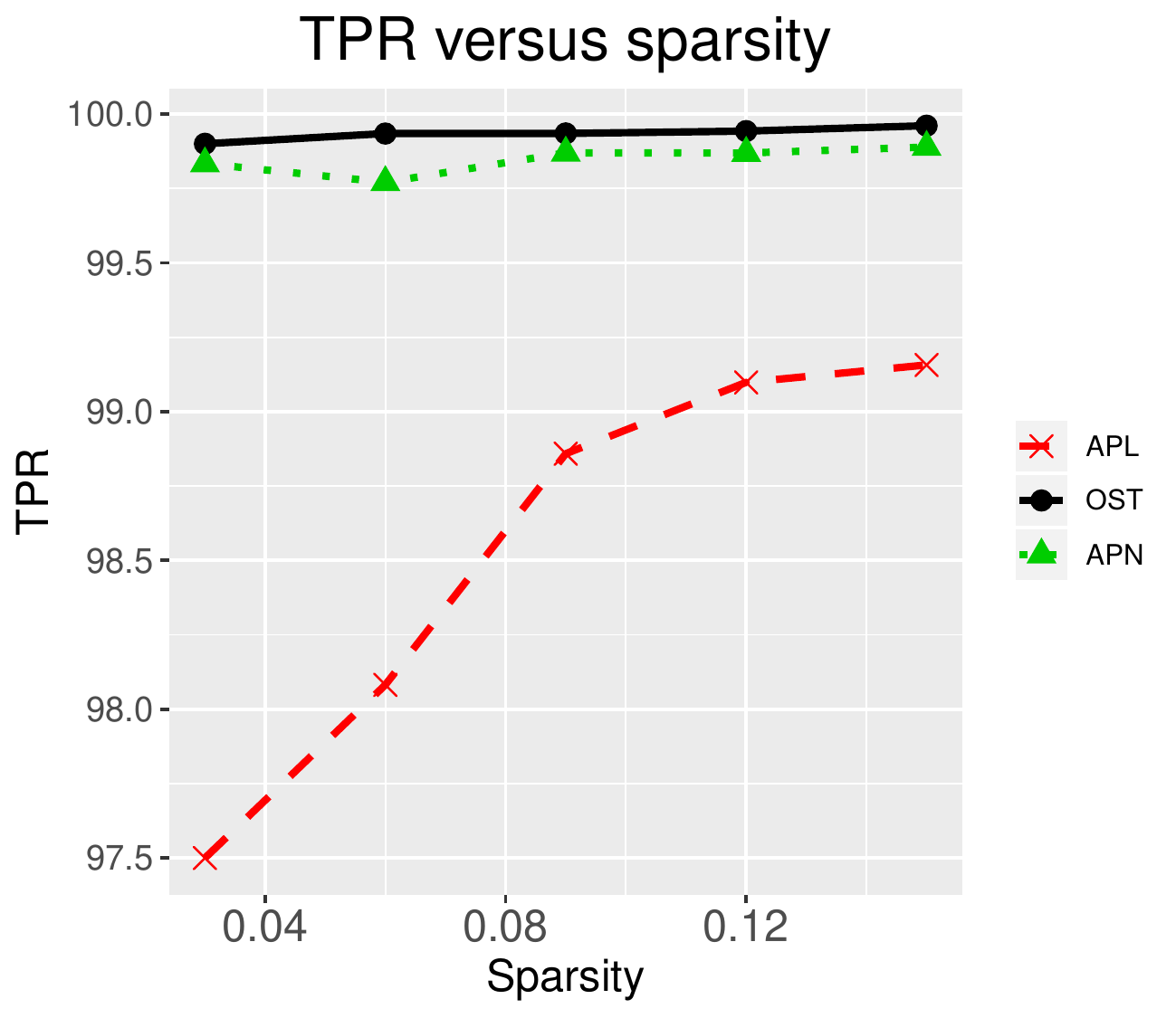}
	\caption{TPR for Model M1. The setting for correlation $\rho$ is 0, 0.2, 0.4, 0.6, 0.8, for degree of freedom $\nu$ is 2, 4, 10, 20, normal distribution, for signal strength $b$ is 0.5, 0.75, 1, 1.25, 1.5, and for sparsity $s$ is 0.03, 0.06, 0.09, 0.12, 0.15. \label{tpr}}
\end{figure}

\begin{table}[ht!]
	\renewcommand\arraystretch{0.8}
	\centering
	\begin{tabular}{ccccc}
		\toprule[2pt]
		&APT&OST&APN&APL\\
		\toprule
		n=100  & 0.094 &0.094& 0.106 &0.110 \tabularnewline
		n=500  & 0.052 &0.052& 0.056 &0.066\tabularnewline
		\toprule[2pt]
	\end{tabular}
	\caption{Type 1 error for testing the first non-zero elements in M1. When the p-value defined in Section \ref{inf} is smaller than $0.05$, we count it. The type-I error is obtained by total counts over 500 replicates.}
	\label{tab: table5}
\end{table}  

\begin{figure}[ht!]
	\centering
	\includegraphics[scale=0.52]{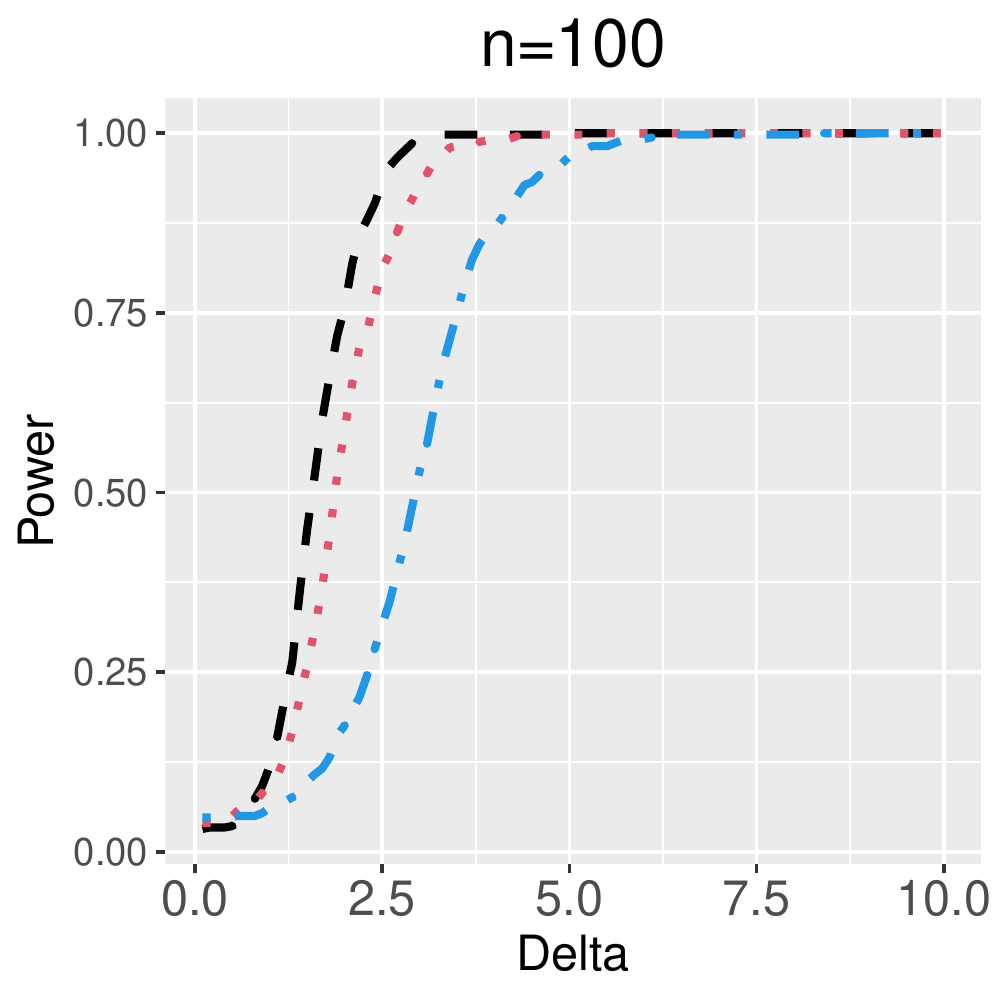}
	\includegraphics[scale=0.52]{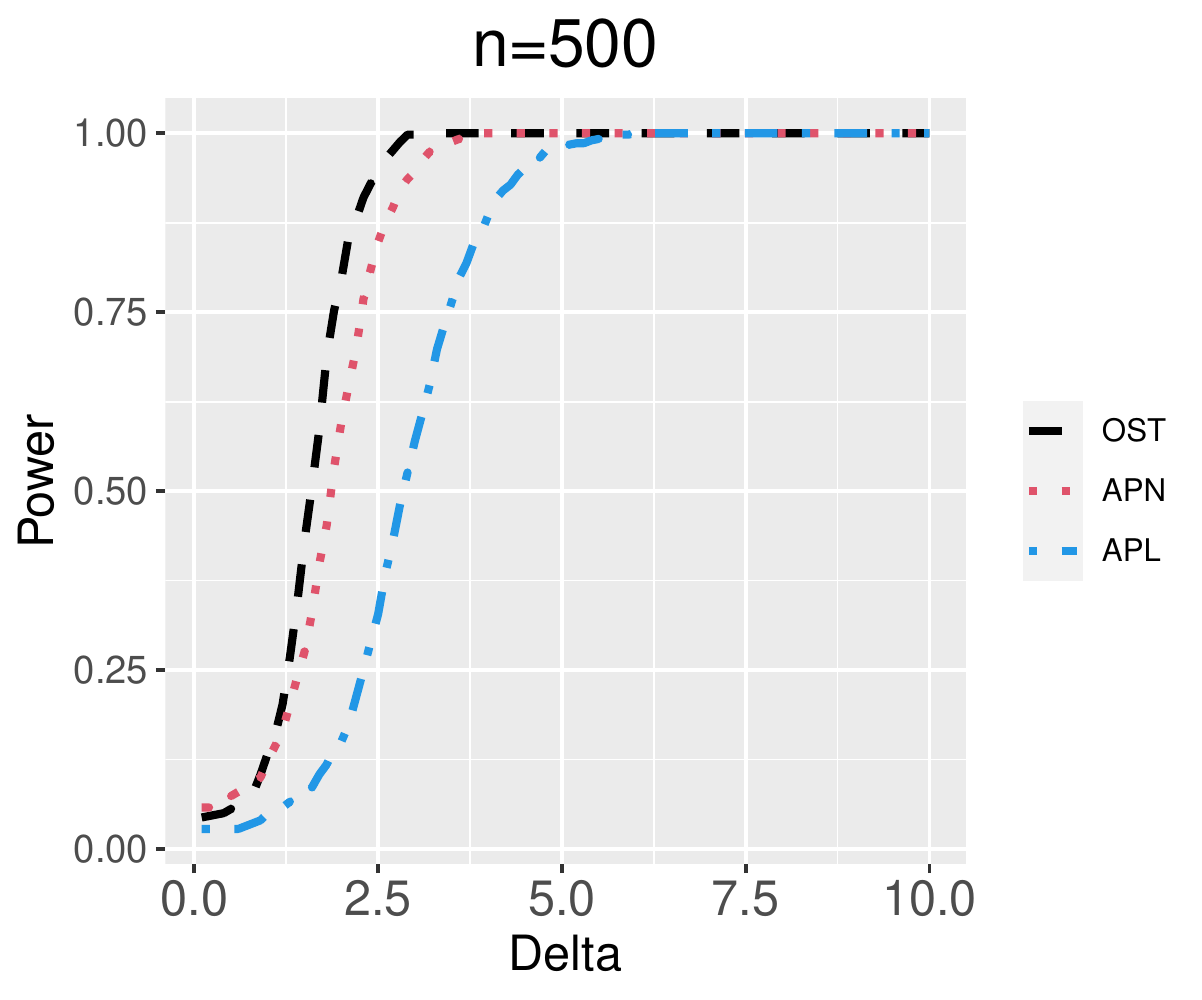}
	\caption{Local power of M1 with $\rho=0.5$, $v=6$, and $b=1$. The two figures show the results of testing the first non-zero elements. Results of OST and APT (not shown) are almost identical. For local alternative test, we varied $\delta$ from 0 to 10 on the horizontal axis. \label{power} }   
\end{figure}
We also use Model M1 to study the proposed hypothesis testing procedure (Section \ref{inf}). We set $\rho=0.5$, $\nu=6$, and $b=1$, and test if the first non-zero element of $\mbB$ equals to the true values or not. The true signal of $\mbB$ is provided in Supplementary Materials (Section \ref{add:numerical}). The p-values in Table \ref{tab: table5} are based on 500 replicates. To obtain the p-value, we count the replicates where test statistic $T$ defined in Section \ref{inf} is greater than $\chi_{k}^{2}(0.95)$ over 500 replicates. 

From Table \ref{tab: table5}, with the increase of sample size, the results of all those methods are close to the significant level 0.05. However, Figure \ref{power} indicates that OST is more powerful than the other two methods. By taking account of heavy tail behavior and covariance information, OST gains much higher power than APL. Statistical inference is difficult for high-dimensional cases. A large sample size is required to guarantee that the empirical p-value converges to the significant level. For our simulation, when $n=500$, the empirical $p$-value is close to the significant level 0.05.


\subsection{Additional numerical studies on hypothesis testing}
In the simulation section, we have seen that fixing $\nu=4$ produces satisfactory results. In this section, we further study how the choice of $\nu$ influences the type-I errors of APT and OST. The model settings are the same as in model M1, except that the true degrees of freedom $\nu^*=6$. We set $n=200$, the sparse coefficients are chosen as the first $2\times 2$ block of $\mbB$, and consider either $p_1=p_2=10$ or $p_1=p_2=32$. We test whether the first element of the  regression coefficient equals 1. 

Table \ref{tab: table3} contains the type-I errors for APT and OST using different $\nu$ in the estimation procedures. For both scenarios, the type-I errors of APT and OST methods using the true $\nu^*=6$ are about the same as the results from the mis-specified $\nu=4$ or estimated $\widehat\nu$ from the algorithm in Section S.2.4. This agree with our suggestion of simply fixing $\nu=4$ and shows the validity of our estimator $\widehat\nu$ for low-dimensional tensor data. Moreover, from the results of $\nu$ being more severely mis-specified as $\nu=0$ or $\nu=10$, we see that the effect of mis-specifying $\nu$ is noticeable for low-dimensional tensor $p_1=p_2=10$ but ignorable for higher dimensions when $p=32\times32=1024$. This simulation experiment agrees with our asymptotic results in Theorem 6: for large enough $p$, a misspecified $\nu$ has a minor influence on the asymptotic variance. As such, we recommend using $\nu=4$ as default for hypothesis testing. When $p$ is small, one may also consider using the estimated $\widehat\nu$ from our proposed algorithm in Section S.2.4. or using bootstrap methods. 

\begin{table}[t!]
	\renewcommand\arraystretch{1.2}
	\resizebox{\textwidth}{!}{
		\centering
		{\scriptsize\begin{tabular}{c|ccccc|ccccc}
				\hline
				&\multicolumn{5}{c|}{$p_1=p_2=10$}&\multicolumn{5}{c}{$p_1=p_2=32$}\\
				$\nu$&0&4&\textbf{6}&10&$\widehat{\nu}$&0&4&\textbf{6}&10&$\widehat{\nu}$\\
				\hline
				{APT}&0.051&0.060&0.060&0.056&0.059&0.059&0.058&0.058&0.059&0.058\\	
				\hline
				{OST}&0.051&0.060&0.060&0.056&0.059&0.059&0.058&0.058&0.059&0.058\\	
				\hline
		\end{tabular}}
	}
	\caption{Type-I error, obtained by counting the proportions of the p-values (defined in Section \ref{inf}) smaller than $0.05$ over 1000 replicates. The columns with $\widehat{\nu}$ are APT and OST using estimated $\nu$ from Section S.3.4.}
	\label{tab: table3}
\end{table}

Next, we compare the powers and the type-I errors of the proposed methods and  competitors such as APN, APL and OLS. From Table \ref{tab:type1}, the type-I errors are controlled around 0.05 for all methods. From Figure \ref{fig:local}, we can see that the power of OLS is almost the same as APL, and APT is more powerful than OLS and APL. Details of the testing procedure and its theoretical justifications are provided in Section \ref{inf}.

\begin{table}[ht!]
	\renewcommand\arraystretch{1.2}
	\centering
	{\begin{tabular}{|ccccc|}
			\hline
			APT&OST&APN&APL&OLS\\
			\hline
			\multicolumn{5}{|c|}{$p_1=p_2=10$}\\
			\hline
			0.060&0.060&0.060&0.061&0.057\\	
			\hline
			\multicolumn{5}{|c|}{$p_1=p_2=32$}\\
			\hline
			0.058&0.058&0.066&0.059&0.049\\	
			\hline
	\end{tabular}}
	\caption{Type-I error based 1000 replicates. }
	\label{tab:type1}
\end{table} 

\begin{figure}[ht!]
	\centering
	\includegraphics[scale=0.38]{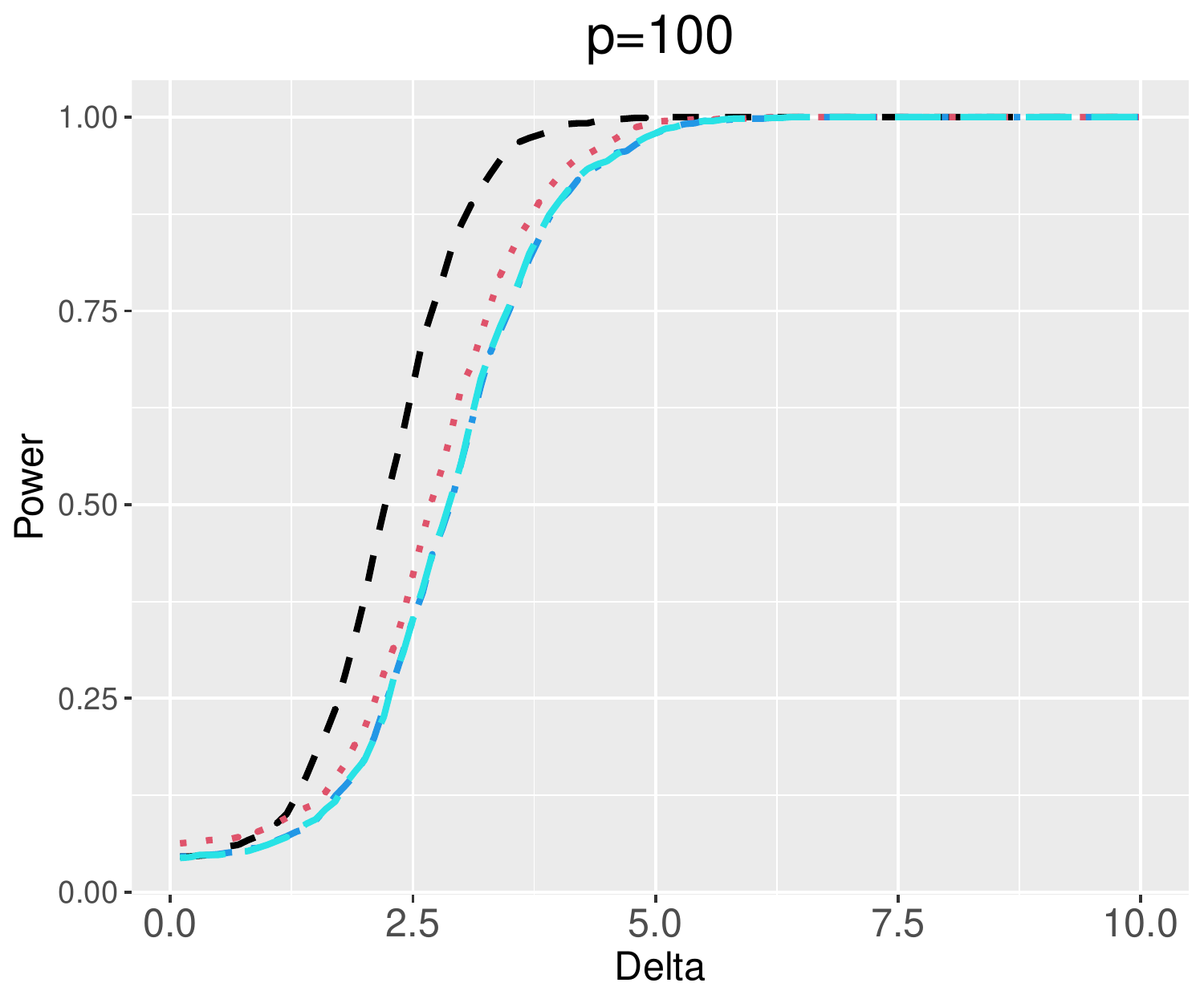}
	\includegraphics[scale=0.38]{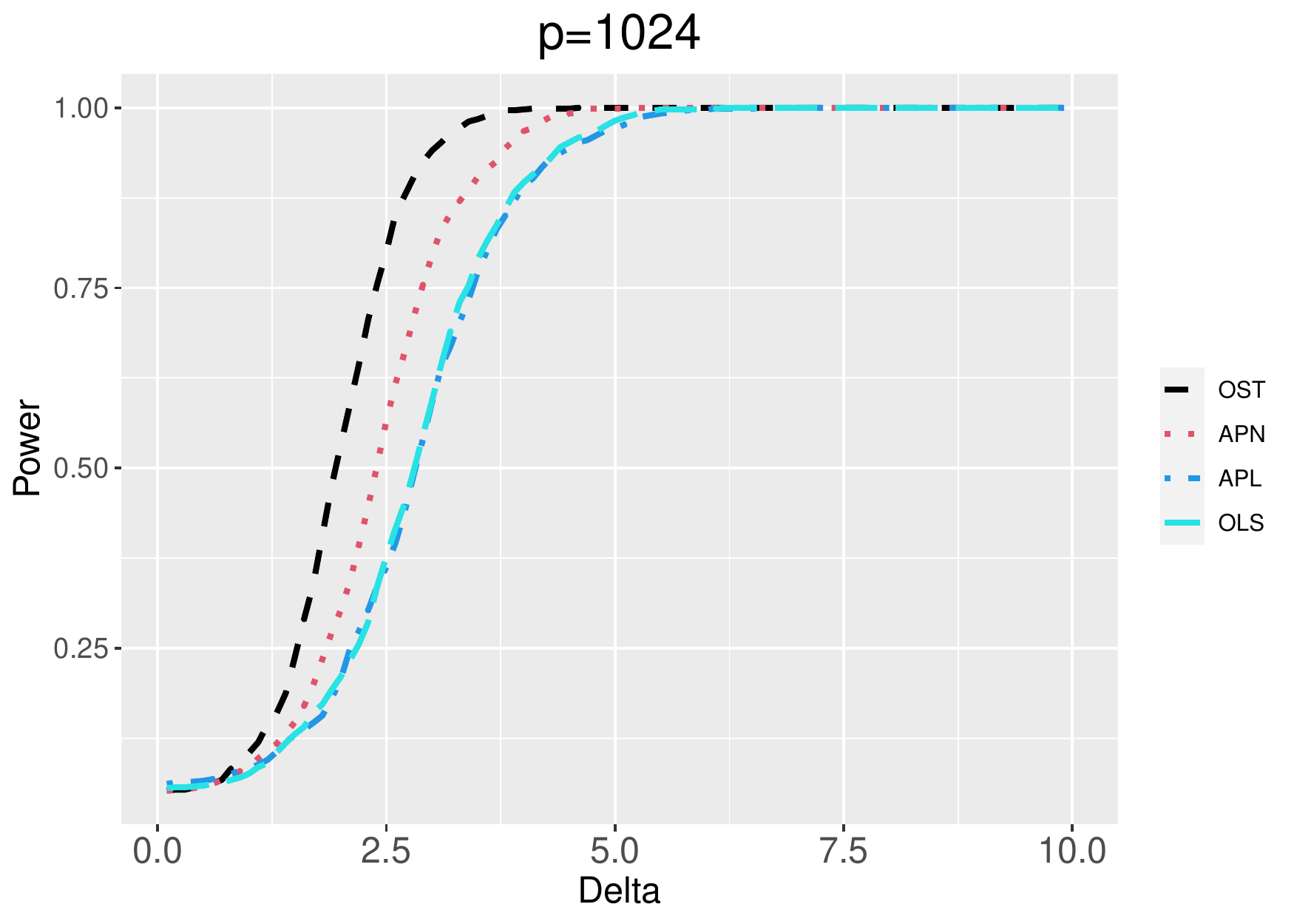}
	\caption{Power for local alternatives. The horizontal axis indicates that $\delta$ value in the local alternatives, given in \eqref{eq:inf}.}
	\label{fig:local}
\end{figure}

\subsection{Sparsity pattern recovery results for M3}
In M3 of the paper, we aim to show the performance of several methods for sparsity pattern recovery. Figure \ref{recover1} shows the sparsity recovery results of OLS, APL, OST, R4, STORE, HOLLR, and APN. The result of HOST is identical to that of OST, so we omit the figures for it. Among all the methods, OST gives the best sparsity recovery result. OLS fails to select any coefficients and gives a vague recovery of the true signal. APL fails to select some true signals, especially for the second and third slices. OST improves the results of APL by considering the heavy tail issue and the covariance information. 
R4 performs similarly to the OLS, which provides an estimation of $\mbB$ with noises and fails to select the non-zero elements. HOLRR and STORE fail to recover the sparsity pattern as a result of the violation for the low rank assumption in $\mbB$. APN performs similarly to APL, which fails to select some non-zero elements, especially for the second and third slices. As a comparison, OST performs better than all of these competing methods as a result of considering the heavy-tail issue. 
\begin{figure}[ht!]
	\centering
	\includegraphics[scale=0.15]{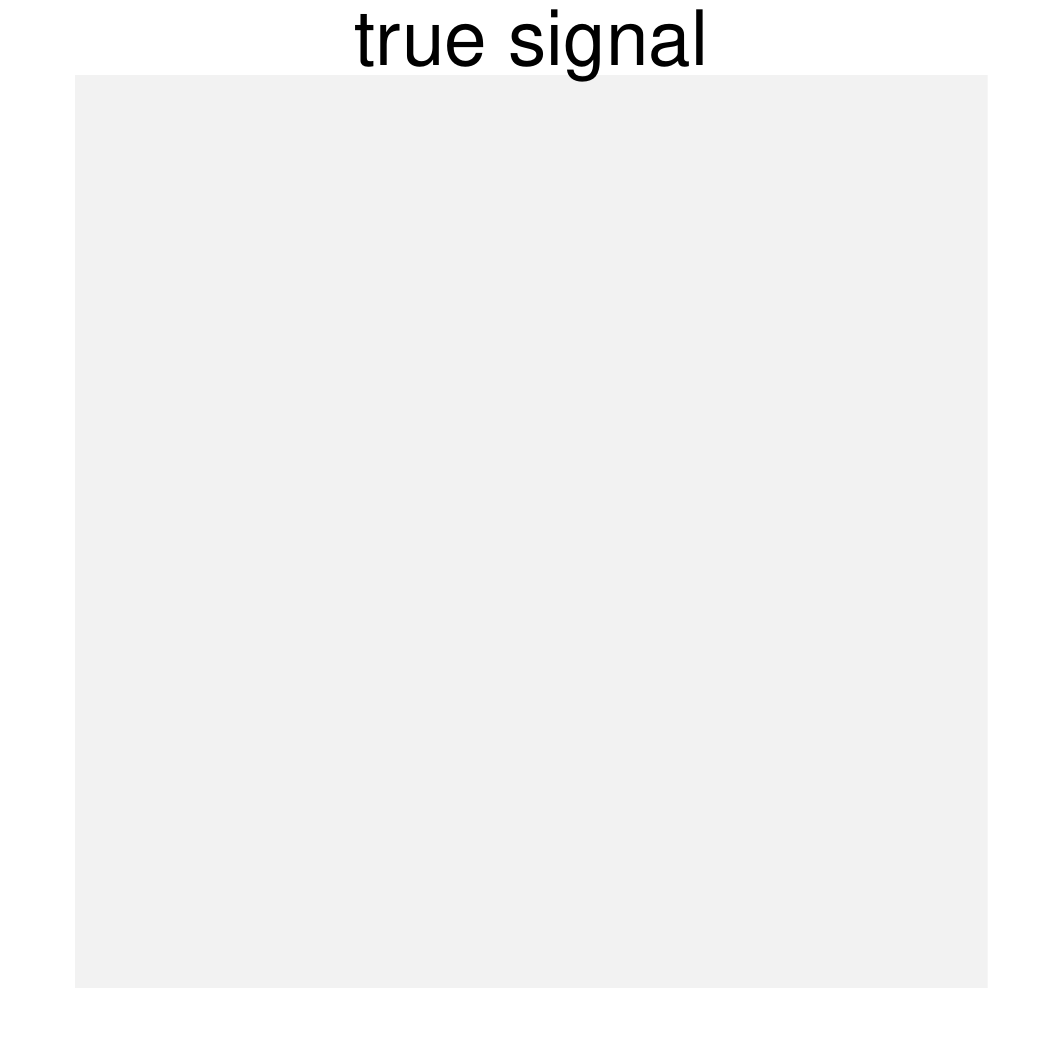}
	\includegraphics[scale=0.15]{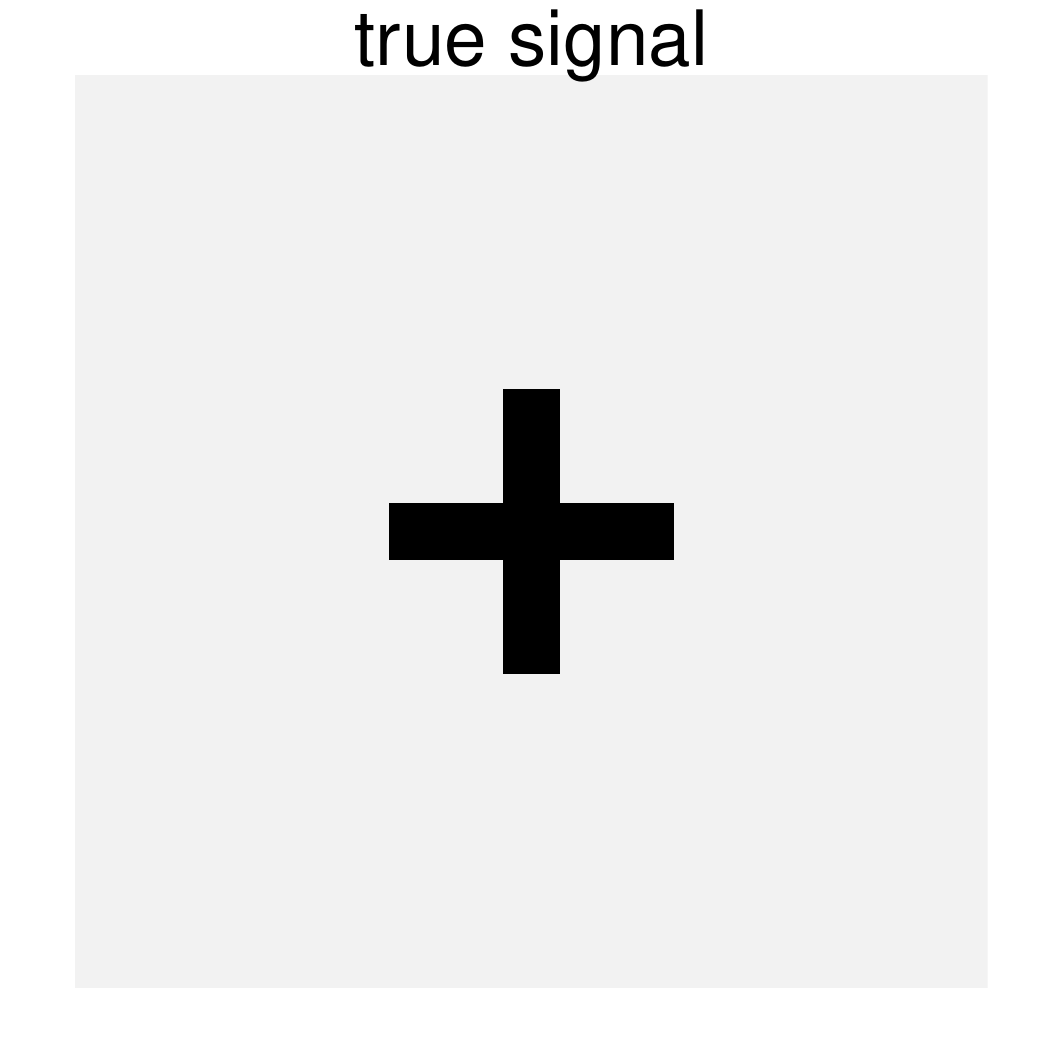}
	\includegraphics[scale=0.15]{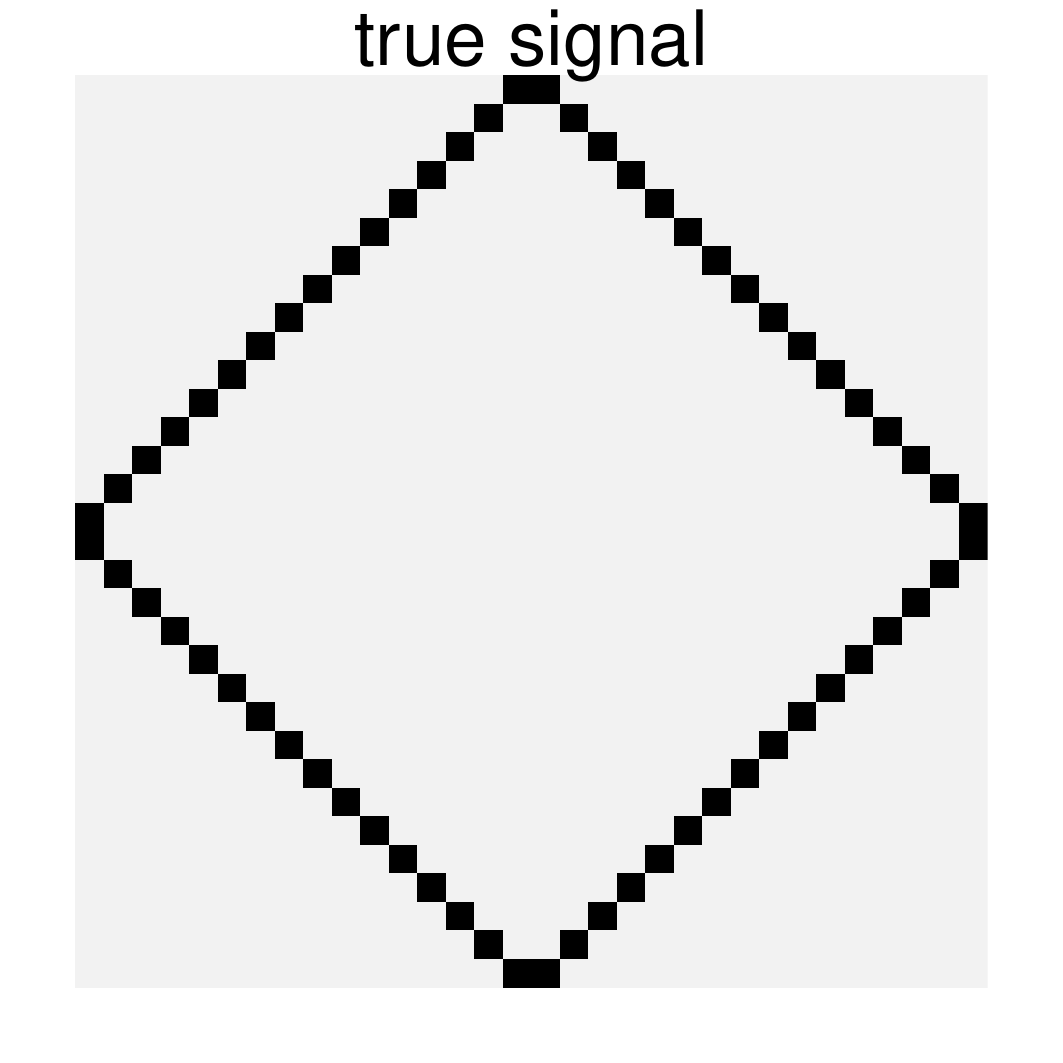}
	\includegraphics[scale=0.15]{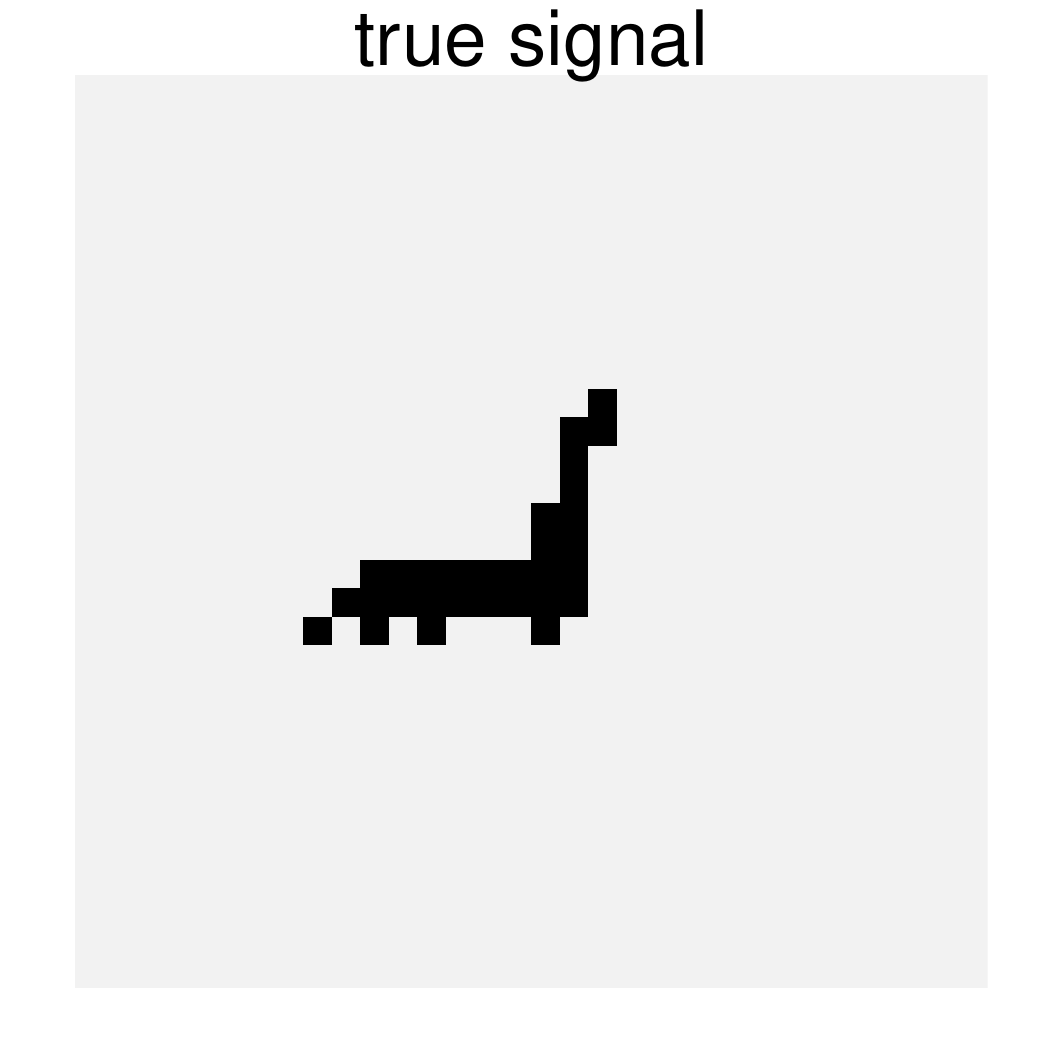}
	\includegraphics[scale=0.15]{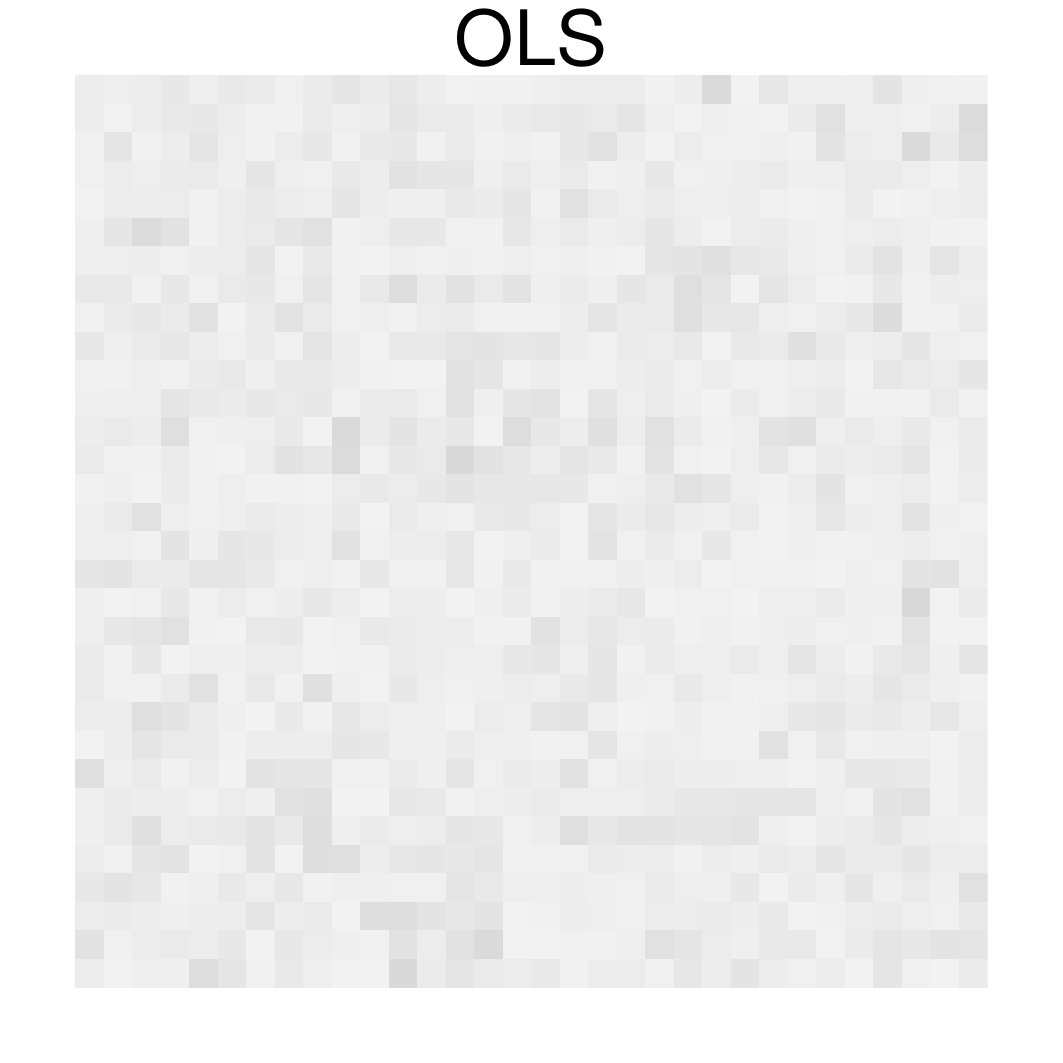}
	\includegraphics[scale=0.15]{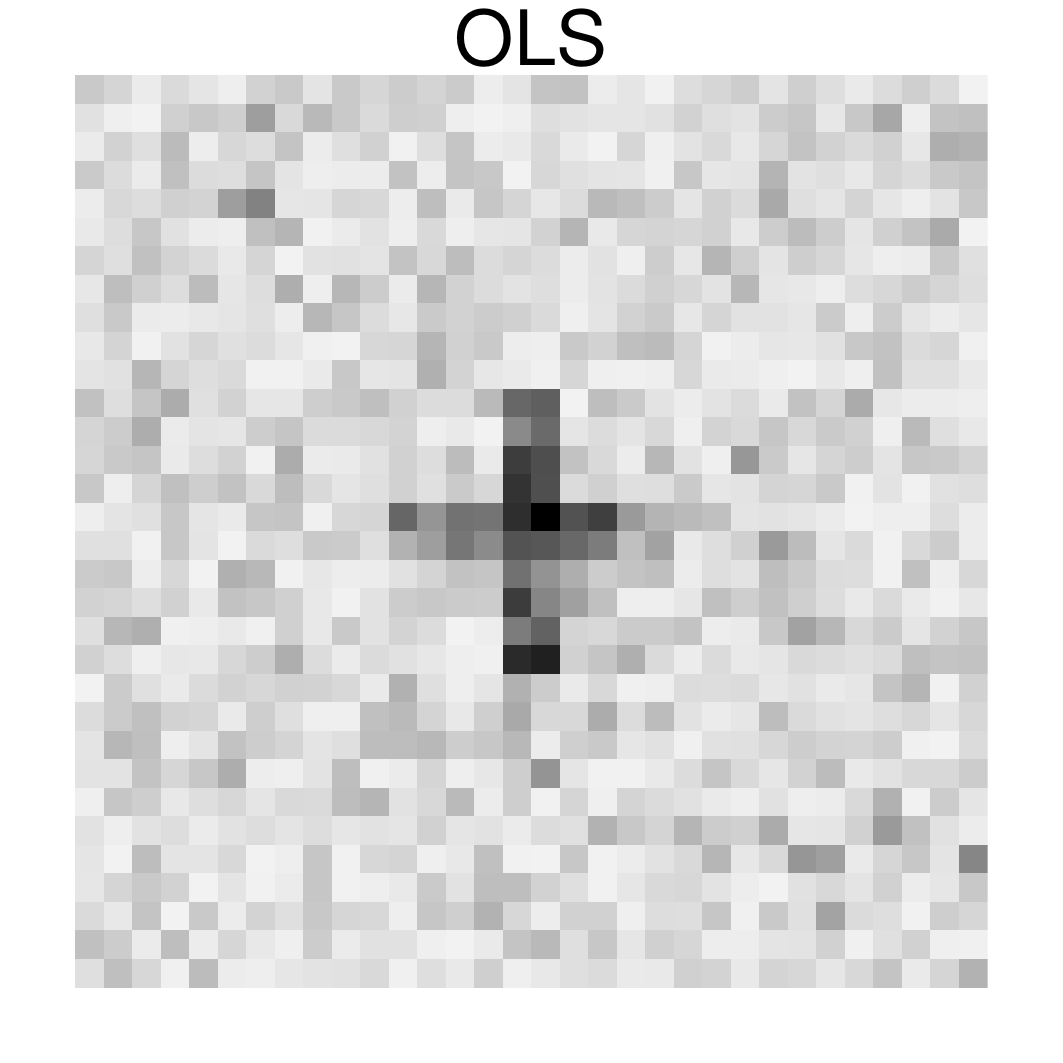}
	\includegraphics[scale=0.15]{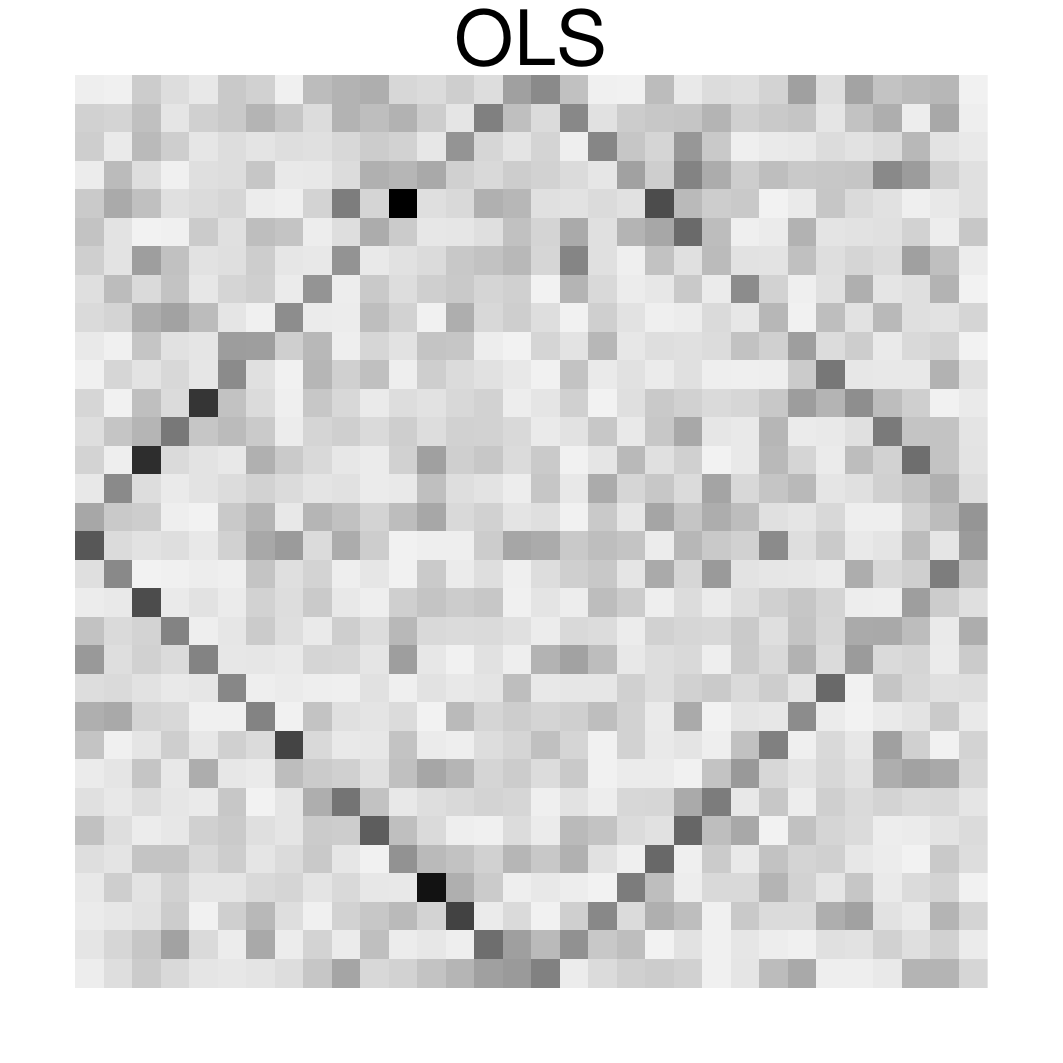}
	\includegraphics[scale=0.15]{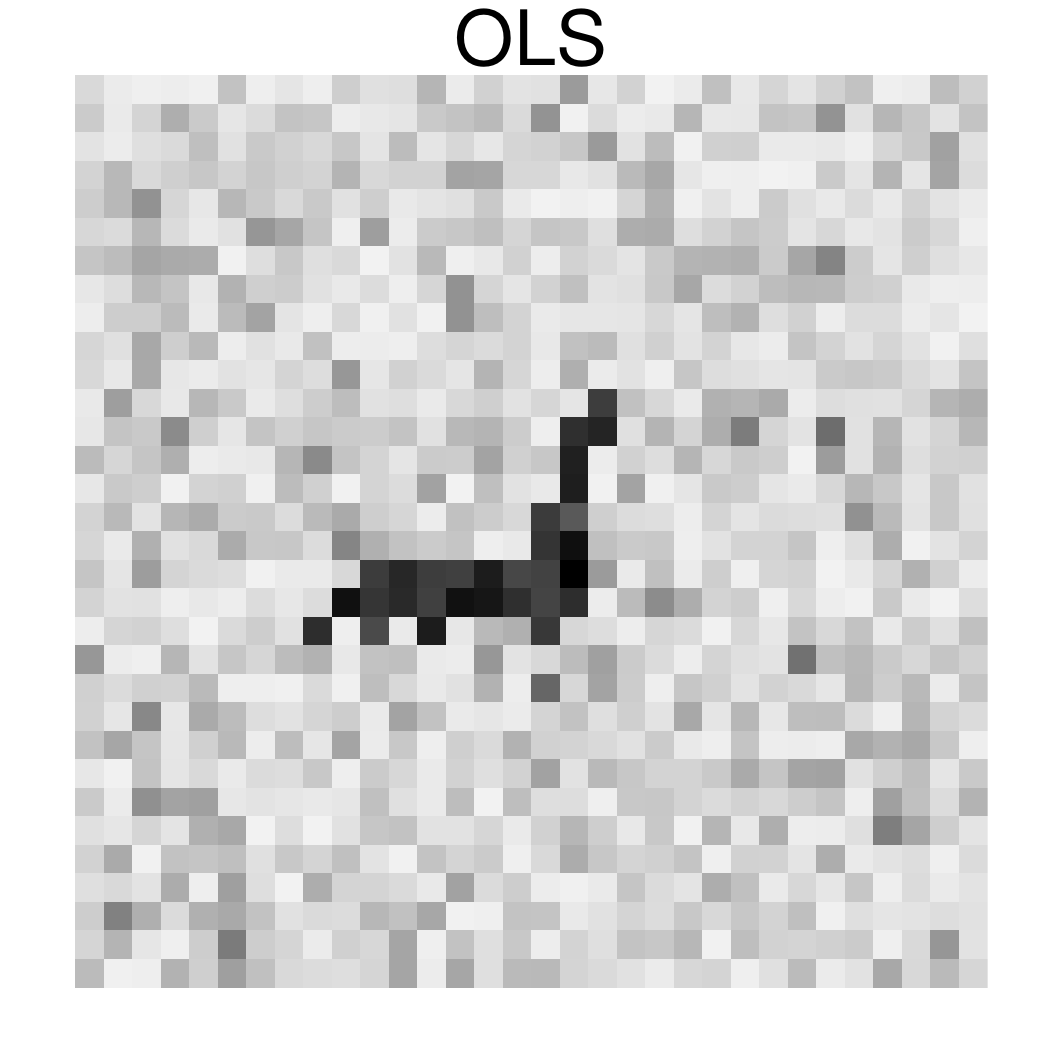}\\
	\includegraphics[scale=0.15]{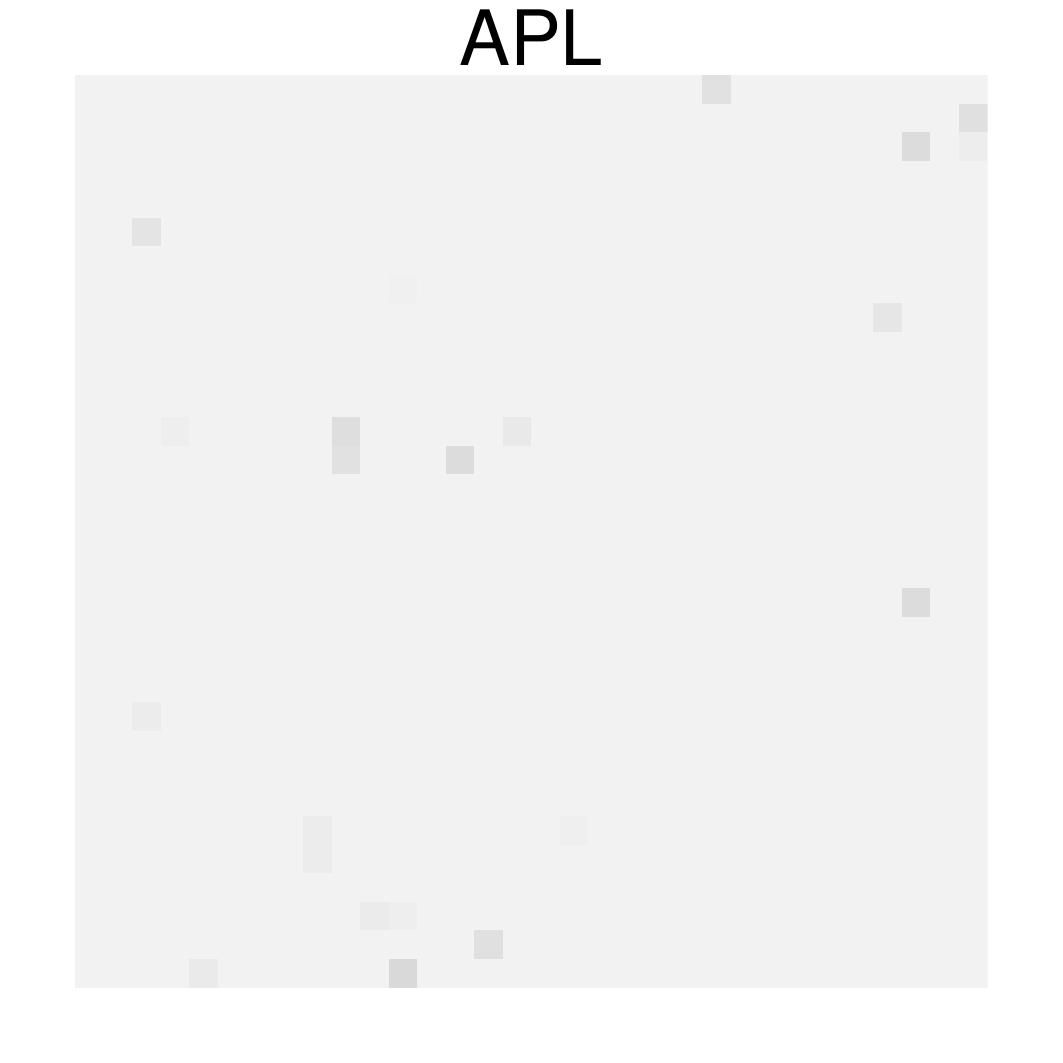}
	\includegraphics[scale=0.15]{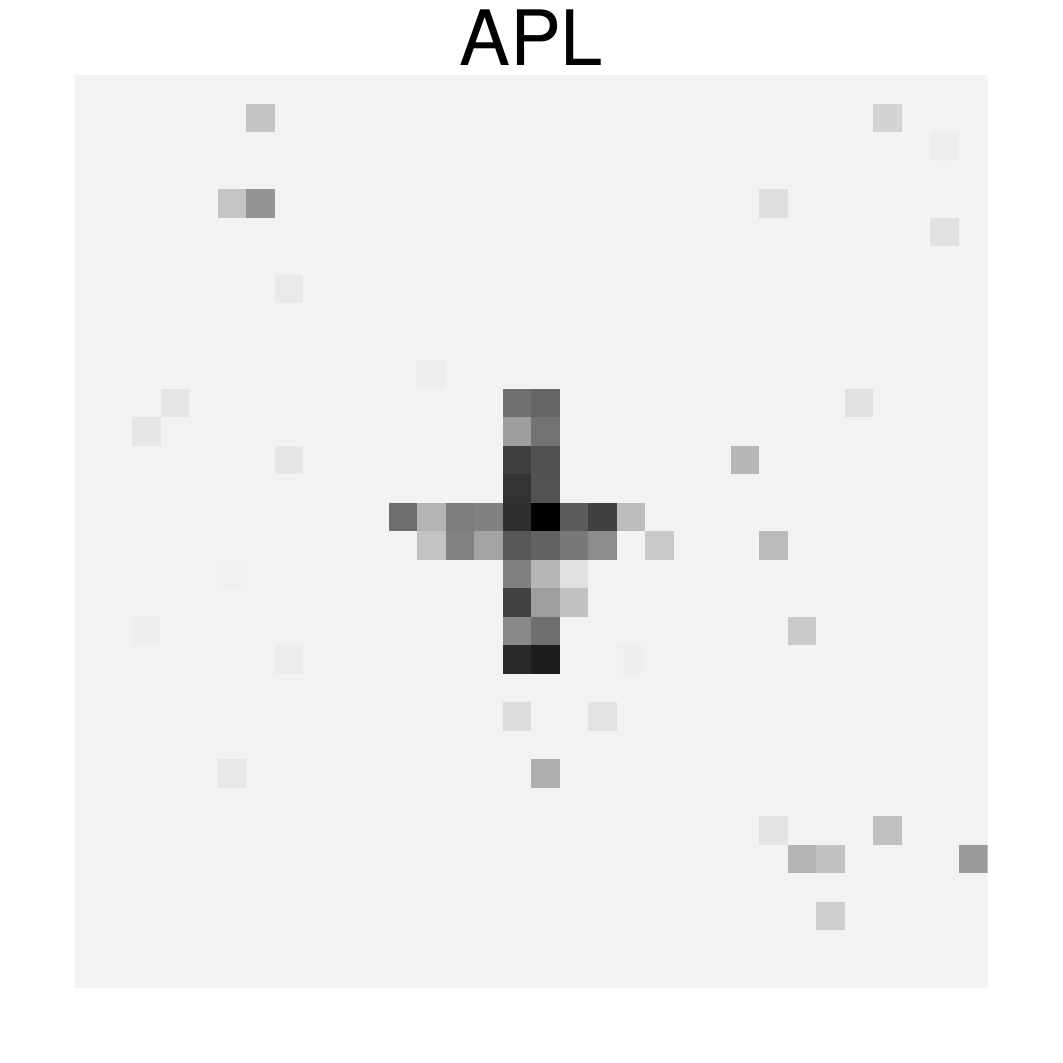}
	\includegraphics[scale=0.15]{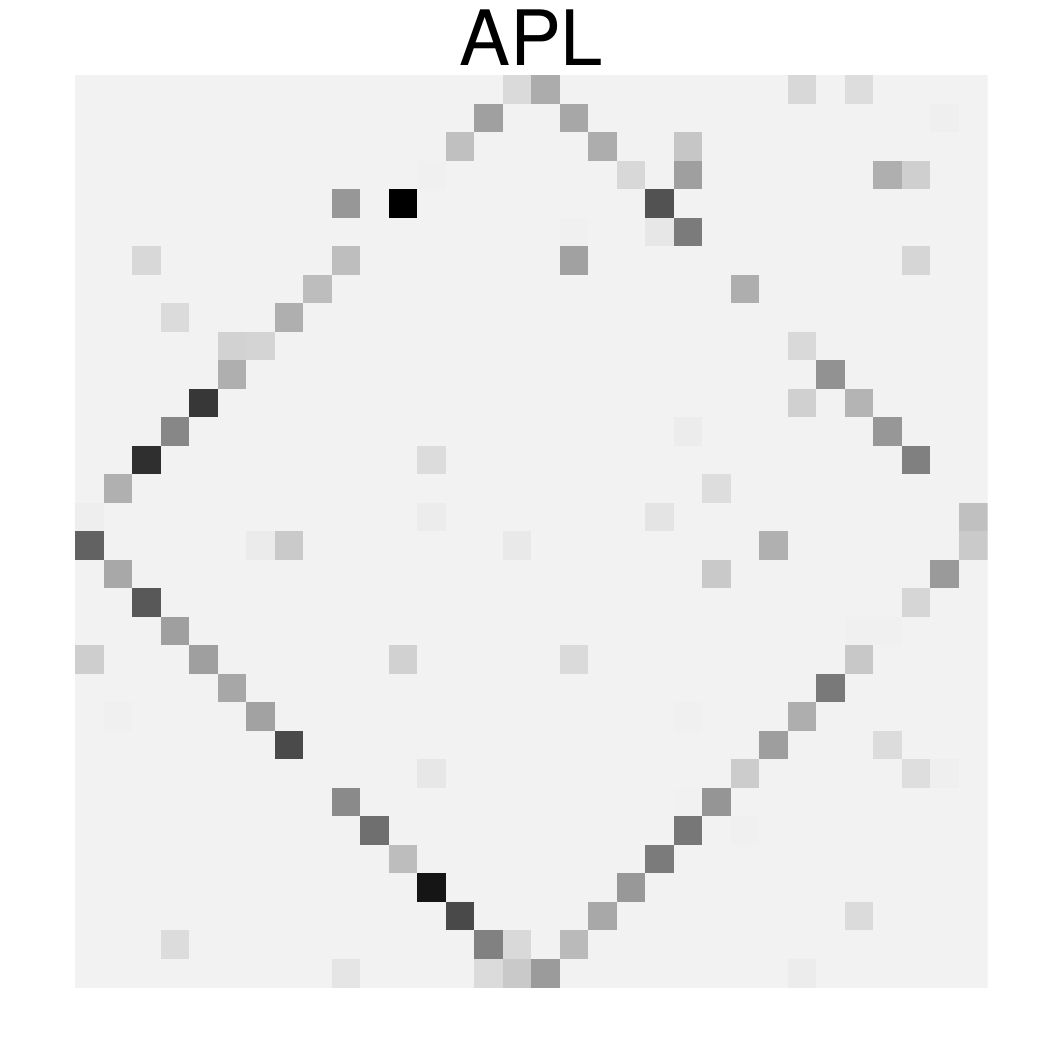}
	\includegraphics[scale=0.15]{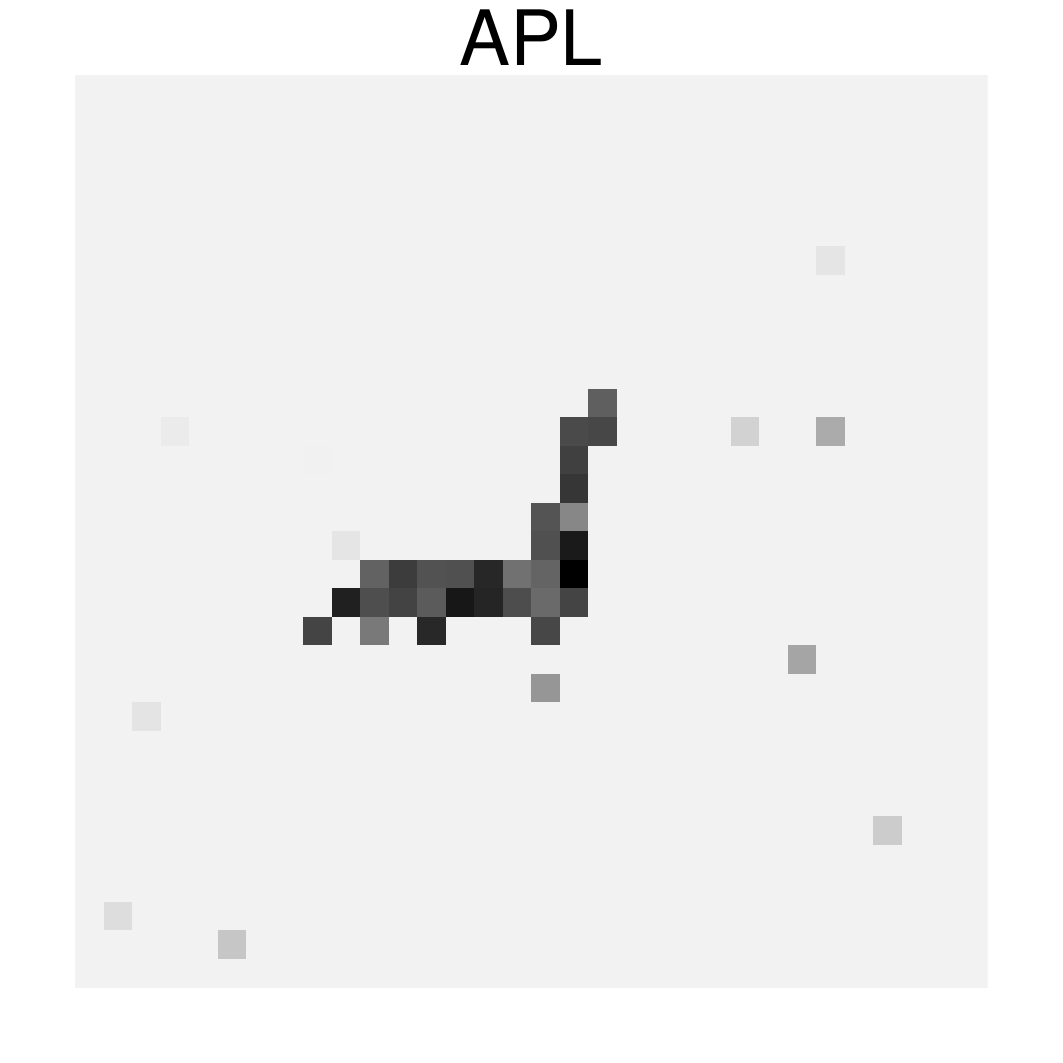}
	\includegraphics[scale=0.15]{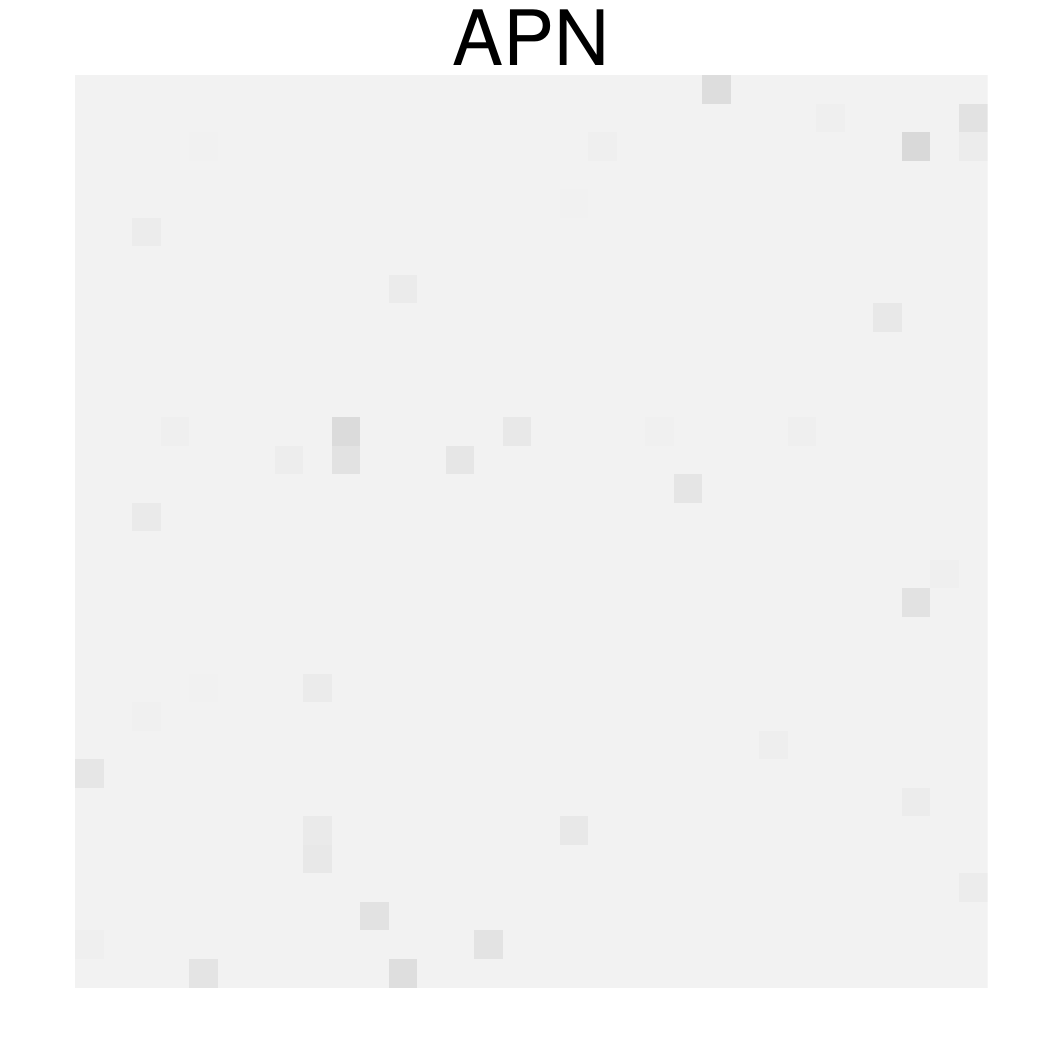}
	\includegraphics[scale=0.15]{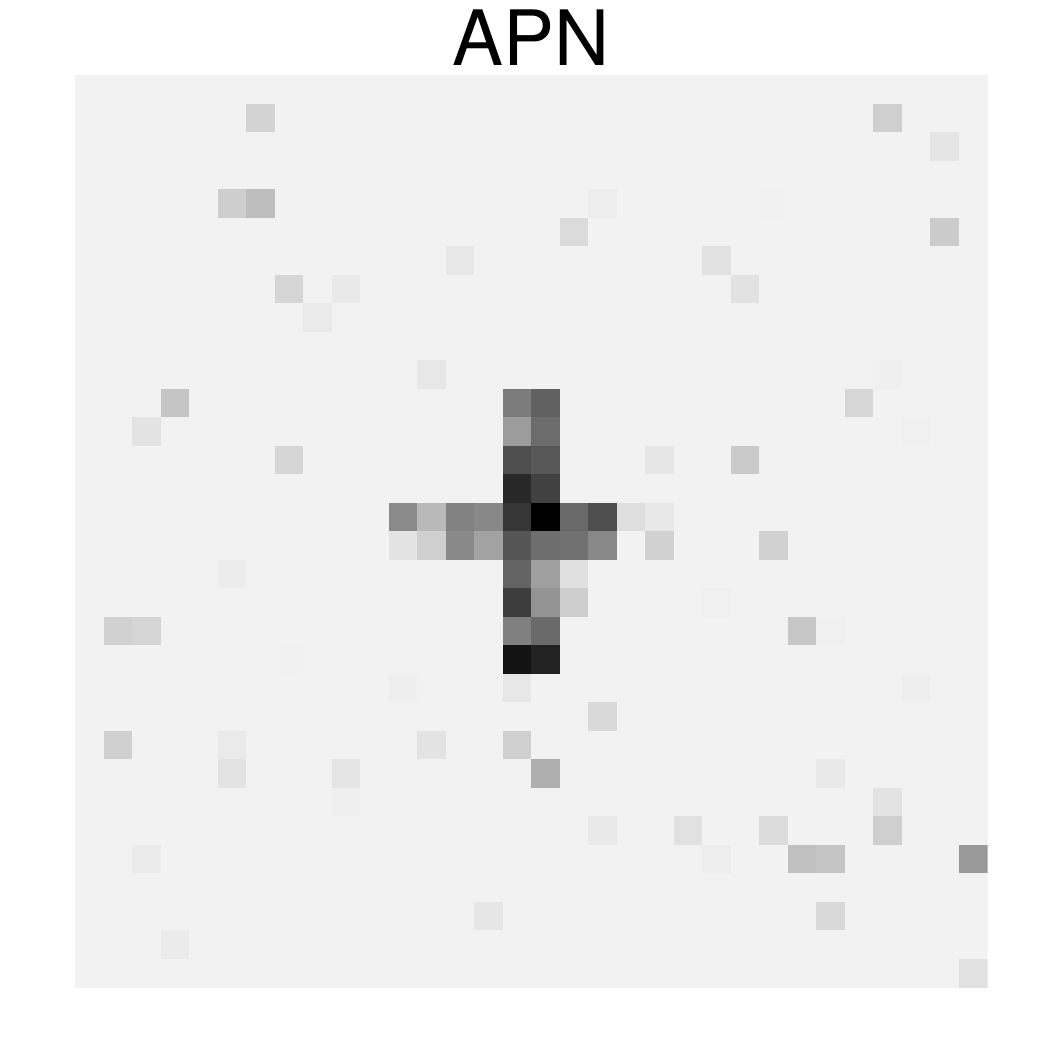}
	\includegraphics[scale=0.15]{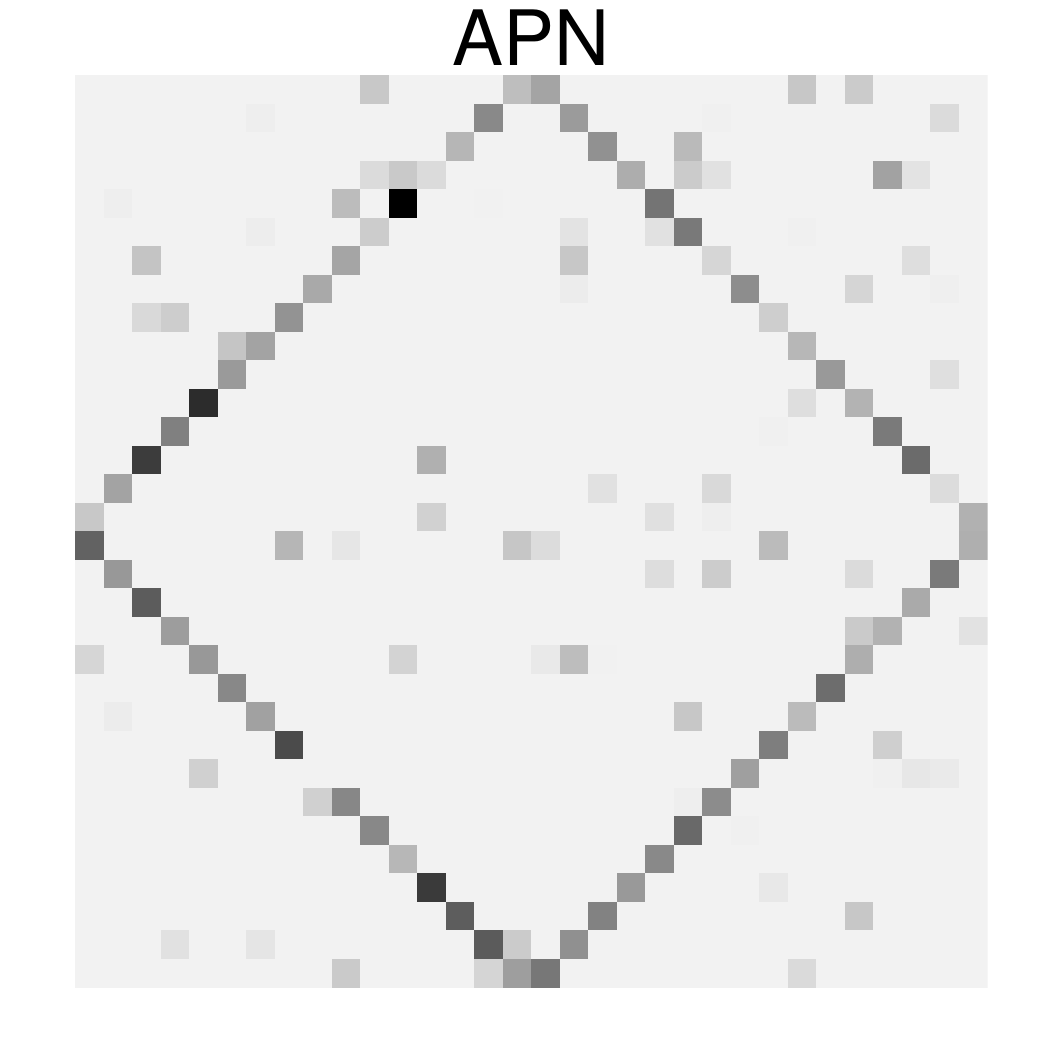}
	\includegraphics[scale=0.15]{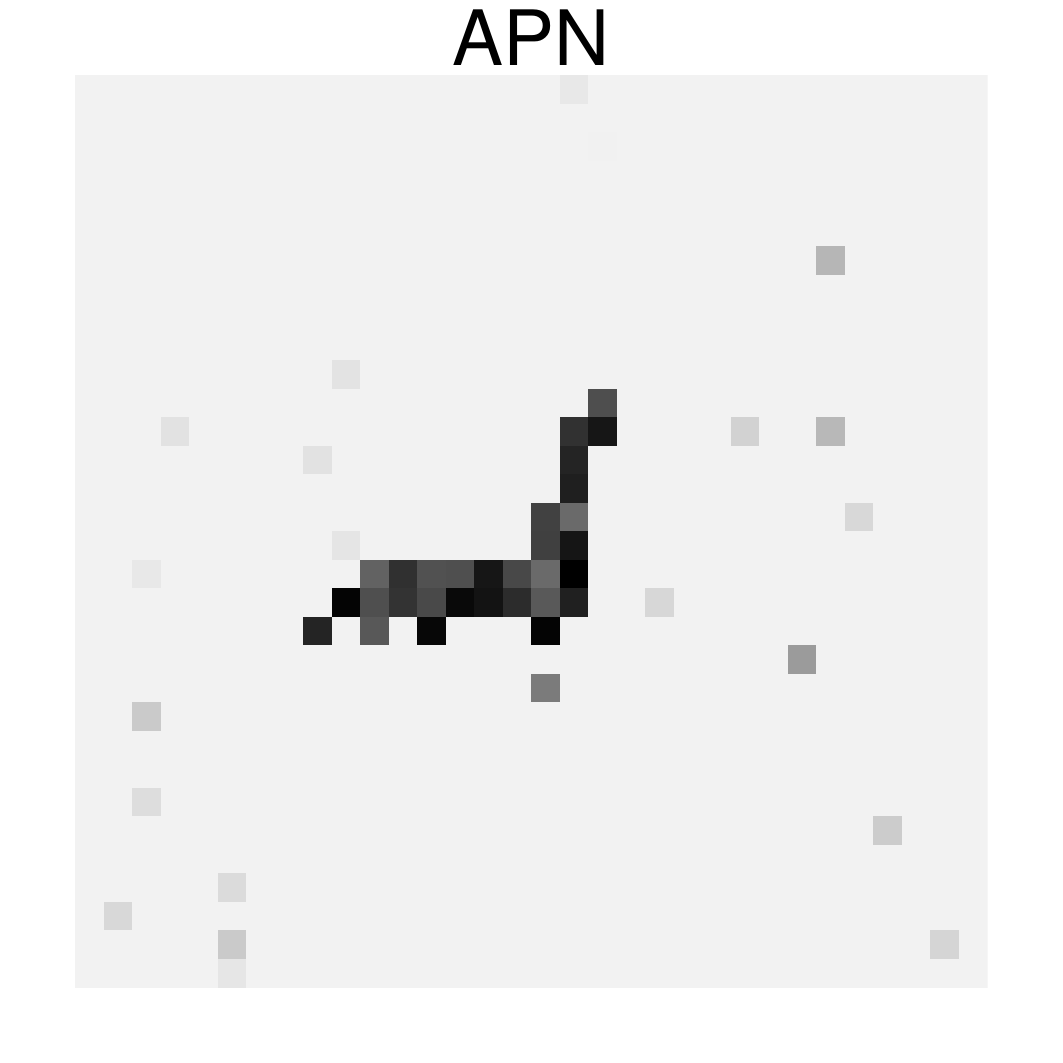}\\
	\includegraphics[scale=0.15]{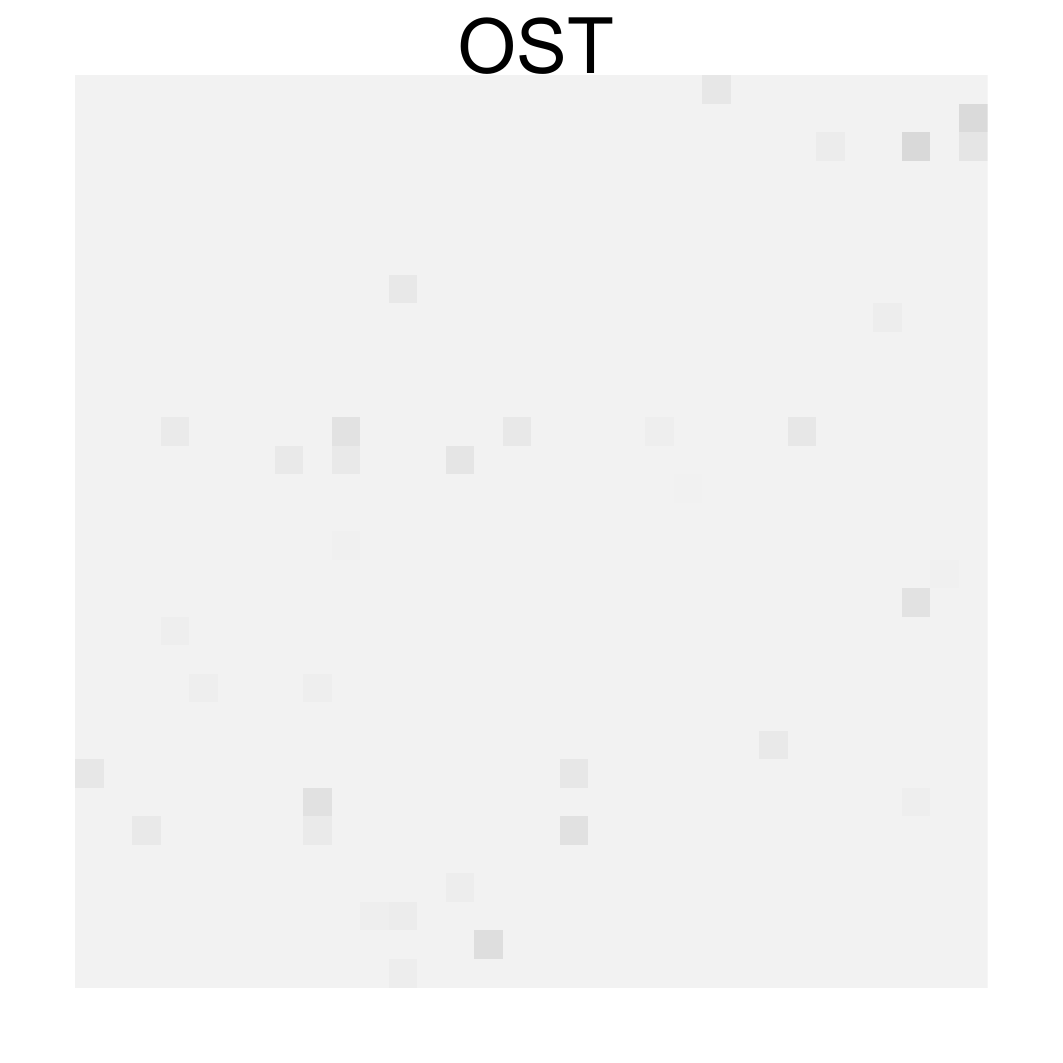}
	\includegraphics[scale=0.15]{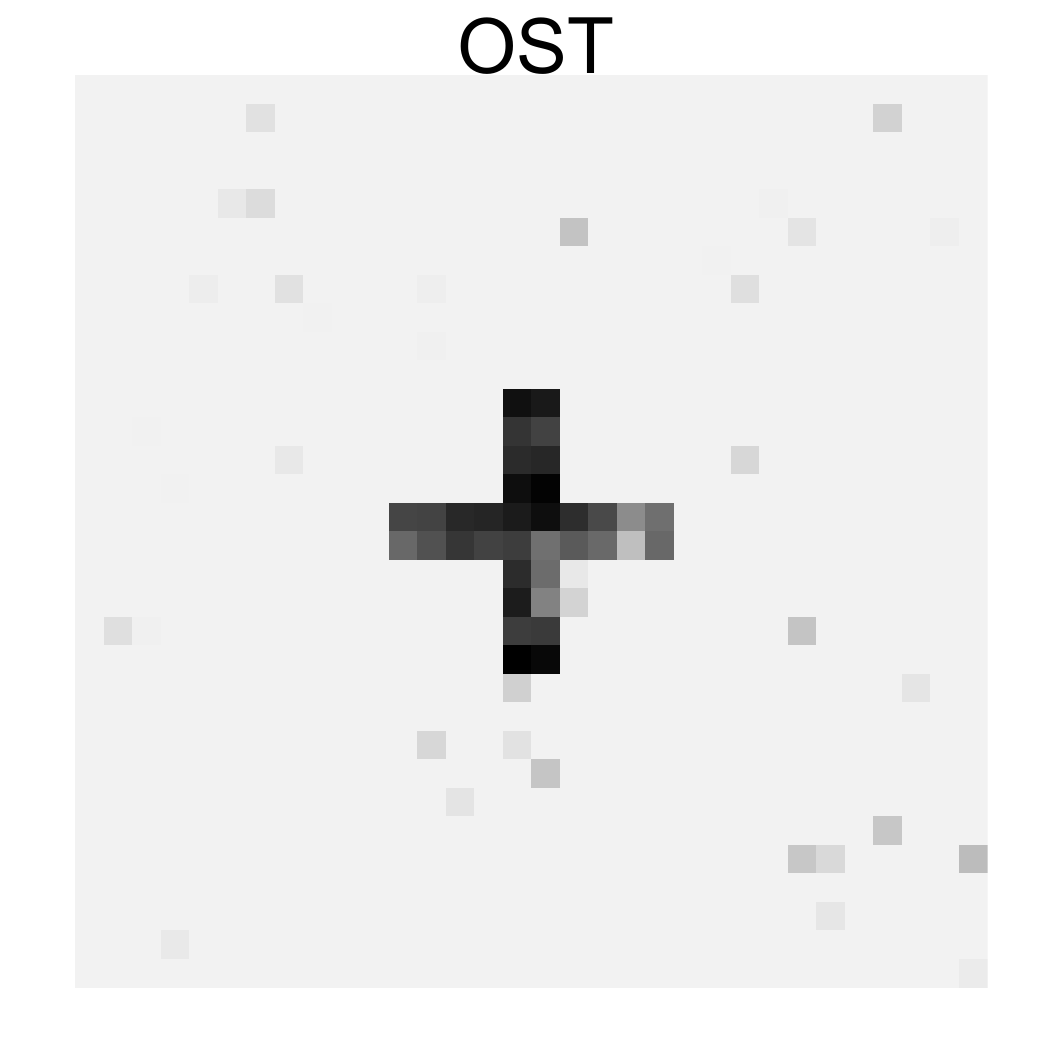}
	\includegraphics[scale=0.15]{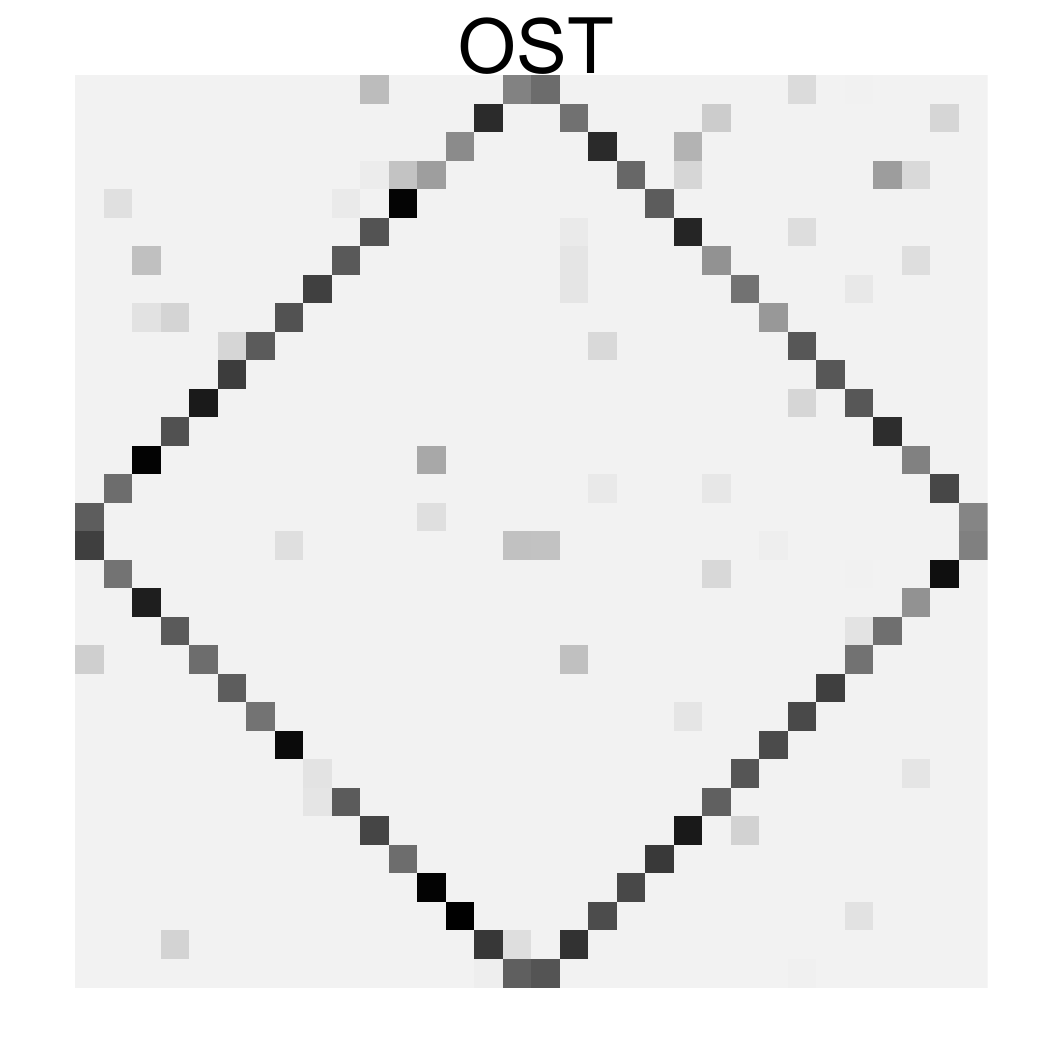}
	\includegraphics[scale=0.15]{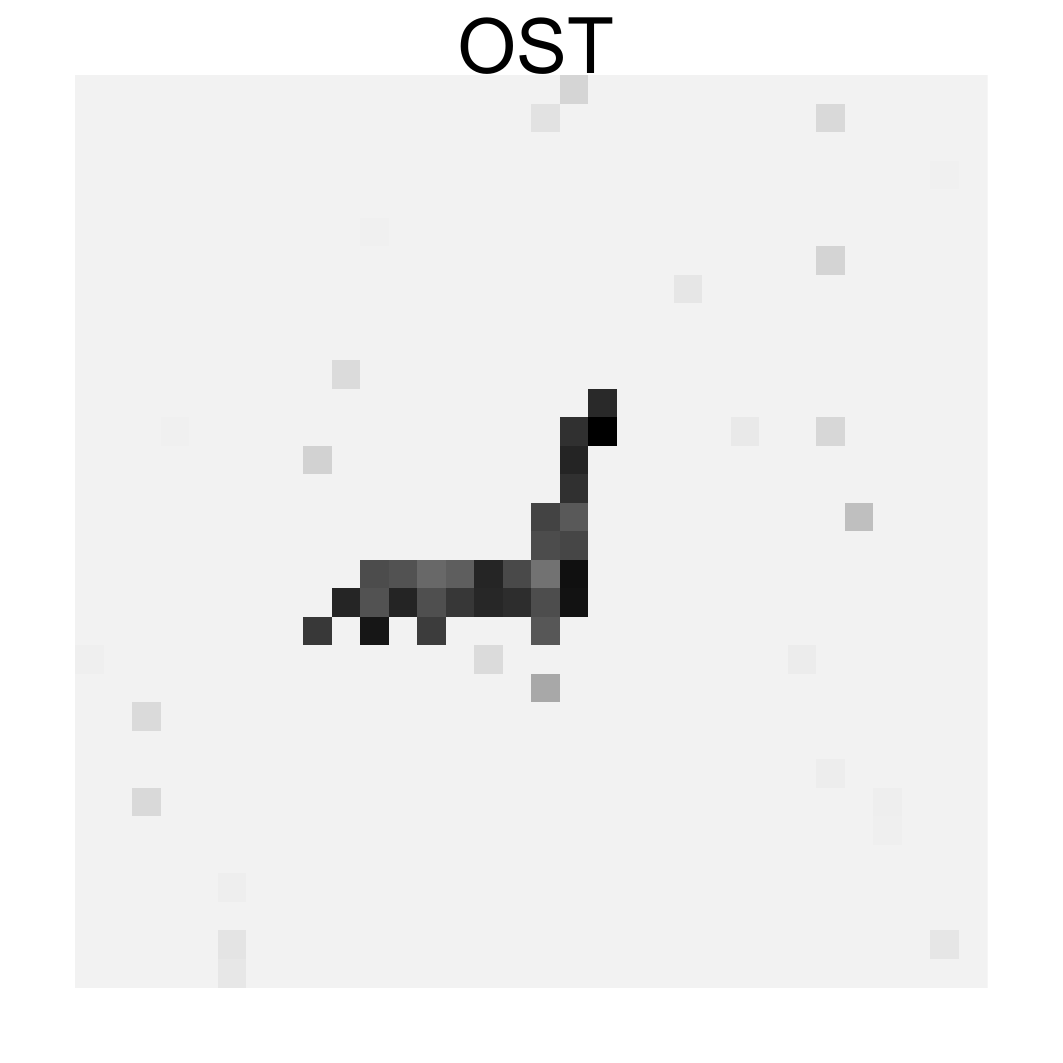}
	\includegraphics[scale=0.15]{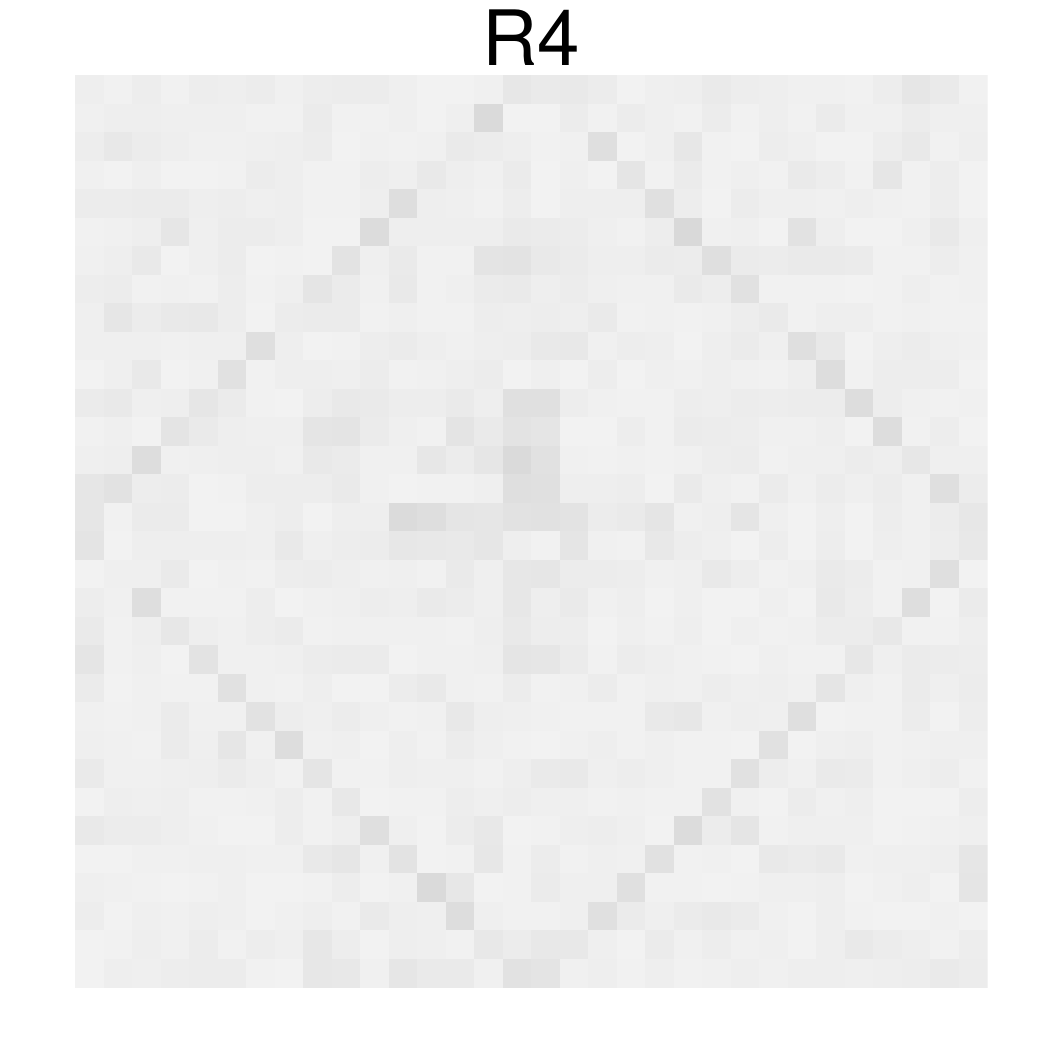}
	\includegraphics[scale=0.15]{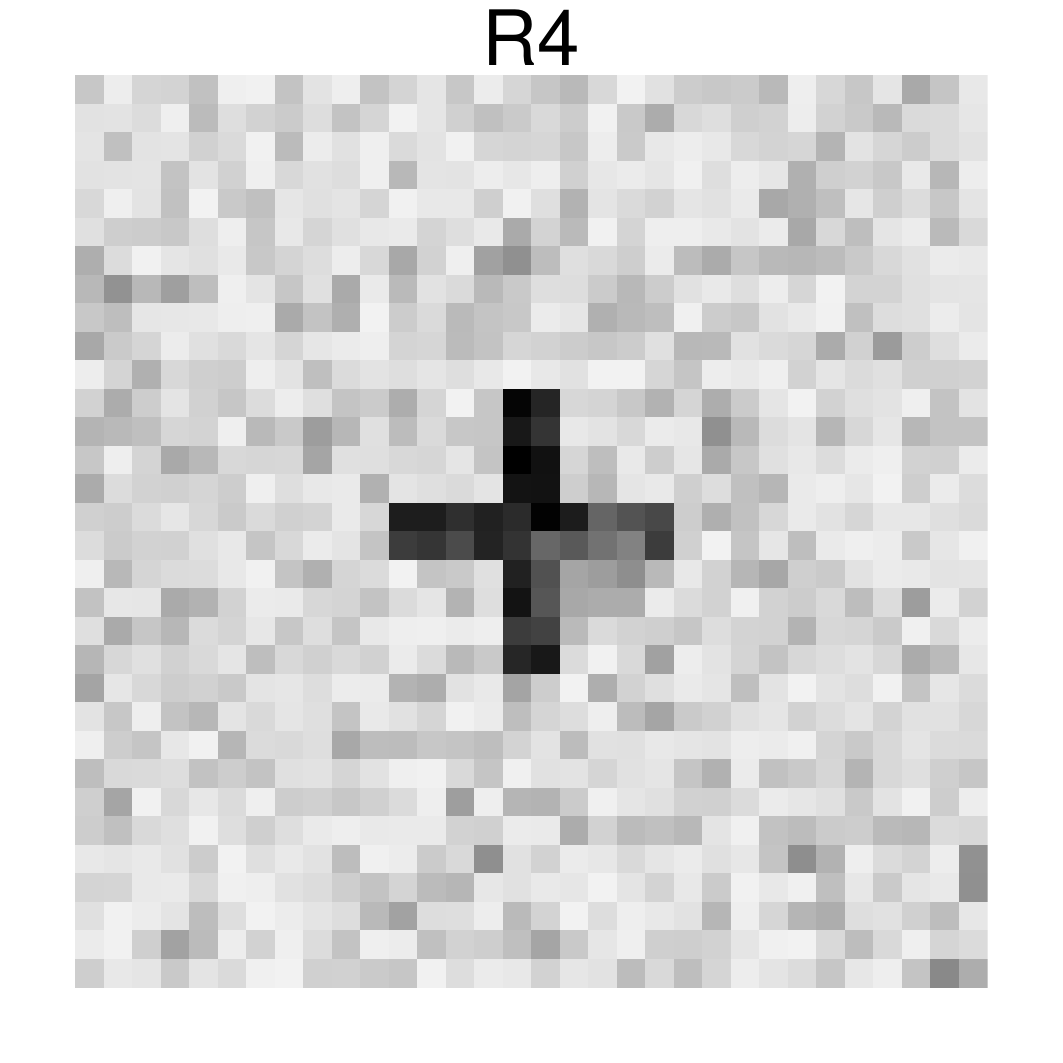}
	\includegraphics[scale=0.15]{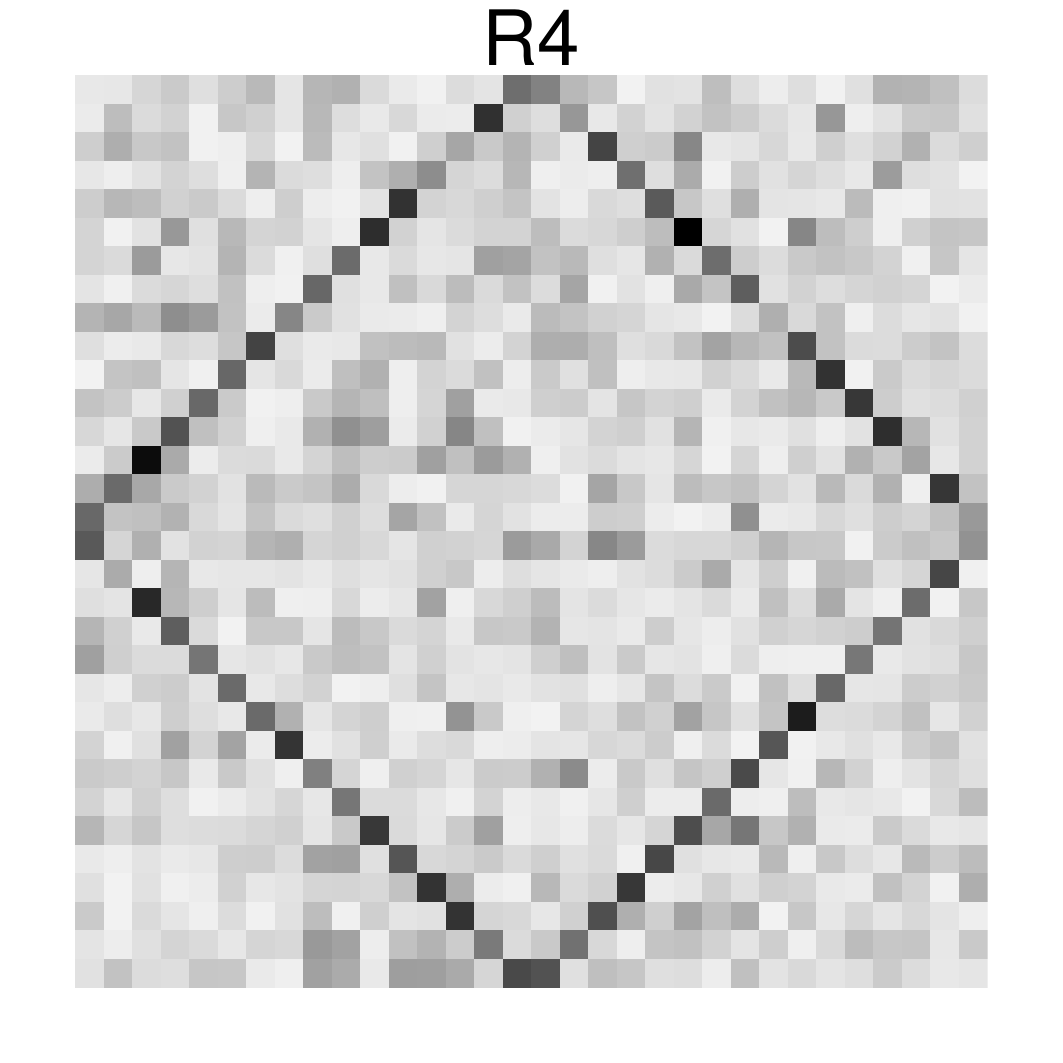}
	\includegraphics[scale=0.15]{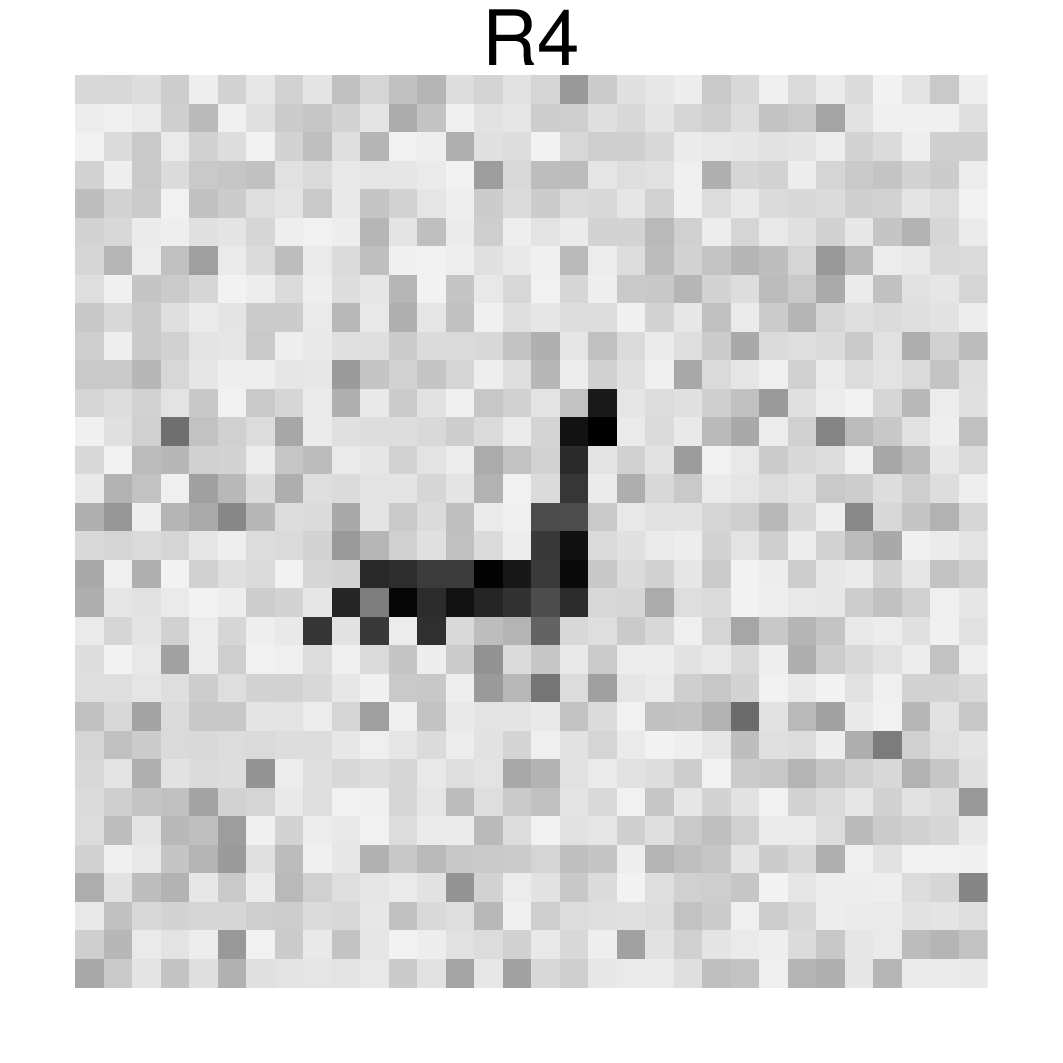}\\
	\includegraphics[scale=0.15]{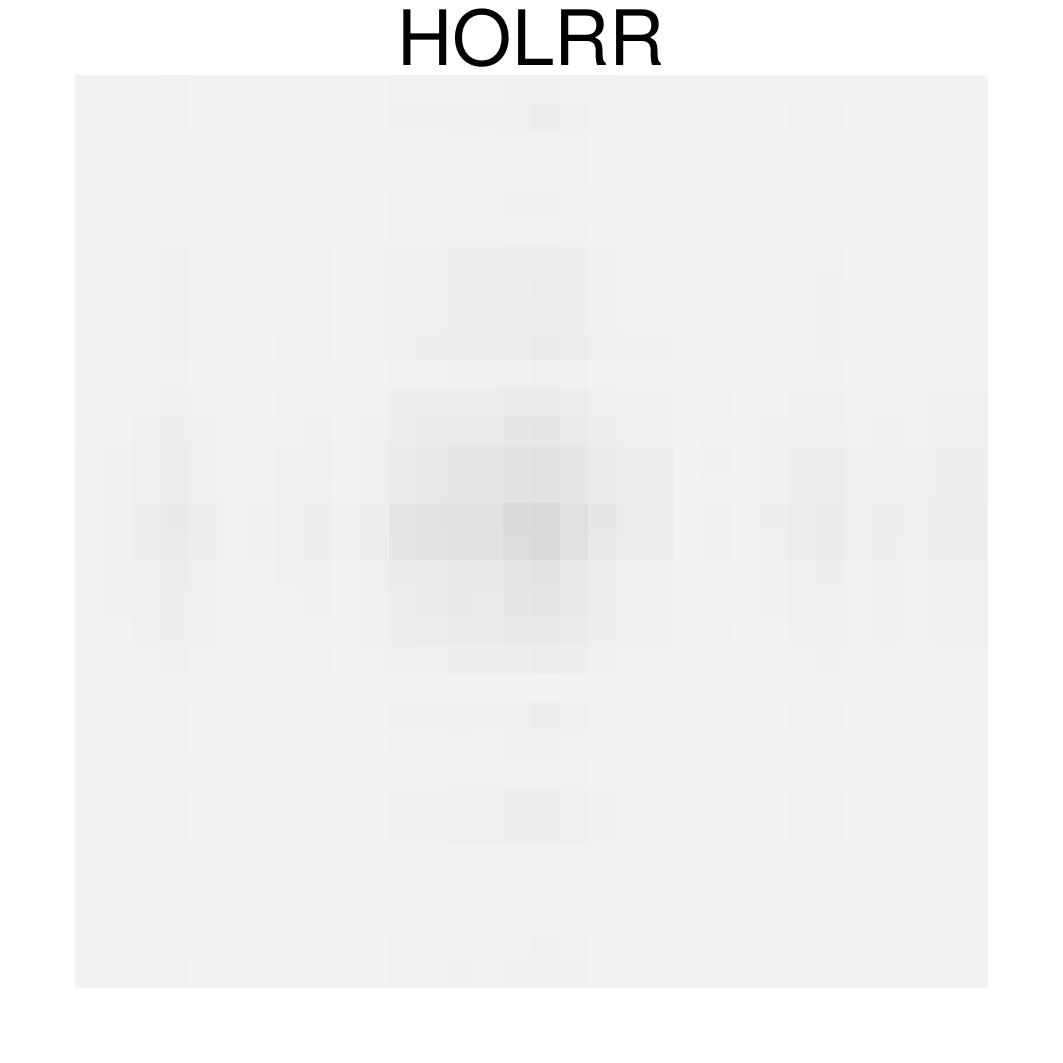}
	\includegraphics[scale=0.15]{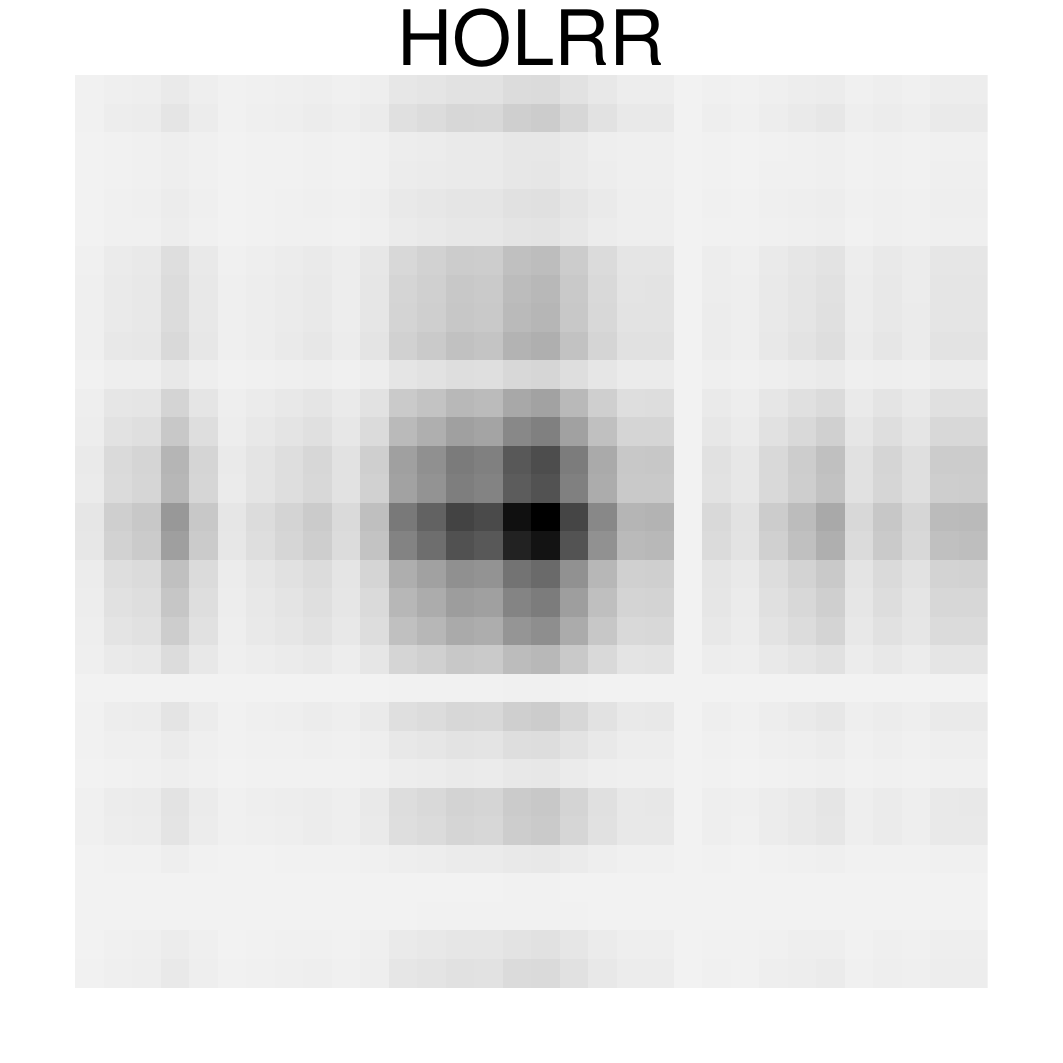}
	\includegraphics[scale=0.15]{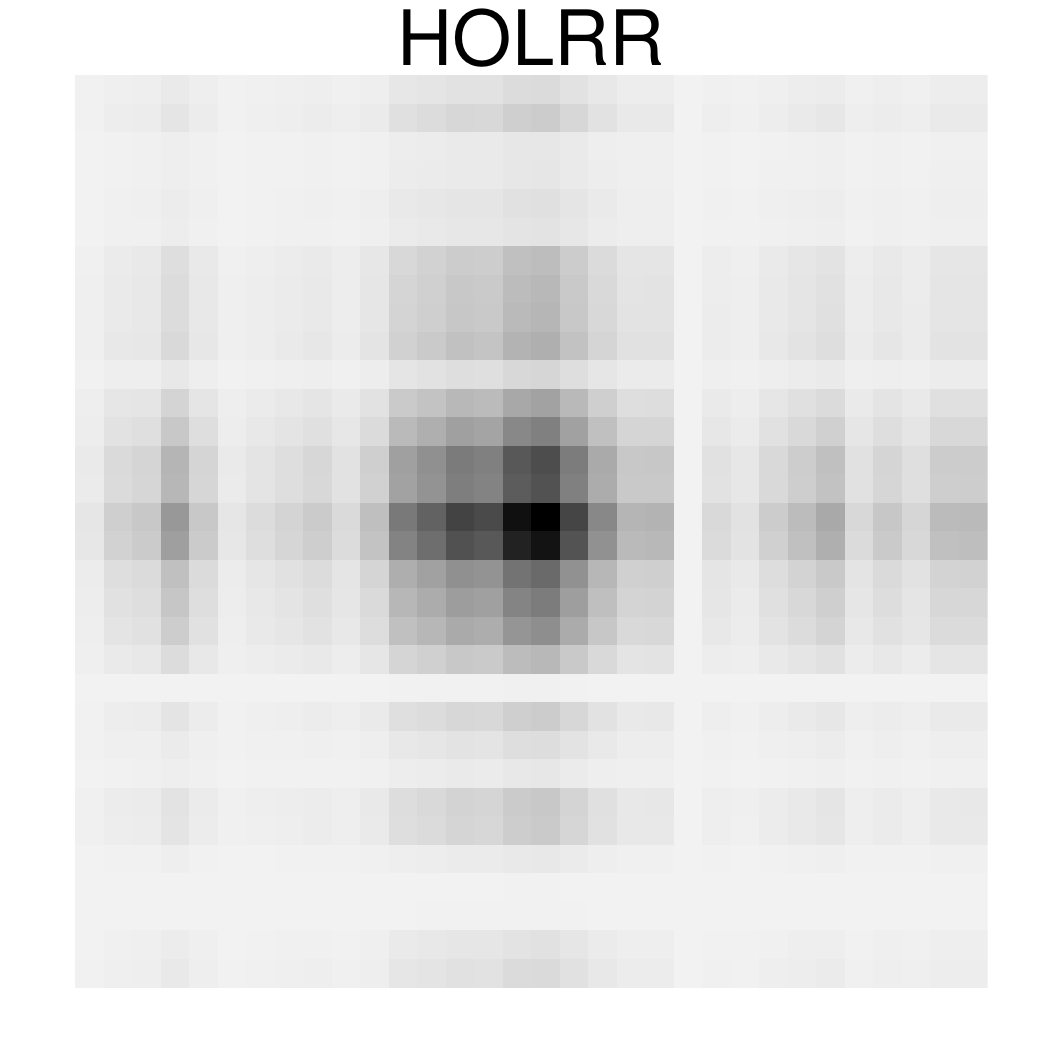}
	\includegraphics[scale=0.15]{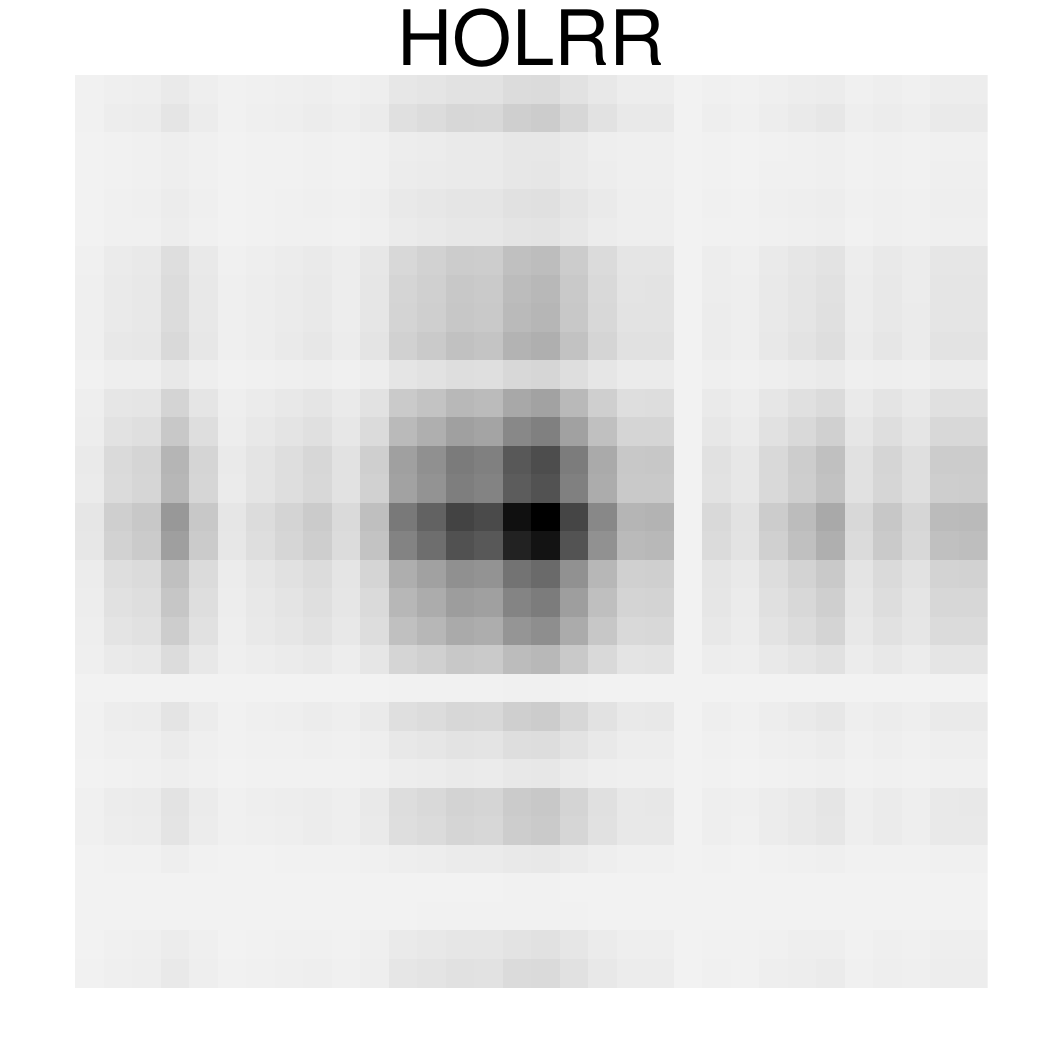}
	\includegraphics[scale=0.15]{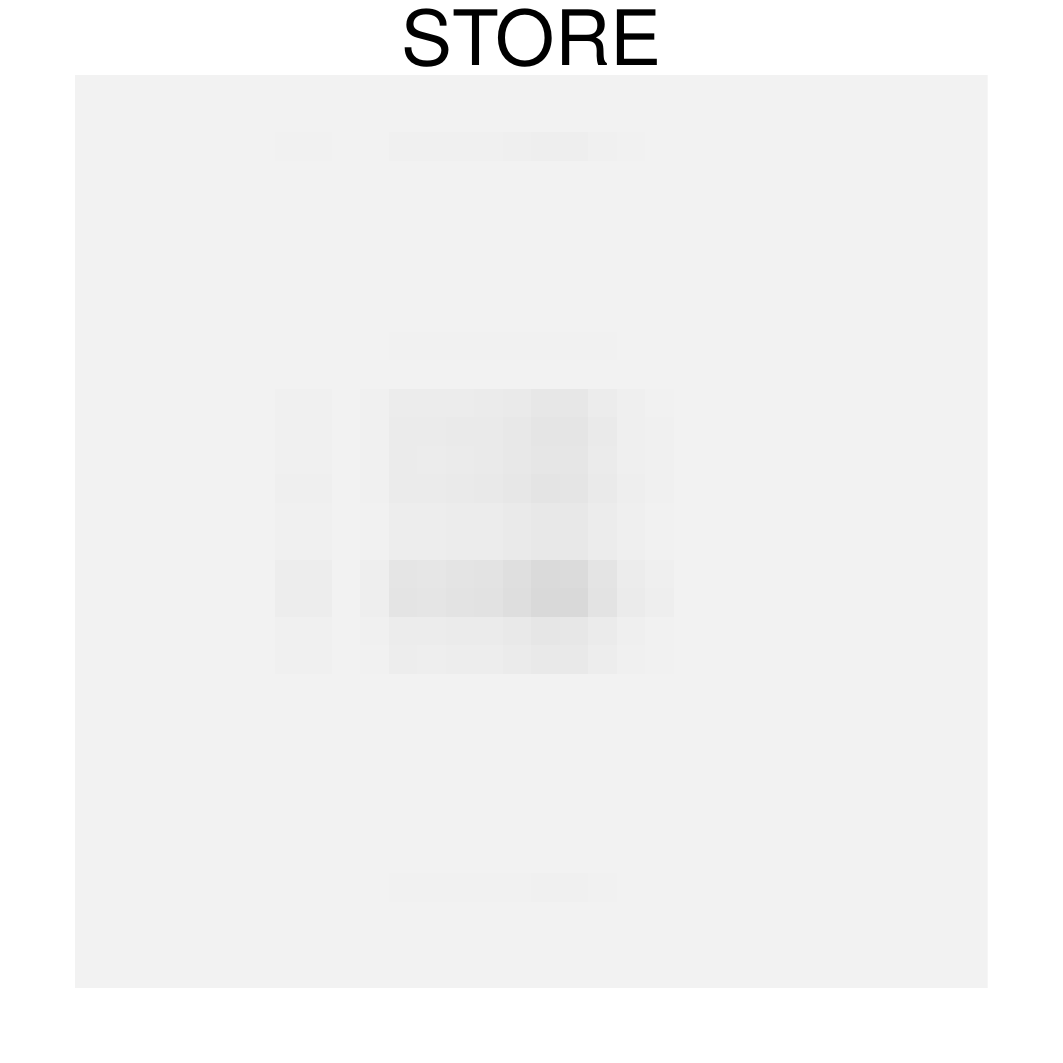}
	\includegraphics[scale=0.15]{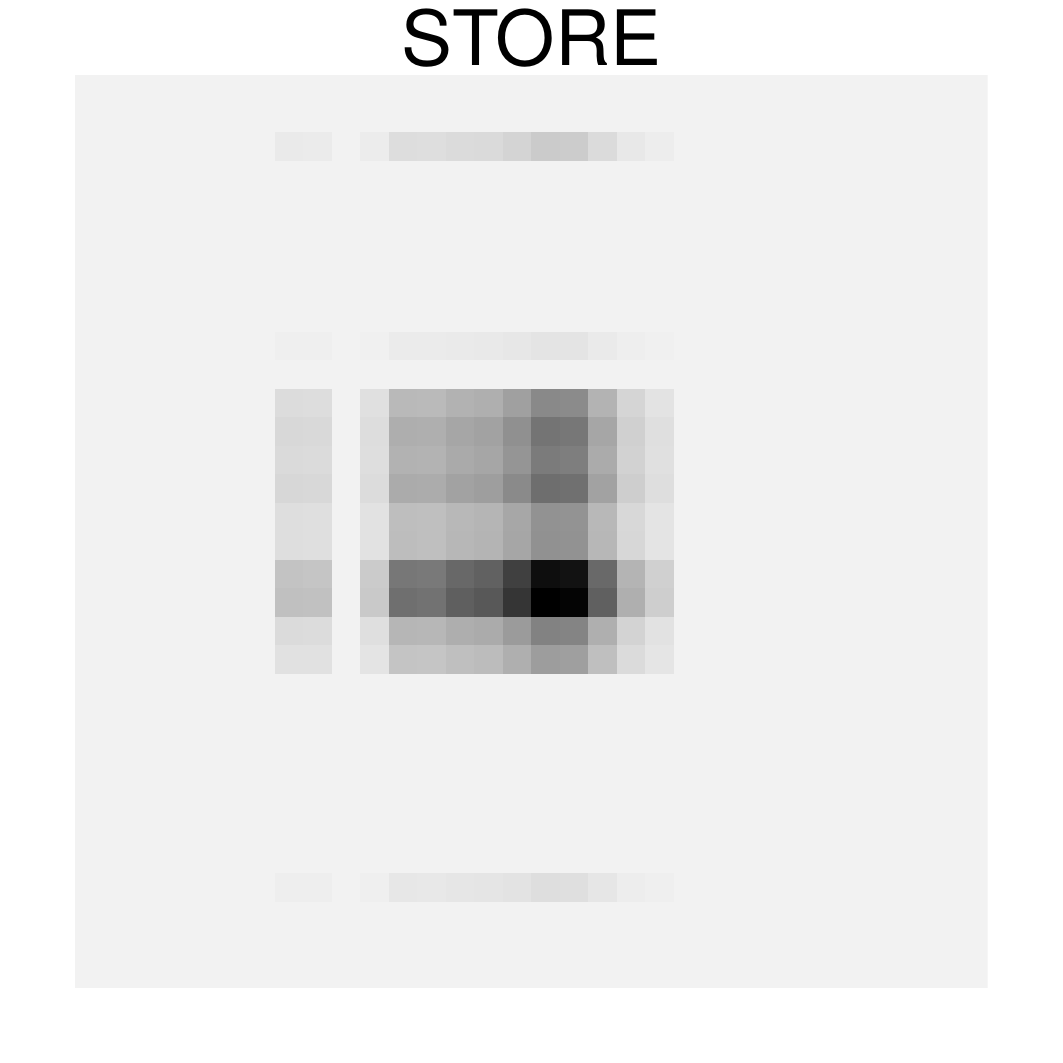}
	\includegraphics[scale=0.15]{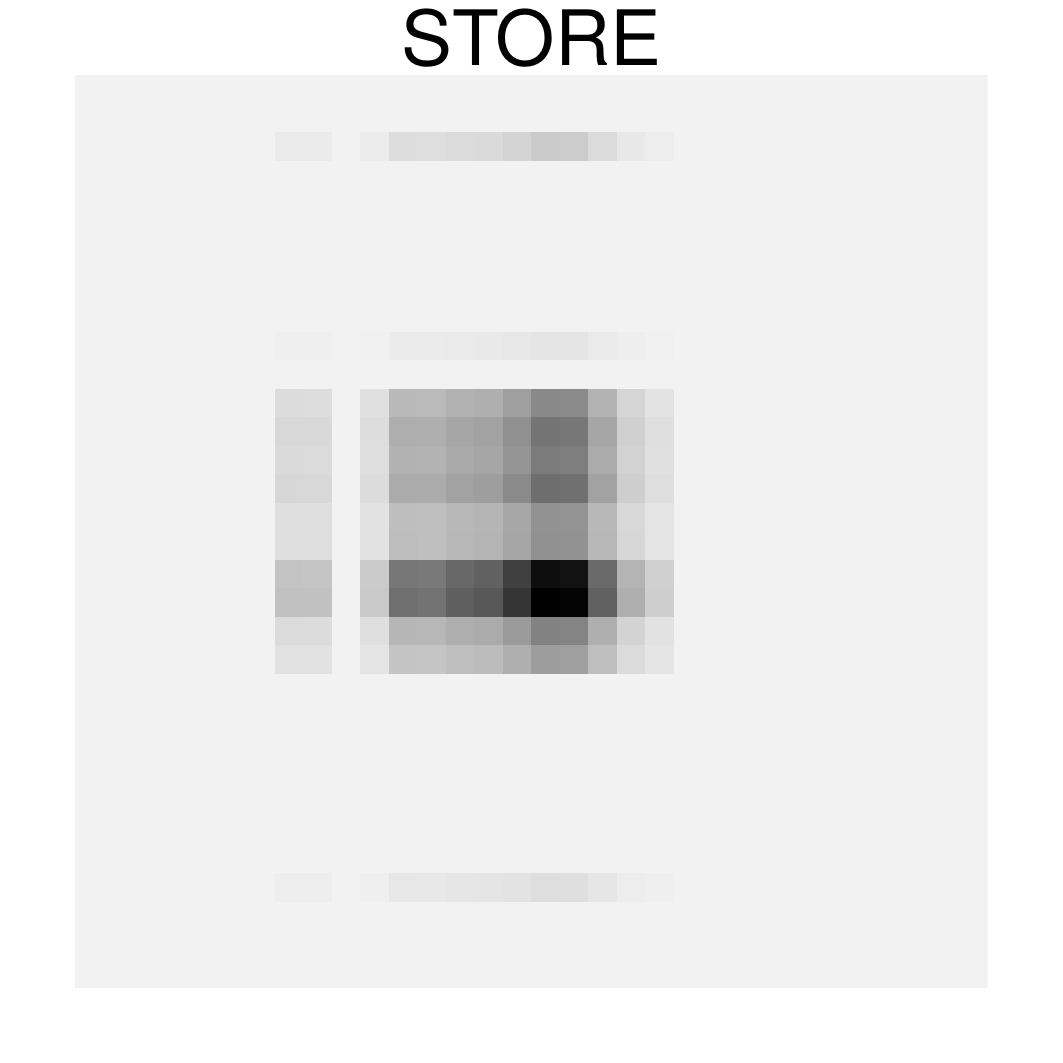}
	\includegraphics[scale=0.15]{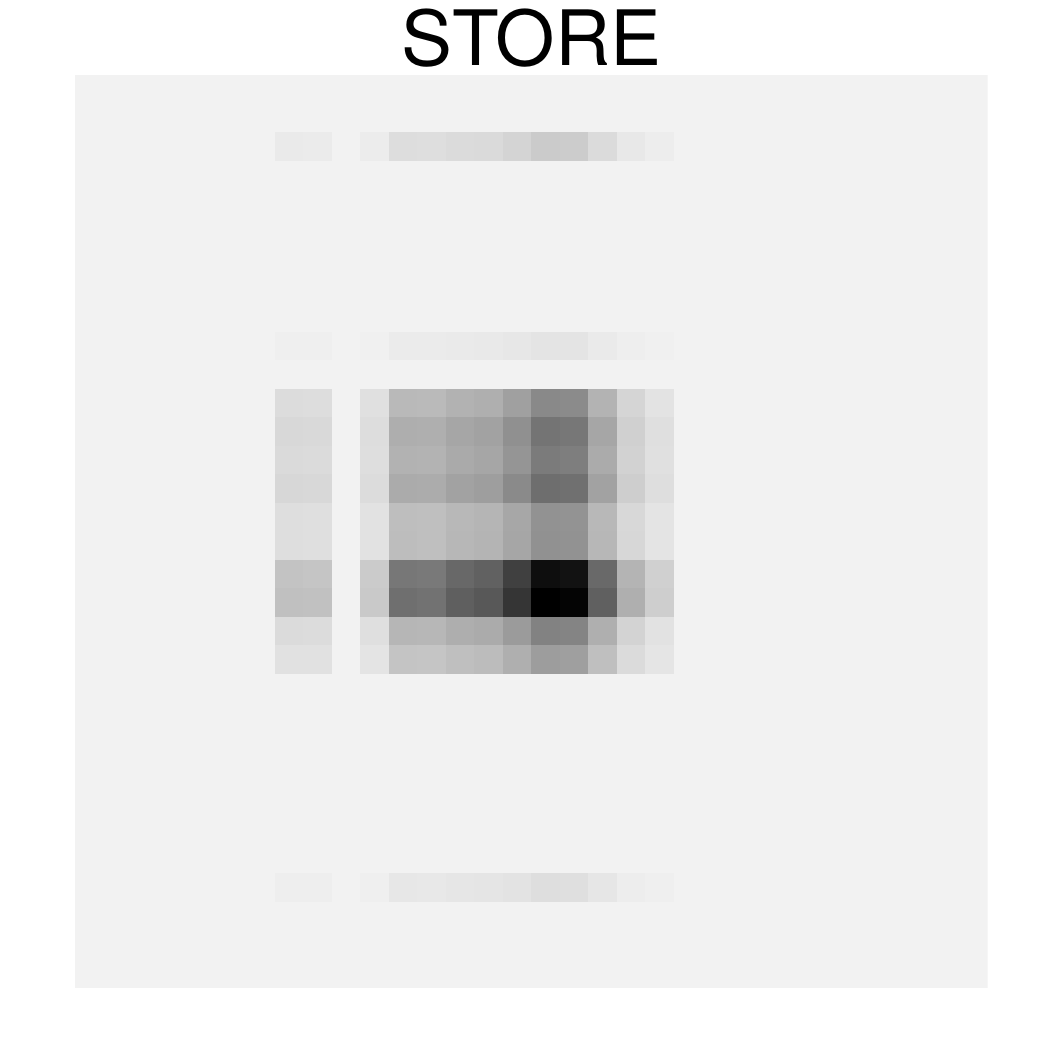}
	\caption{Sparsity pattern recovery results for M3. Reported values are the absolute values of the estimated coefficients. The background light gray represents zero, the larger the absolute value is, the darker the pixel is.}
	\label{recover1}
\end{figure}

\end{document}